\documentclass [a4paper]{article}

\usepackage[english]{babel}
\usepackage{graphicx}
\usepackage{amsmath}
\usepackage{amsthm}
\usepackage{amssymb}
\usepackage{mathtools}
\usepackage{enumitem}
\usepackage[utf8]{inputenc}
\usepackage{jheppub}
\usepackage{tensor} 
\usepackage{blkarray} 
\usepackage{slashed}
\usepackage{upgreek}
\usepackage{dutchcal}
\usepackage{upgreek} 
\usepackage[mathscr]{eucal}
\usepackage{xcolor} 
\usepackage{hyperref} 
\usepackage{tikz-cd}
\newcommand{\on}{\operatorname}
\newcommand{\tf}{\tfrac}
\DeclareMathOperator{\gr}{\mathcal{R}}
\newcommand{\di}{\slashed D}
\newcommand{\inv}[1]{^{-#1}}

\newcommand{\minot}{-1 \leftrightarrow 2}
\newcommand{\mc}{\mathcal}
\newcommand{\ms}{\mathscr}
\newcommand{\mb}{\mathbb}
\newcommand{\mf}{\mathfrak}
\newcommand{\ol}{\overline}

\newcommand{\slot}{\;\cdot\;}
\newcommand{\gm}{{\mc G}} 
\newcommand{\la}{\langle}
\newcommand{\ra}{\rangle}
\newcommand{\dt}{\left.\tfrac{d}{dt}\right|_{t=0}}
\renewcommand{\above}[1]{\stackrel{\mathclap{#1}}{=}}
\newcommand{\lbl}[1]{\stepcounter{equation}\tag{\theequation}\label{#1}}

\newcommand{\gld}{\ms L}
\newcommand{\ld}{L}
\newcommand{\G}{\mathcal{G}}
\newcommand{\dil}{\sigma}
\newcommand{\dilino}{\rho}
\newcommand{\Gino}{\psi}
\newcommand{\eom}[1]{ E\hspace{-.5mm}L[#1] }
\newcommand{\inner}[1]{\langle #1 \rangle}
\newcommand{\Ginobild}[2]{(\bar\Gino_{#1}\gamma_{#2}\Gino^{#1})}
\newcommand{\ginobilu}[2]{(\bar\Gino_{#1}\gamma^{#2}\Gino^{#1})}

\newcommand{\connection}{\Gamma}
\newcommand{\Spin}{S(C_+)}
\newcommand{\Tr}{\on{Tr}}
\newcommand{\Ric}{\ms R}
\newcommand{\af}[1]{{#1^*}}
\newcommand{\pb}[1]{\{#1\}}
\newcommand{\bv}{Q_{BV}}
\newcommand{\brst}{Q_{B}}
\newcommand{\vari}[2]{\frac{\delta #1}{\delta #2}}
\newcommand{\cmeTerm}[2]{\vari{#1}{\af{\phi_i}} \vari{#2}{\phi^i}}

\title{Geometry of supergravity and the Batalin--Vilkovisky formulation of the $\ms N=1$ theory in ten dimensions}
\author[a]{Julian Kupka,}
\emailAdd{j.kupka@herts.ac.uk}
\author[a]{Charles Strickland-Constable,}
\emailAdd{c.strickland-constable@herts.ac.uk}
\author[b]{and Fridrich Valach}
\emailAdd{fridrich.valach@matfyz.cuni.cz}
\affiliation[a]{Department of Physics, Astronomy and Mathematics,
University of Hertfordshire, College Lane, Hatfield, AL10 9AB, United Kingdom}
\affiliation[b]{Mathematical Institute, Faculty of Mathematics and Physics, Charles University, Prague 186 75, Czech Republic}
\abstract{
We provide full details of a BV formulation of $\ms N=1$ supergravity in ten dimensions, to all orders in fermions, built from the generalised geometry description of the theory. 
In contrast to standard treatments, 
we introduce neither the degrees of freedom corresponding to orthonormal frames for the metric nor the local Lorentz symmetries that remove them again. 
Instead, we observe that the field space has a fibred structure, with the fermionic degrees of freedom spanning the fibres. 
We explain in detail how this geometric picture allows one to understand simultaneous variations of spinorial quantities and the metric with respect to which the spinor bundles are defined. 
This leads to additional terms in certain commutators on field space which account for the Lorentz transformation terms appearing in the calculation of the supersymmetry algebra.
Unencumbered by the Lorentz degrees of freedom, 
we provide an efficient and full demonstration that 
our action satisfies the classical master equation.}

\allowdisplaybreaks

\begin{document}
\maketitle

\section{Introduction}
$\ms N = 1$ supergravity in ten dimensions coupled to super Yang--Mills theory plays an important role in the patchwork of superstring and supergravity theories. It describes the low-energy behaviour of heterotic and Type I superstring theories for appropriate choices of gauge group (see e.g.~\cite{Green:1987mn}), and gives rise to many lower-dimensional supergravity theories, with up to half-maximal supersymmetry, through dimensional reduction. 
The bosonic field content of the theory consists of the metric $g$, Kalb--Ramond two-form $B$, dilaton $\varphi$ and gauge fields $A$, the latter providing much phenomenological interest as a possible higher-dimensional origin of the gauge fields of the standard model during the first superstring revolution~\cite{Candelas:1985en}. 
These are accompanied by their fermionic superpartners: the spin-$\frac32$ gravitino $\uppsi$ and spin-$\frac12$ dilatino $\uprho$ and gauginos $\upchi$. 

Although the theory was introduced four decades ago \cite{Bergshoeff:1981um, Chapline:1982ww, Dine:1985rz}, it is only more recently that its hidden (generalised) geometric structure has been appreciated~\cite{Coimbra:2014qaa}, with the bosonic fields being combined into a generalised metric tensor $\gm$ and a density $\sigma^2$, while the fermions naturally become spinor fields $\rho$, $\psi$ associated to it, the latter combining both the gravitino and gauginos into a single geometric object. 
However, the structure of the action and supersymmetry transformations has only been understood at the lowest order in fermions, while the higher-order fermion terms remained elusive.\footnote{Though see~\cite{Jeon:2012hp} for a double field theory approach to higher-order fermion terms in the type II case.} 
Recently \cite{Kupka:2024xur, Kupka:2024vrd}, 
we showed that the framework of generalised geometry also gives a much simpler picture of these terms, and the full action for $\ms N = 1$, $D=10$ supergravity coupled to super Yang--Mills theory can be concisely written as
\begin{equation}
      \begin{aligned}
        S_0 &= \smash{\int_M} \mc R \dil^2 + \bar \Gino_{\alpha} \slashed D \Gino^{\alpha}+\bar\dilino \slashed D\dilino + 2\bar\dilino D_{\alpha} \Gino^{\alpha} - \tf1{768} \dil\inv2 (\bar\Gino_{\alpha}\gamma_{abc} \Gino^{\alpha}) (\bar\dilino \gamma^{abc} \dilino)\\
        &\qquad\qquad-\tf1{384}\dil^{-2}(\bar\Gino_{\alpha} \gamma_{abc} \Gino^{\alpha}) (\bar\Gino{\beta} \gamma^{abc} \Gino^{\beta}),
      \end{aligned}
    \end{equation}
featuring only two higher-order terms. (For notation and more details see section \ref{sec:sugra-all-orders}.) This was shown to be invariant under local supersymmetry transformations, which similarly only feature three higher-order fermion terms.

In this work, we present a key application of these results: the Batalin--Vilkovisky (BV) formulation of $\ms N=1$, $D=10$ supergravity. Thus far, most works on the BV formulation of supergravity have been restricted to the $D=4$ (see \cite{Baulieu:1990uv,Cattaneo:2025wbz}) and $D=3$ (see \cite{Cattaneo:2024sfd}) cases, whereas, to the best of our knowledge, the BV action for higher-dimensional supergravity in the component formalism\footnote{For a pure spinor superfield formulation (around flat space) see \cite{Cederwall:2010tn}.} has not yet appeared in the literature.
Our approach follows the insights gained from the BV analysis of a simplified supergravity model dubbed ``dilatonic supergravity" \cite{Kupka:2024tic}. 
This purely topological model is obtained by taking the generalised-geometric formulation of supergravity and choosing the generalised metric to lie in a specific --- namely trivial --- orbit of the structure group. 
Since this admits no pertubations, the procedure effectively removes all generalised metric (and gravitino) degrees of freedom, while leaving other formal aspects of the theory (such as local supersymmetry or the on-shell closure of the algebra) intact. The resulting BV formulation of the dilatonic theory thus displays most of the features expected from a supergravity theory.

Nevertheless, passing from the analysis of this topological theory to the full physical theory is quite non-trivial. One complication stems from the more subtle structure of the field space. Usually, supergravity is formulated in a vielbein formalism, whereas we are working in a full (generalised) metric formalism and hence Lorentz transformations are \emph{not} part of our gauge symmetries. On one hand this eliminates the need for (anti-)ghosts corresponding to such transformations, while on the other hand it necessitates a more geometric view of the fermionic spinor degrees of freedom as sections of suitable spinor bundles. Since the very definition of these spinor bundles relies on an underlying choice of generalised metric, one is forced to regard the field space as (the total space of) the vector bundle
\begin{equation}\label{bundle}
  \text{space of all supergravity fields}\to\text{space of bosonic fields},
\end{equation}
whose fibres are given by the possible fermionic configurations for a given bosonic one, the latter including the choice of a generalised metric.\footnote{In fact this bundle structure becomes even more pronounced in the generalised-geometric formulation, as the vector index of the field $\psi$ is valued in the bundle $C_-$ which also depends on the generalised metric.} Thus the fermionic fields $\rho$ and $\psi$ do not correspond to coordinates of the field space but only to coordinates on the individual fibres of the bundle \eqref{bundle}. 

As usual, symmetry transformations (such as local supersymmetry) correspond to vector fields on the field space, which is now the total space of \eqref{bundle}. However, in order to write this vector field explicitly in terms of $\delta \gm$, $\delta \sigma$ and $\delta \rho$, $\delta\psi$, respectively, one needs to split the tangent space at a given field configuration $(\gm,\sigma,\rho,\psi)$ into a horizontal and vertical part (in the sense of the vector bundle \eqref{bundle}). This splitting is provided by a natural connection, described in section \ref{subsec:field-space}. Importantly though, this connection is \emph{not} flat but rather has its curvature proportional to the (algebraic) commutator of generalised metric deformations, schematically
\begin{equation}
  F(\delta_1\G, \delta_2 \G) \sim [\delta_1\G, \delta_2\G],
\end{equation}
which we calculate in appendix \ref{app:curv}. Since the symmetry transformations of the fermions are described by the vertical parts of the corresponding symmetry vector fields, in calculating the commutator of two symmetries acting on fermions one picks up an extra term involving the curvature (see appendix \ref{app:commutator}), which precisely cancels the Lorentz-transformation-looking terms which seemingly emerge in the calculation.

Using these results we calculate the symmetry algebra of the $\ms N=1$ supergravity (see section \ref{sec:closure} and appendix \ref{sec:closurecomputation}) and in particular obtain the following remarkably simple form of the supersymmetry algebra on fermions:
\begin{align}
  \begin{aligned}
    [\delta_{\epsilon_1}, \delta_{\epsilon_2}] \dilino &= \delta_{V}\dilino + \delta_{\zeta}\dilino -\tf12 \slashed V \eom{\dilino}, \\
    [\delta_{\epsilon_1}, \delta_{\epsilon_2}] \Gino_\alpha &= \delta_{V}\Gino_\alpha + \delta_{\zeta}\Gino_\alpha + (\tf14 \epsilon_{[2} \bar \epsilon_{1]} - \tf12 \slashed V ) \eom{\Gino}_\alpha,
  \end{aligned}
\end{align}
where the generalised diffeomorphism parameter $V$ and the supersymmetry parameter $\zeta$ are given in terms of $\epsilon_{1/2}$ and $\dilino$ in \eqref{vandzeta}, and $\eom{\slot}$ are the expressions \eqref{eq:fieldeom} whose vanishing constitutes the equations of motion of the corresponding field.

These simplifications carry on to the BV analysis, and together with the insights from the dilatonic supergravity \cite{Kupka:2024tic} lead to the following BV form of the $\ms N=1$ supergravity, proposed in our recent work \cite{Kupka:2025hln}:
\begin{equation}\label{introaction}
  \begin{aligned}
    S&=\smash{\int_M}\mc R\sigma^2+\bar\psi_{\alpha}\slashed D\psi^{\alpha}+\bar\rho\slashed D\rho+2\bar\rho D_{\alpha}\psi^{\alpha}\\
      &\qquad\quad-\tfrac1{768}\sigma^{-2}(\bar\psi_{\alpha}\gamma_{abc}\psi^{\alpha})(\bar\rho\gamma^{abc}\rho) - \tfrac1{384}\sigma^{-2}(\bar\psi_{\alpha}\gamma_{abc}\psi^{\alpha})(\bar\psi_{\beta}\gamma^{abc}\psi^{\beta})\\
      &\qquad\quad+\sigma^*[\ms L_\xi \sigma-\tfrac18 \sigma^{-1}(\bar \rho e)]+\gm^*_{a\beta}[(\ms L_\xi\gm)^{a\beta}+\tfrac12\sigma^{-2}(\bar e\gamma^a\psi^{\beta})]\\
      &\qquad\quad+\bar\rho^*[\ms L_\xi \rho+\di e+\tfrac1{192}\sigma^{-2}(\bar\psi_\beta\gamma_{abc}\psi^\beta)\gamma^{abc}e]\\
        &\qquad\quad+\bar\psi^*_{\beta}[(\ms L_\xi \psi)^{\beta}+D^{\beta}e+\tfrac18\sigma^{-2}(\bar\psi^{\beta}\rho)e-\tfrac18\sigma^{-2}(\bar\psi^{\beta}\gamma_ae)\gamma^a\rho]\\
        &\qquad\quad+\bar e^*[\ms L_\xi e+\tfrac1{16}\sigma^{-2}(\bar e\gamma_a e)\gamma^a\rho] + \la \xi^*,\ms D\!f+\tfrac12\ms L_\xi\xi\ra-\tfrac18\xi^{*}_a\sigma^{-2}(\bar e\gamma^a e)\\
        &\qquad\quad+\tfrac12f^*(\ms L_\xi f+\tfrac18\sigma^{-2}(\bar e\gamma_a e)\xi^a-\tfrac16\la \xi,\ms L_\xi\xi\ra)\\
        &\qquad\quad-\tfrac1{64}\sigma^{-2}(\bar e\gamma_a e)(\bar\psi^*_\beta\gamma^a\psi^{*\beta})-\tfrac1{32}\sigma^{-2}(\bar e\psi^*_{\beta})(\bar e\psi^{*\beta})-\tfrac1{64}\sigma^{-2}(\bar e\gamma_a e)(\bar\rho^*\gamma^a\rho^*),
  \end{aligned}
\end{equation}
where $e$, $\xi$, and $f$ denote the ghosts for supersymmetry, generalised diffeomorphisms, and gauge-for-gauge transformations, respectively, and the antifields are denoted by a star.
We provide a full verification that this action indeed satisfies the classical master equation
\begin{equation}
  \{S,S\}=0.
\end{equation}
In doing so, we again need to address the subtleties originating from the fibred structure of the field space --- for instance, identifying the antifields with sections of appropriate bundles over the spacetime manifold (these are the starred variables above) again requires the use of a connection on the field space. Consequently, the canonical BV Poisson bracket, written out in terms of variables $\gm,\sigma,\rho,\psi,e,\xi,f,\gm^*,\sigma^*,\rho^*,\psi^*,e^*,\xi^*,f^*$ again features terms coming from the curvature (see formulas \eqref{eq:CME-with-curvature} and \eqref{eq:CME-with-curvature-concise} and appendix \ref{app:cotangent}) and these need to be taken into account when calculating $\{S,S\}$.

This paper is structured as follows: 
In section \ref{sec:sugra-all-orders} we give an introduction to the generalised geometry framework and review how it provides a formulation of $\ms N=1$, $D=10$ supergravity to all orders in fermions. Furthermore, we describe the fibred structure of the field space and its curvature and leverage it to compute the closure of the supersymmetry algebra. In section \ref{sec:BV-action} we clarify the structure of the space of BV fields and then present the BV action for $\ms N =1$, $D=10$ supergravity. Section \ref{sec:non-trivial-checks} details the calculations showing that this action indeed satisfies the classical master equation. 
In section~\ref{sec:conclusion} we summarise our results and comment on possible future directions. 
Numerous appendices contain a review of the general construction of BV actions as well as details of our conventions and derivations of technical results quoted in the main text. 


\section{Geometry of \texorpdfstring{$\ms N = 1$}{N=1} supergravity to all orders in fermions} \label{sec:sugra-all-orders}
In this section, we review the structure of $D=10$, $\ms N=1$ supergravity including higher-order fermion terms as found in \cite{Kupka:2024vrd,Kupka:2024xur}. We first introduce the notion of generalised geometry in the language of Courant algebroids, then clarify the makeup of the field space and finally present the full action and its local supersymmetry algebra.
\subsection{Courant algebroids}\label{sec:gen-geo}
Generalised geometry is best captured via the notion of \textit{Courant algebroids} \cite{liu1997manin, Severa:1998let}, which naturally encodes the structure of the bosonic symmetries of supergravity. A Courant algebroid over a manifold $M$ is defined by
\begin{itemize}
  \item a vector bundle $E \to M$, whose elements are called \emph{generalised vectors}
  \item an $\mathbb{R}$-bilinear bracket $[\slot, \slot]$ on the space of sections $\Gamma(E)$ of $E$
  \item a non-degenerate\footnote{The non-degeneracy allows us to identify $E^* \cong E$. We will use this in the remainder of the paper without further note.} symmetric bilinear pairing $\inner{\slot,\slot}$ on the fibres of $E$
  \item and the vector bundle map $a\colon E \to TM$ called the \textit{anchor map},
\end{itemize}
together with the conditions that for all sections $V, W, U \in \Gamma(E)$ and functions $f\in C^\infty(M)$
\begin{enumerate}[label=$\roman*$)]
  \item the bracket satisfies the \textit{Jacobi identity}: $[V, [W, U]] = [[V, W], U] + [W, [V, U]]$
  \item the bracket has a derivative property on functions: $[V, fW] = f[V,W] + (a(V)f)W$
\item the bracket preserves the pairing: $a(V)\inner{W,U} = \inner{[V, W],U} + \inner{W, [V, U]}$
\item the symmetric part of the bracket is given by the pairing: $[V, W] + [W, V] = \ms D \inner{V, W}$
\end{enumerate}
where we used the map $\ms D\colon C^\infty(M) \to \Gamma(E)$ defined by
\begin{equation} \label{eq:action-of-msD}
  \inner{V, \ms D f} \coloneqq a(V)f.
\end{equation}
Denoting the transpose of $a$ by $a^*$, it follows from the axioms that $a\circ a^* = 0$, allowing us to define the chain complex
\begin{equation}
  0 \longrightarrow T^*M \overset{a^*}\longrightarrow E \overset{a}\longrightarrow TM \longrightarrow 0.
\end{equation}
We call the Courant algebroid $E$ \textit{exact} if the above sequence is exact and \textit{transitive} if $a$ is surjective.
It is also easy to show that the anchor necessarily intertwines the brackets on $E$ and $TM$, i.e.
\begin{equation}
  a([V,W])=[a(V),a(W)].
\end{equation}

\subsection{Generalised Lie derivative}
The bracket defines a \textit{generalised Lie derivative} $\gld_V$ for any generalised vector $V\in \Gamma(E)$ acting on another vector $W\in \Gamma(E)$ as
\begin{equation}
  \gld_VW \coloneqq [V, W]
\end{equation}
and on functions $f\in C^\infty$ by the ordinary Lie derivative
\begin{equation}\label{gldonfunctions}
  \gld_V f \coloneqq \ld_{a(V)}f .
\end{equation}
Due to the pairing, we have $E^*\cong E$ and hence the above definition extends to arbitrary generalised tensors $\Gamma(E^{\otimes n})$ by requiring $\gld_V$ to satisfy the Leibniz rule. Integrating the action of the generalised Lie derivative we obtain \emph{generalised diffeomorphisms} (also called \emph{Courant algebroid automorphisms}).

One can in fact extend the action of the generalised Lie derivative to an even wider class of objects. We will discuss next the case of $\nu$-densities, and in a little while we shall see how $\ms L_V$ can be defined for the spinorial supergravity fields.

To finish, we note that it follows from the axioms that the generalised Lie derivative has a non-trivial kernel since
\begin{align}
  \gld_{\ms D f} = 0, \quad \forall f \in C^\infty(M). \label{eq:diffeo-kernel}
\end{align}
This kernel is responsible for the appearance of a ghost for ghost in the BV formulation, since the generalised diffeomorphisms defined by $V$ and $V+\ms D f$ act equivalently. 

\subsection{\texorpdfstring{$\nu$}{nu}-densities}
Recall that \emph{$\nu$-densities} on $M$ (for $\nu\in\mb R$) are sections of the line bundle which is associated to the tangent bundle $TM$ via the representation $A\mapsto |\det(A)|^{-\nu}$. In other words, any choice of local coordinates on $M$ produces a particular $\nu$-density, and changing the coordinates results in multiplying this density by $|\on{Jac}|^{-\nu}$. As a consequence of the absolute value, the bundle of $\nu$-densities is always trivial, though in general the trivialisation is non-canonical. Furthermore, there is a well-defined notion of a positive/negative half-density.

Because of the link between this line bundle and $TM$, the (ordinary) Lie derivative admits an action on $\nu$-densities. Following \eqref{gldonfunctions}, we thus also have an action of the generalised Lie derivative on any $\nu$-density $\mu$ by
\begin{equation}
  \gld_V \mu \coloneqq \ld_{a(V)}\mu .
\end{equation}

The product of a $\nu_1$-density and a $\nu_2$-density is a $(\nu_1+\nu_2)$-density. The most important for us will be $\nu=1$ and $\nu=\frac12$, in which cases we talk about \emph{densities} and \emph{half-densities}, respectively. Densities are natural objects over which one can integrate (assuming either a compact support or a fast-enough fall-off at infinity), and this is true even on not orientable manifolds. Correspondingly, there is a natural inner product on half-densities, $\sigma_1,\sigma_2\mapsto \int_M\sigma_1\sigma_2$. We will denote the line bundle of half-densities over $M$ by
\begin{equation}
  H\to M
\end{equation}
and the space of its everywhere nonvanishing sections will be denoted
\begin{equation}
  \ms H^*:=\{\sigma\in\Gamma(H)\mid \sigma \text{ everywhere nonzero}\}.
\end{equation}

\subsection{Local model of generalised geometry}\label{subsec:localmodel}
Ten-dimensional $\ms N=1$ supergravity coupled to super-Yang--Mills is captured by the transitive Courant algebroid case, which is locally modeled \cite{Severa:1998let} over a contractible open set $U\subseteq M$ by
\begin{equation}\label{loc}
  TU \oplus T^*U \oplus (\mathfrak g \times U),
\end{equation}
where $\mathfrak g$ is a Lie algebra with an invariant bilinear symmetric pairing, which we will denote simply $\Tr$. The rest of the structure is given by
\begin{align}
    a(v + \lambda + s) &\coloneqq v, \nonumber\\
    \inner{v_1 + \lambda_1 + s_1, v_2 + \lambda_2 + s_2} &\coloneqq \lambda_1(v_2) + \lambda_2(v_1) + \Tr(s_1 s_2), \\
    [v_1 + \lambda_1 + s_1, v_2 + \lambda_2 + s_2] &\coloneqq \ld_{v_1}v_2 + (\ld_{v_1} \lambda_2 - i_{v_2}d\lambda_1 + \Tr(s_2 d s_1)) + (\ld_{v_1} s_2 - \ld_{v_2} s_1 + [s_1, s_2]_{\mathfrak g}).\nonumber
\end{align}
Here, we have written the generalised vectors locally as a formal sum of a vector field, a 1-form field and a $\mf g$-valued function, and $\ld$ is the usual Lie derivative. These local models can then be glued together by Courant automorphisms to give interesting global models. One can in particular construct transitive Courant algebroids associated to principal $G$-bundles with vanishing first Pontryagin class.\footnote{Generalising the framework to setups with non-vanishing Pontryagin class requires working within the framework of shifted symplectic geometry. We are grateful to Pavol \v Severa, from whom we learned this fact.}

\subsection{Bosonic fields}
We will now describe the bosonic field content of the $\ms N=1,$ $D=10$ supergravity in terms of structures on a Courant algebroid $E\to M$. We start with a \textit{generalised metric}, which is defined to be a vector bundle endomorphism $\G\colon E \to E$ that is symmetric and satisfies $\G^2 = \on{id}$. Consequently (or equivalently), we have an orthogonal decomposition
\begin{equation}
  E = C_+ \oplus C_-  
\end{equation}
into $\pm1$ eigenbundles of $\G$. In addition we will require the following conditions on $\gm$ (or equivalently $C_+$): 
\begin{enumerate}[label=$\roman*)$]
  \item the induced inner product $\inner{\cdot,\ \cdot}|_{C_+}$ has signature $(1,9)$ \label{ite:signature}
  \item $C_+$ admits a spin structure\footnote{Strictly speaking, one should include the choice of the spin structure as part of the field data. We will ignore this minor point here.} \label{ite:spin-struct}
  \item the anchor restricted to $C_+$ is an isomorphism, i.e.\ $a|_{C_+}\colon C_+ \smash{\overset{\cong}\longrightarrow} TM$. \label{ite:iso-a}
\end{enumerate}
The assumption \ref{ite:spin-struct} ensures the existence of a spinor bundle $\Spin$ to encode the fermionic fields, and the assumption \ref{ite:signature} identifies the correct spin representations appropriate to the $\ms N=1$, $D=10$ theory.

The last assumption \ref{ite:iso-a} is not necessary for the setup below (including the generalised-geometric form of the symmetries and the BV action) to work and make sense;\footnote{In fact, one can even relax slightly the signature requirement on $C_+$, either to $(5,5)$ or $(9,1)$.} it is only important in order to recover the standard bosonic field content of the $\ms N=1$, $D=10$ theory. Other setups are thus also meaningful and can lead to interesting modifications and toy models, cf.\ the case of \emph{dilatonic supergravity} when $\gm=1$ \cite{Kupka:2024tic}. However, in order to not stray afar from our main theme, in the present work we will from now on require condition \ref{ite:iso-a} as well.

To see how to recover the standard field content, we note that \ref{ite:iso-a} in conjunction with \ref{ite:signature} imply both that $\dim M=10$ and that the Courant algebroid is transitive. Following \cite{Coimbra:2014qaa}, any $C_+$ in a transitive Courant algebroid can be written in the local model from subsection \ref{subsec:localmodel} as
\begin{align} \label{eq:C-plus}
  C_+ = \{ v + (i_vg + i_vB -\tf12 \Tr(Ai_vA)) + i_vA \ |\  v \in TM \},
\end{align}
where $g$ is a Lorentzian metric, $B\in \Omega^2(M)$ the Kalb--Ramond field, and $A\in \Omega^1(M, \mf g)$ a gauge field. The complement $C_-$ is then given by $C_- = (C_+)^\perp$.

There are now two (essentially equivalent) ways to include the dilaton $\varphi$:
\begin{itemize}
  \item following \cite{CSW1} one can enhance the generalised $O(p,q)$-structure given by $\inner{\cdot\ ,\ \cdot}$ to an $\mathbb{R}^+ \times O(p,q)$-structure by appropriate weightings by $\det T^*M$; the bosonic field content (including $\varphi$) then corresponds to the reduction of the structure group to $O(9,1)\times O(p-9,q-1)$
  \item following \cite{Siegel:1993th} one can encode the dilaton in terms of a half-density.
\end{itemize}
We will follow here the second approach. More concretely, we will supplement the generalised metric $\G$ by a nowhere vanishing half-density $\dil \in \ms H^*$. The standard dilaton $\varphi$ is then recovered via
\begin{align}
  \dil^2 = \sqrt{|g|} e^{-2\varphi}, \label{eq:standard-dilaton}
\end{align}
where $\sqrt{|g|}$ stands for the metric density. 

It will often be convenient to introduce local orthonormal frames of $E$ adapted to $C_+$ and $C_-$
\begin{equation}
  e_A = \{e_a,\ e_\alpha \},\quad \inner{e_A, e_B} = \eta_{AB},
\end{equation}
where $e_a$ is a frame of $C_+$ and $e_\alpha$ of $C_-$ and $\eta$ is given by an appropriate $\pm 1$-diagonal matrix. It is easy to check from the Courant algebroid axioms that the corresponding \emph{structure functions}
\begin{align} \label{eq:structure-functions}
 c_{ABC} \coloneqq \inner{[e_A, e_B], e_C}
\end{align}
are completely antisymmetric as consequence of the constancy of $\eta_{AB}$.

\subsection{Fermionic fields}
By our assumptions regarding the existence of spinors and the signature of $C_+$, we have Majorana--Weyl spinor bundles associated to $C_+$, denoted by $\Spin_\pm$.
The fermionic fields are given by the \emph{(generalised) dilatino} and \emph{gravitino} \cite{CSW1,Coimbra:2014qaa}
\begin{align} \label{eq:spinor-content}
  \dilino \in \Gamma(\Pi \Spin_+ \otimes H), \quad \Gino \in \Gamma(\Pi \Spin_- \otimes C_- \otimes H),
\end{align}
where $\Pi$ denotes the parity shift. Following \cite{Kupka:2024tic, Kupka:2025hln} (see also \cite{Bergshoeff:1988nn}), we have weighted the spinor fields by half-densities to simplify their supersymmetry transformations and the full action \eqref{eq:S0}.

Note in particular the form of the gravitino as a spinor over $C_+$ (i.e.\ admiting a Clifford action only by $C_+$ and \emph{not} $C_-$) together with a vector index in $C_-$ \cite{CSW1}. This combination is central to the findings in \cite{Kupka:2024vrd,Kupka:2024xur}, since it massively reduces the number of possible higher fermion terms in the action and supersymmetry variations compatible with generalised geometry.

To recover the fermionic fields in the standard description, note that assumption \ref{ite:iso-a} implies that in the local model we can identify $C_+ \cong TM$ and $C_- \cong TM \oplus (\mf g \times M)$. Using the choice of half-density \eqref{eq:standard-dilaton} we can then write (cf.\ \cite{Kupka:2024vrd})
\begin{equation}
  \dilino = \sqrt[4]{2}\dil \uprho, \quad \Gino = \sqrt[4]{2}\dil \uppsi + \tf1{\sqrt[4]{2}}\dil \upchi,
\end{equation}
where $\uprho$ is the standard positive chirality dilatino, $\uppsi$ the standard negative chirality vector-spinor gravitino, and $\upchi$ the standard $\mf g$-valued negative chirality gaugino.\footnote{The factors of $\sqrt[4]{2}$ are conventional and chosen for later convenience.}

\subsection{Generalised connections and unique operators}
Here we again follow \cite{CSW1,Coimbra:2014qaa}. First, we introduce \emph{generalised connections} on $E$ as $\mathbb{R}$-bilinear maps
\begin{equation}
  \begin{split}
    D\colon \Gamma(E) \times \Gamma(E) &\to \Gamma(E)\\
    (V, W) &\mapsto D_V W,
  \end{split}
\end{equation}
satisfying
\begin{equation}
  D_{f V}W = f D_V W,\qquad D_V(fW)=fD_VW + (a(V)f) W, \qquad D\inner{\slot,\slot} =0
\end{equation}
for all functions $f\in C^\infty(M)$ and generalised vectors $V, W \in \Gamma(E)$. In the last equation we have, similarly to $\ms L_V$, extended the connection to arbitrary tensors by requiring $D_V$ to satisfy the Leibniz rule and to act on functions as
\begin{equation}
  D_Vf:=a(V)f.
\end{equation}
Note that this is not merely a connection on a vector bundle but a genuine generalisation thereof. 
Finally, we note that generalised connections also naturally act on $\nu$-densities via the prescription
\begin{equation}\label{connectionondensities}
  D_V\mu :=L_{a(V)}\mu-\nu(D_AV^A)\mu.
\end{equation}

As usually, after a choice of orthonormal frame $\{e_A\}$, we can define the connection coefficients $\Gamma$ via
\begin{equation}
  D_{e_A} e_B \coloneqq \connection\indices{_A^C_B} e_C.
\end{equation}
Due to the compatibility with the inner product, these are the components of an $\mf o(p, q)$-valued form, i.e.\ $\Gamma_{ABC} = \Gamma_{A[BC]}$.

Given a choice of generalised metric $\G$ and half-density $\dil$, we define a \emph{Levi-Civita connection for $\G$ compatible with $\dil$} (sometimes denoted $D\in LC(\gm,\sigma)$) to be a generalised connection $D$ such that
\begin{align} \label{eq:GLC-1}
  D\G =0, \quad D\dil =0,
\end{align}
and $D$ is torsion free, i.e.\ in any orthonormal frame we have
\begin{align} \label{eq:GLC-2}
  \connection_{[ABC]} = - \tf13 c_{ABC}.
\end{align}
Note that the compatibility with $\dil$ implies
\begin{align} \label{eq:divergence}
   \on{div} V \coloneqq D_AV^A =\dil\inv2 \gld_V \dil^2,
\end{align}
while the compatibility with $\G$ amounts to $D_V$ preserving $C_\pm$. Thus Levi-Civita connections naturally act also on sections of $\Spin$.

Importantly, Levi-Civita connections can always be constructed, but \emph{are not unique} \cite{Garcia-Fernandez:2016ofz} (the last part can be seen by a counting argument for the components of $\Gamma$ vs the number of equations in \eqref{eq:GLC-1} and \eqref{eq:GLC-2}).

The non-uniqueness can be used as a guiding principle to build interesting so-called \emph{unique operators} that do not depend on the undetermined connection components, i.e.\ they are operators constructed out of $\gm$, $\sigma$, and $D\in LC(\gm,\sigma)$ which nevertheless depend only on $\gm$ and $\sigma$ \cite{Siegel:1993th,CSW1,Coimbra:2014qaa,Garcia-Fernandez:2016ofz}. For instance, denoting the orthogonal projection of $V$ onto $C_\pm$ by $V_\pm$, the operators
\begin{equation}
  D_{V_+}\colon \Gamma(C_-)\to\Gamma(C_-),\qquad D_{V_-}\colon \Gamma(C_+)\to\Gamma(C_+)
\end{equation}
are independent of the representative $D\in LC(\gm,\sigma)$. In particular whenever acting on objects associated to $C_+$ (such as $\Spin$), the operator $D_{V_-}$ is again unique in the above sense. Similarly, the trace $D_AV^A$ is unique due to \eqref{eq:divergence}.

One can also construct less obvious unique operators. For example, when acting on
\begin{equation}\label{fermionbundles}
  \Gamma(\Spin \otimes H)\quad \text{or} \quad\Gamma(\Spin\otimes C_-\otimes H),
\end{equation}
the Dirac operator 
\begin{equation}
  \di:=\gamma^aD_a
\end{equation}
is unique.\footnote{In fact, this operator turns out to only depend on $\gm$, i.e.\ it is independent of $\sigma$. Note that this $\sigma$-independence requires the half-density line bundle $H$ to be present in \eqref{fermionbundles}.} (Recall that $e_A$, $e_a$, and $e_\alpha$ denote the orthonormal frames of $E$, $C_+$, and $C_-$, respectively). Thus, on the fermionic fields \eqref{eq:spinor-content} of supergravity, we have the unique operators
\begin{equation}
  \di \dilino,\quad \di \Gino_\alpha, \quad D_\alpha \dilino, \quad D_\alpha \Gino^\alpha.
\end{equation}
Furthermore, by composition we can build further unique objects such as the \emph{generalised Ricci tensor} $\Ric_{a\alpha}(\gm,\sigma)\in \Gamma(C_+\otimes C_-)$ given by
\begin{equation}
  [\di, D_\alpha] \dilino = \tf14 \Ric_{a \alpha} \gamma^a \dilino
\end{equation}
and the \emph{generalised Ricci scalar} $\Ric(\gm,\sigma)\in C^\infty(M)$ defined as
\begin{equation}
  (\di^2 + D_\alpha D^\alpha ) \dilino = -\tf18 \Ric \dilino.
\end{equation}
These generalised curvatures have been introduced and studied in detail in numerous works, see \cite{Siegel:1993th,CSW1,Garcia-Fernandez:2013gja,Coimbra:2014qaa,Garcia-Fernandez:2016ofz,Severa:2016lwc,Jurco:2016emw,sv2,Kupka:2024vrd}.

If we express the pair $(\G,\sigma)$ in terms of the usual fields $g,B,A,\varphi$ in the physically relevant case \eqref{eq:C-plus}, 
we get \cite{Coimbra:2014qaa}
\begin{equation}
  \int \dil^2 \Ric = \int \sqrt{|g|}e^{-2\varphi}(R + 4 |\nabla \varphi|^2 - \tf1{12}H_{\mu\nu\rho}H^{\mu\nu\rho} + \tf14 \Tr F_{\mu\nu}F^{\mu\nu}),
\end{equation}
which coincides with the bosonic part of the action of $\ms N=1$ supergravity.
\subsection{Structure of the classical field space} \label{subsec:field-space}
Here we expand upon the discussion of the field space structure from \cite{Kupka:2025hln}. We start by noting that the fields $\psi$ and $\rho$ are sections of bundles which are associated to a generalised metric $\gm$. This means that the classical field space has the structure of (the total space of) a vector bundle
  \begin{equation}\label{fieldspacebundle}
    \ms S_0\to \ms M\times \ms H^*,
  \end{equation}
  where the first factor $\ms M$ is the space of all generalised metrics on $E$, the second factor $\ms H^*$ is the space of everywhere non-vanishing half-densities, and the fibre of $\ms S_0$ at the point $(\gm,\sigma)$ is given by
  \begin{equation}
    \Gamma(\Pi \Spin_+ \otimes H)\oplus\Gamma(\Pi \Spin_- \otimes C_- \otimes H).
  \end{equation}
  Although this might seem like a minor point, we will soon see that it has important consequences, e.g.\ for conceptual understanding of the supersymmetry and generalised diffeomorphism field variations. Concretely, it shows that $\gm(m)$, $\sigma(m)$, $\rho(m)$, and $\psi(m)$, for $m\in M$, \emph{cannot} be regarded as coordinates on the field space. It is only $\gm(m)$ and $\sigma(m)$ which give coordinates on the base of $\ms S_0\to \ms M\times \ms H^*$, while $\rho(m)$ and $\psi(m)$ only correspond to coordinates along the \emph{individual} fibres. For instance, it is meaningless to talk about expressions such as the supersymmetry transformation formula $\delta \rho=\di \epsilon+\dots$ without invoking any further structure on the bundle $\ms S_0$.\footnote{After all, since supersymmetry and generalised diffeomorphisms deform the generalised metric, the deformed fermions will be sections of a different spinor bundle, which makes it impossible (without a further structure) to compare them to the original fermions, in order to give meaning to the infinitesimal difference $\delta\rho$.}

  Fortunately, there is an additional structure naturally present on the bundle $\ms S_0$, namely a connection. This allows us to identify infinitesimally close fibres and thus give concrete meaning to the (supersymmetry and generalised diffeomorphism) variation formulas. To describe this connection, we first note that generalised metric $\gm$ is completely determined in terms of its $+1$-eigenbundle $C_+$. Suppose now we deform $\gm$ infinitesimally, arriving at $\gm'=\gm+\delta\gm$ corresponding to $C_+'$. The orthogonal projection $E\to C_+$ now restricts to a map $C_+'\to C_+$. Since the change was infinitesimal, it is easy to see that this map is in fact an isometry (between $\la \slot,\slot\ra|_{C_+}$ and $\la \slot,\slot\ra|_{C_+'}$). The identification $C_+\cong C_+'$ then lifts to an identification of the corresponding spinor bundles $S_\pm(C_+)$ and $S_\pm(C_+')$. Similarly, we obtain a natural identification of $C_-$ and $C_-'$. Putting things together we thus have an identification of nearby fibres of the bundle $\ms S_0$, i.e.\ a connection.

  An important feature of this connection is that it has nonzero curvature. This is calculated in Appendix \ref{app:curv}, with the following result. Suppose we have two different infinitesimal changes of $\gm$, called $\delta_1\gm$ and $\delta_2\gm$. Then the curvature associates to these two directions a map of the fibre to itself. Concretely,
  \begin{align}
    \begin{aligned} \label{eq:field-space-curvatures}
      F(\delta_1\gm,\delta_2\gm)\rho&=\tfrac1{16}[\delta_1\gm,\delta_2\gm]_{ab}\gamma^{ab}\rho,\\
      F(\delta_1\gm,\delta_2\gm)\psi^\alpha&=\tfrac1{16}[\delta_1\gm,\delta_2\gm]_{ab}\gamma^{ab}\psi^\alpha+\tfrac14[\delta_1\gm,\delta_2\gm]^\alpha{}_\beta\psi^\beta.
    \end{aligned}
  \end{align}
  
  Another thing that the connection allows us to do is to give meaning to the generalised Lie derivative of the spinor fields, i.e.\ sections of $\Spin$. First, it should be stressed that the action of a generalised diffeomorphism $\Phi$ on spinors is unambiguous --- in other words $\Phi$ induces a natural map $S(C_+)\to S(\Phi(C_+))$.\footnote{This can be readily seen by viewing sections of $S(C_+)$ as equivariant maps from the bundle of $\gm$-orthonormal frames (which is the subbundle of the frame bundle of $E$) to the spinor representation; in order to produce a section of $S(\Phi(C_+))$ from a section of $S(C_+)$ we simply precompose this equivariant map with $\Phi^{-1}$, regarded as a map between the bundle of $\Phi(\gm)$-orthonormal and $\gm$-orthonormal frames.} In contrast, the definition of the generalised Lie derivative requires considering the difference between the original spinor field and the infinitesimally deformed spinor field, which is in general problematic as these are sections of two different bundles. It is precisely at this point that we use the above connection on the field space, identifying the two bundles. This results in the following easy formula for the generalised Lie derivative $\ms L_V$ acting on spinors $\chi\in\Gamma(\Spin)$: 
  \begin{equation} \label{eq:diffeo-on-spinor}
    \ms L_V\chi=D_V\chi+\tfrac12(D_aV_b)\gamma^{ab}\chi.
  \end{equation}
  This is the natural analogue of the classical construction \cite{kosmann1971derivees} and its geometric interpretation \cite{bourguignon1992spineurs} (for the generalised-geometric case with $\gm=1$ see \cite{Kupka:2024tic}). On spinor half-densities $\lambda\in\Gamma(\Spin\otimes H)$ one similarly gets\footnote{This follows easily from writing $\lambda=\chi\sigma$ and using \eqref{eq:diffeo-on-spinor} and \eqref{connectionondensities}.}
  \begin{equation} \label{eq:diffeo-on-spinor-half-density}
    \ms L_V\lambda=D_V\lambda+\tfrac12(D_aV_b)\gamma^{ab}\lambda+\tfrac12(D_A V^A)\lambda.
  \end{equation}

  The reason that the above approach via the connection on the field space is not usually encountered in the literature stems from the fact that most of the treatments of supergravity have been formulating the theory in terms of the vielbeine instead of the metric tensor (or generalised metric) directly. This even occasionally leads to the misleading ``common wisdom'' that `supergravity requires working with the vielbein degrees of freedom' --- which as we saw above and will further witness below is by no means the case. Although the vielbein approach does replace the bundle structure of the field space \eqref{fieldspacebundle} with a simpler product bundle and hence leads to a flat connection instead of the curved one, at the same time it unnecessarily inflates the symmetry structure of the theory due to the presence of redundant degrees of freedom and their accompanying local Lorentz symmetries. Apart from increasing the size of the BV space, this results in significantly more complicated and lengthy expressions in the symmetry algebra, e.g.\ compare the expression (4.18) for the commutator of two supersymmetries in the original work \cite{Bergshoeff:1981um} in the abelian case with the succinct formula \eqref{eq:closurefermions} below. As we will see, the simplicity of the present formalism arises in part precisely from taking the above bundle structure of the field space into account.

\subsection{Classical action and its symmetries}\label{action-symmetries}
In \cite{Kupka:2024xur, Kupka:2024vrd} it was found that the action of $\ms N = 1$ supergravity \cite{Bergshoeff:1981um, Chapline:1982ww, Dine:1985rz} to all orders in fermions can be compactly formulated using generalised geometry. On the field space $\ms S_0$ as described in section \ref{subsec:field-space} the action takes the form
\begin{equation} \label{eq:S0}
      \begin{aligned}
        S_0 &= \smash{\int_M} \mc R \dil^2 + \bar \Gino_{\alpha} \slashed D \Gino^{\alpha}+\bar\dilino \slashed D\dilino + 2\bar\dilino D_{\alpha} \Gino^{\alpha} - \tf1{768} \dil\inv2 (\bar\Gino_{\alpha}\gamma_{abc} \Gino^{\alpha}) (\bar\dilino \gamma^{abc} \dilino)\\
        &\qquad\qquad-\tf1{384}\dil^{-2}(\bar\Gino_{\alpha} \gamma_{abc} \Gino^{\alpha}) (\bar\Gino{\beta} \gamma^{abc} \Gino^{\beta}).
      \end{aligned}
    \end{equation}
    This is invariant (by construction) under generalised diffeomorphisms parametrised by $V\in\Gamma(E)$,
    \begin{equation}\label{symdiff}
      \delta_V \text{(field)}=\ms L_V\text{(field)}.
    \end{equation}
    Looking at the local identification \eqref{loc} we see that this unifies the infinitesimal diffeomorphisms (parametrised by a vector field), gauge transformations of the $B$-field (parametrised by a 1-form), and gauge transformations of $A$ (parametrised by a $\mf g$-valued function).
    
    Less trivially, \eqref{eq:S0} is also invariant under the local supersymmetry transformations \cite{Kupka:2024xur, Kupka:2024vrd} parametrised by $\epsilon\in\Gamma(\Pi \Spin_-\otimes H)$
\begin{equation}
  \begin{aligned} \label{eq:susyvariations}
        \delta_\epsilon \G_{ab} & =\delta_\epsilon\G_{\alpha\beta}=0,\quad \delta_\epsilon\G_{a\beta}=\delta_\epsilon\G_{\beta a}=\tfrac12\dil^{-2}\bar \epsilon\gamma_a\Gino_{\beta} \\
        \delta_\epsilon\dil& =\tfrac18\dil^{-1}(\bar\dilino\epsilon)                                                                                                              \\
        \delta_\epsilon\dilino& =\di\epsilon+\tfrac1{192}\dil^{-2}(\bar\Gino_{\alpha}\gamma_{cde}\Gino^{\alpha})\gamma^{cde}\epsilon\\
        \delta_\epsilon\Gino_{\alpha}    & =D_{\alpha}\epsilon+\tfrac18\dil^{-2}(\bar\Gino_{\alpha}\dilino)\epsilon+\tfrac18\dil^{-2}(\bar\Gino_{\alpha}\gamma_c\epsilon)\gamma^c\dilino.
    \end{aligned}
\end{equation}
  Both \eqref{eq:S0} and \eqref{eq:susyvariations} can be seen as extensions of the results from \cite{CSW1,Coimbra:2014qaa} to all orders in fermions.
  
  As explained in the preceding subsection, both \eqref{symdiff} and \eqref{eq:susyvariations} in fact involve the connection on $\ms S_0$. Concretely, ($\delta \gm$, $\delta \sigma$) and ($\delta\dilino$, $\delta\Gino$) express the horizontal and vertical parts, respectively, of the infinitesimal transformation $\delta$, seen as a tangent vector in the field space $\ms S_0$ given by $V$ or $\epsilon$ (see the figure, where the horizontal and vertical spaces are marked by blue and olive, respectively).
  
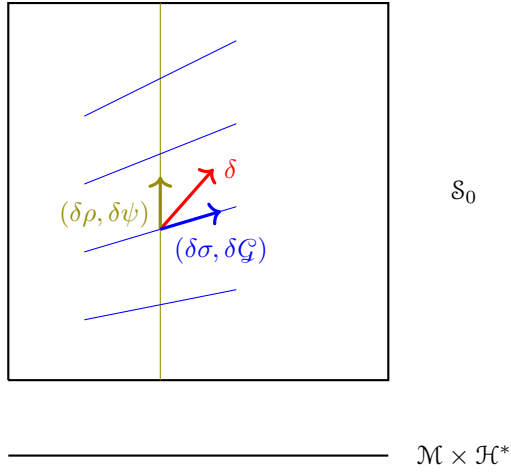
\begin{center}
  \begin{tikzpicture}
    \draw[thick] (0,0) -- (5,0) -- (5,5) -- (0,5) -- (0,0) node[xshift=6cm,yshift=2.5cm] {$\ms S_0$};
    \draw[thick] (0,-1) -- (5,-1) node[xshift=1cm] {$\ms M\times\ms H^*$};
    \draw[olive] (2,0) -- (2,5);
    \draw[blue] (2-1,1-.2) -- (2+1,1+.2);
    \draw[blue] (2-1,2-.3) -- (2+1,2+.3);
    \draw[blue] (2-1,3-.4) -- (2+1,3+.4);
    \draw[blue] (2-1,4-.5) -- (2+1,4+.5);     
    \draw[very thick,red,->] (2,2) -- (2.7,2.8) node[anchor=west] {$\delta$};
    \draw[very thick,olive,->] (2,2) -- (2,2.7) node[anchor=east,yshift=-.5cm] {$(\delta\rho,\delta\psi)$};
    \draw[very thick,blue,->] (2,2) -- (2.8,2.24) node[anchor=north,yshift=-.2cm] {$(\delta\sigma,\delta\gm)$};
  \end{tikzpicture}
  \end{center}
  
  We remark that with the help of \eqref{eq:fierzG1_1} one can rewrite the supersymmetry transformations of the fermions as
  \begin{equation}\label{rewriting}
    \begin{split}
        \delta_\epsilon \dilino &= \di \dilino + \tf14 \delta \G_{a \alpha} \gamma^a \Gino^\alpha,\\
        \delta_\epsilon \Gino_\alpha &= D_\alpha \epsilon + \tf18 \dil\inv2 (\bar \Gino_\alpha \dilino) \epsilon - \tf14 \delta \G_{\alpha a} \gamma^a \dilino,
    \end{split}
  \end{equation}
  which we will be very convenient in the calculation of the supersymmetry algebra in appendix \ref{app:commutator-fermoins}.
  
  Also note that $\delta\gm_{AB}$ should be strictly speaking read as $(\delta\gm)_{AB}$, i.e.\ we are comparing components of the deformed $\gm$ with the original $\gm$ in an orthonormal frame associated to the latter generalised metric. The vanishing of $\delta\gm_{ab}$ and $\delta\gm_{\alpha\beta}$ in \eqref{eq:susyvariations} and the fact that $\delta \gm_{a\beta}=\delta\gm_{\beta a}$ follows from the variation of the condition $\gm^2=1$ and symmetry of $\gm$. Thus we have a natural identification for the tangent space
  \begin{equation}\label{tangent_to_generalised_metrics}
    T_\gm\ms M\cong \Gamma(C_+\otimes C_-).
  \end{equation}
  
Finally, using in particular the variation formulas \eqref{eq:connection-on-spinor-variation}, the supergravity action \eqref{eq:S0} gives the equations of motion \cite{Kupka:2024vrd}
\begin{align*}
      \eom{\G}_{a\alpha} &\coloneqq\gr_{a\alpha}+\,\dil^{-2}(\tfrac12\bar\Gino_\beta\gamma_aD_\alpha\Gino^\beta+\bar\Gino_\alpha\gamma_aD_\beta\Gino^\beta-\bar\Gino_\beta\gamma_aD^\beta\Gino_\alpha+\tfrac12\bar\dilino\gamma_aD_\alpha\dilino-\tfrac12\bar\Gino_\alpha D_a\dilino\\
      &\qquad\qquad\qquad\qquad+\tfrac14\bar\dilino\gamma_{ab}D^b\Gino_\alpha-\tfrac14\bar\Gino_\alpha\gamma_{ab}D^b\dilino) = 0,\\
      \eom{\dil} &\coloneqq\gr+\,\dil^{-2}(2\bar\Gino^\alpha D_\alpha \dilino+2\bar\dilino D_\alpha \Gino^\alpha)+\dil^{-4}[\tfrac1{768}(\bar\Gino_\alpha\gamma_{cde}\Gino^\alpha)(\bar\dilino\gamma^{cde}\dilino) \lbl{eq:fieldeom}\\
      &\qquad\qquad\qquad\qquad+\tfrac1{384}(\bar\Gino_\alpha\gamma_{cde}\Gino^\alpha)(\bar\Gino_\beta\gamma^{cde}\Gino^\beta)]=0,\\
      \eom{\dilino} &\coloneqq\di\dilino+D_\alpha\Gino^\alpha-\tfrac1{768}\dil^{-2}(\bar\Gino_\alpha\gamma_{cde}\Gino^\alpha)\gamma^{cde}\dilino = 0,\\
      \eom{\Gino}^{\alpha}&\coloneqq\di\Gino^\alpha-D^\alpha\dilino-\dil^{-2}[\tfrac1{768}(\bar\dilino\gamma_{cde}\dilino)\gamma^{cde}\Gino^\alpha+\tfrac1{192}(\bar\Gino_\beta\gamma_{cde}\Gino^\beta)\gamma^{cde}\Gino^\alpha] = 0.
\end{align*}
The equations of motion play a vital role in the computation of the supersymmetry algebra, which only closes on-shell.

\subsection{Algebra of local symmetries} \label{sec:closure}
Essentially by construction one has
\begin{equation}\label{simplecommutators}
  [\delta_{V_1},\delta_{V_2}]=\delta_{\ms L_{V_1}V_2},\qquad [\delta_V,\delta_\epsilon]=\delta_{\ms L_V\epsilon}.
\end{equation}
As is usually the case, the most involved part of the algebra is the commutator of local supersymmetry transformations. We present the full details of the computation in Appendix \ref{sec:closurecomputation} and only state the results here for brevity.\footnote{It is in fact quite remarkable that the generalised geometry formalism allows us to display the full computation in just a few pages.} On the bosonic fields, the algebra closes into generalised diffeomorphism and a supersymmetry transformation as expected from a local supersymmetry algebra
\begin{equation}\label{susycommutatorbosonic}
  [\delta_{\epsilon_1}, \delta_{\epsilon_2}] = \delta_{V} + \delta_{\zeta}.
\end{equation}
We have introduced the (field-dependent) generalised vector $V\in\Gamma(C_+)$ and supersymmetry transformation $\zeta\in\Gamma(\Pi \Spin_-\otimes H)$
\begin{equation}\label{vandzeta}
  V^a \coloneqq \tfrac14 \dil^{-2} \bar \epsilon_2 \gamma^a \epsilon_1, \quad \zeta \coloneqq -\tfrac12 \slashed V \dilino + \delta_{\epsilon_1} \epsilon_2 - \delta_{\epsilon_2} \epsilon_1.
\end{equation}
The last two terms in the supersymmetry transformation are present because the parameters $\epsilon_{1/2}$ generally are field-dependent.
In other words, the general meaning of the symmetry parameters $V$ and $\epsilon$ is that they are functions on the field space $\ms S_0$, valued in the corresponding vector space $\Gamma(E)$ and $\Gamma(\Pi\Spin_-\otimes H)$, respectively.\footnote{That is, they are sections of the vector bundles over $\ms S_0$ whose fibres are given by $\Gamma(E)$ and $\Gamma(\Pi\Spin_-\otimes H)$. If the entire symmetry algebra closed off-shell and there were no ghosts-for-ghosts, the direct sum of these two vector bundles would form a Lie algebroid (its anchor map would encode the assignment of a vector on $\ms S_0$ to either $V$ or $\epsilon$).} The formulas \eqref{symdiff} and \eqref{eq:susyvariations} then describe (the horizontal and vertical parts of) a vector field on the field space $\ms S_0$, representing the symmetry in question. In particular, the commutator of symmetries is to be understood as the commutator of these vector fields.

On the fermions, the algebra only closes on-shell, i.e.\ up to equations of motion:
\begin{align}
  \begin{aligned} \label{eq:closurefermions}
    [\delta_{\epsilon_1}, \delta_{\epsilon_2}] \dilino &= \delta_{V}\dilino + \delta_{\zeta}\dilino -\tf12 \slashed V \eom{\dilino}, \\
    [\delta_{\epsilon_1}, \delta_{\epsilon_2}] \Gino_\alpha &= \delta_{V}\Gino_\alpha + \delta_{\zeta}\Gino_\alpha + (\tf14 \sigma^{-2}\epsilon_{[2} \bar \epsilon_{1]} - \tf12 \slashed V ) \eom{\Gino}_\alpha,
  \end{aligned}
\end{align}
where $\eom{\slot}$ denotes the equation of motion \eqref{eq:fieldeom} of the corresponding field. The appearance of the $\eom\slot$ terms implies that quadratic (and potentially also higher) antifield terms need to be present in the BV action (see appendix \ref{sec:review-BV}).

Note that if one were to compute the commutator of supersymmetries naively, by simply using the formulas \eqref{eq:susyvariations} twice in a row on $\rho$ or $\psi$ (which would correspond to taking the expressions in \eqref{dilatinocommutator} and \eqref{gravitino-start-derivation} with the last curvature term $F$ ignored), one would arrive at the right-hand side of \eqref{eq:closurefermions} with additional terms taking the form of local Lorentz transformations. However, we stress that we are working in a full metric formalism and thus Lorentz transformations are \emph{not} gauge symmetries of the action \eqref{eq:S0}.
The reason they seem to appear is because the naive calculation does not correctly account for the fact that the supersymmetry variations are to be understood as vector fields expanded with respect to a ``covariant" basis of horizontal and vertical vector fields on the fibred field space with respect to its natural connection, as in appendix~\ref{app:commutator}.  
The curvature term in the correct formula for the commutator~\eqref{commutator} written with respect to this basis then precisely cancels these Lorentz transformation terms, so that they do not appear in~\eqref{eq:closurefermions} in the end. 
They could be regarded as an artifact of the deformation of the spinor bundle $\Spin$ due to the variation of the generalised metric $\G$ and are not deformations of the spinor fields themselves.

\section{BV action of \texorpdfstring{$\ms N = 1$}{N=1} supergravity} \label{sec:BV-action}
In this section we present the full BV action for $D=10$, $\ms N = 1$ supergravity, expanding the results of \cite{Kupka:2025hln}. We first clarify the structure of the BV field space (extending the construction from subsection \ref{subsec:field-space}) and then write down the corresponding BV action. We will verify the classical master equation in the subsequent section.
\subsection{BV field space and classical master equation} \label{subsec:BV-field-space}
Following the general philosophy outlined in appendix \ref{sec:review-BV}, we construct the BV space $\ms F_{BV}$ in stages, starting from the physical field space $\ms S_0$ given by the bosonic fields $\G, \dil$ and the fermionic fields $\Gino, \dilino$. We introduce a degree-shifted ghost for each symmetry of our theory, namely the fermionic generalised diffeomorphism ghost $\xi$ and the bosonic ghost for supersymmetry $e$. Due to the non-trivial kernel of the action of generalised diffeomorphisms \eqref{eq:diffeo-kernel}, we also need to introduce the bosonic ghost for ghost $f$. From now on, we collectively call the physical fields in $\ms S_0$, together with the ghosts and ghosts for ghosts, simply \emph{fields}. Similarly to before, the space of fields forms the total space of the vector bundle
\begin{equation}
  \ms S \to \ms M \times \ms H^*,
\end{equation}
whose fibre at $(\G, \dil)$ is given by
\begin{equation}
  \Gamma(\Pi \Spin_+ \otimes H) \times \Gamma(\Pi \Spin_- \otimes C_- \otimes H) \times \Gamma(\Pi \Spin_- \otimes H)[1] \times \Gamma(E)[1] \times C^\infty(M)[2],
\end{equation}
where by $[n]$ we denote the degree shift, which corresponds to the ghost number. More explicitly, the field content is
\begin{equation}\label{fields}
  \begin{aligned}
    \G &\in \ms M, \\
    \dil &\in \ms H^*, \\
    \dilino &\in \Gamma(\Pi \Spin_+ \otimes H), \\
    \Gino &\in \Gamma(\Pi \Spin_- \otimes C_- \otimes H), \\
    \xi &\in \Gamma(E)[1], \\
    e &\in \Gamma(\Pi \Spin_- \otimes H)[1], \\
    f &\in C^\infty(M)[2].
  \end{aligned}
\end{equation}
To complete the BV space, we now need to include the conjugate ``momenta'' to the fields, the so-called \emph{antifields}. This is achieved by setting
\begin{equation}
  \ms F_{BV} \coloneqq T^*[-1] \ms S.
\end{equation}

More concretely, recall from subsection \ref{subsec:field-space} that there is a natural way to identify $C_+$ subbundles associated to ``nearby'' generalised metrics $\gm$. Just as before, this produces a connection on the vector bundle $\ms S\to \ms M\times\ms H^*$, and hence a splitting of $T\ms S$ into a horizontal and vertical part, i.e. 
\begin{equation}
  T\ms S\cong \pi^* T(\ms M\times \ms H^*)\oplus \pi^*\ms S,
\end{equation}
where $\pi\colon \ms S \to \ms M \times \ms H^*$ is the projection onto the base, and $\pi^* T(\ms M\times \ms H^*)$ and $\pi^*\ms S$ are pullbacks of the bundles $T(\ms M\times \ms H^*)\to \ms M \times \ms H^*$ and $\ms S\to \ms M \times \ms H^*$ along this projection. This in turn gives the identification
\begin{equation}
  T^*[-1] \ms S \cong \pi^* T^*[-1](\ms M \times \ms H^*) \oplus \pi^* \ms S^*[-1].
\end{equation}
We can thus describe the fibre of $T^*[-1]\ms S$ at a given configuration
\begin{equation}
  (\G, \dil, \Gino, \dilino, \xi, e, f)\in \ms S
\end{equation}
by the fields
\begin{equation}\label{antifields}
  \begin{aligned}
    \af\G &\in T^*_\G[-1] \ms M \cong \Gamma(C_+ \otimes C_- \otimes H^2)[-1], \\
    \af\dil &\in T^*_\sigma[-1]\ms H^*\cong\Gamma(H)[-1], \\
    \af\dilino &\in \Gamma(\Pi\Spin_- \otimes H)[-1], \\
    \af\Gino &\in \Gamma(\Pi\Spin_+ \otimes C_- \otimes H)[-1], \\
    \af\xi &\in \Gamma(E\otimes H^2)[-2], \\
    \af e &\in \Gamma(\Pi \Spin_+ \otimes H)[-2], \\
    \af f &\in \Gamma(H^2)[-3].
  \end{aligned}
\end{equation}
Note the presence of (the powers of) $H$, which arises due to the fact that the pairing of coordinates and momenta in the infinite-dimensional space of fields is realised via an integral, e.g.\ the pairing of $\xi$ and $\xi^*$ is $\int_M\xi^\alpha\xi^*_\alpha$.
The form of $\af\G$ follows from \eqref{tangent_to_generalised_metrics}.
Furthermore, we used the identifications
\begin{equation}
  E^*\cong E,\qquad C_{\pm}^* \cong C_{\pm},\qquad \Spin_{\pm}^* \cong \Spin_{\mp}.
\end{equation}

For future reference, we collect the fermion number $\mathbb{Z}_2$, the ghost number $\mathbb{Z}$, the spinor chiralities, and the overall parity of the BV fields in the table below
\begin{center}
  \begin{tabular}{c|c|c|c|c|c|c|c}
    & $\G$ & $\dil$ & $\dilino$ & $\Gino$ & $\xi$ & $e$ & $f$  \\
    \hline
    Chirality & &  & $+$ & $-$ & & $-$ &   \\
    \hline
    Fermion $\mathbb{Z}_2$ & $[0]$ & $[0]$ & $[1]$ & $[1]$ & $[0]$ & $[1]$ & $[0]$ \\
    \hline
    Ghost $\mathbb{Z}$ & 0 & 0 & 0 & 0 & 1 & 1 & 2 \\
    \hline
    Parity & even & even & odd & odd & odd & even & even \\
    \hline \hline
    & $\af\G$ & $\af\dil$ & $\af\dilino$ & $\af\Gino$ & $\af\xi$ & $\af e$ & $\af f$ \\
    \hline
    Chirality & &  & $-$ & $+$ & & $+$ &   \\
    \hline
    Fermion $\mathbb{Z}_2$ & $[0]$ & $[0]$ & $[1]$ & $[1]$ & $[0]$ & $[1]$ & $[0]$ \\
    \hline
    Ghost $\mathbb{Z}$ & $-1$ & $-1$ & $-1$ & $-1$ & $-2$ & $-2$ & $-3$ \\
    \hline
    Parity & odd & odd & even & even & even & odd & odd
  \end{tabular}
\end{center}

Lastly, let us describe the Poisson bracket on $\ms F_{BV}$. Although this comes canonically from the cotangent bundle structure of $T^*[-1]\ms S$, its concrete description in terms of fields \eqref{fields} and antifields \eqref{antifields} is slightly subtle (recall for instance that the very identification of antifields with sections of the appropriate bundles in \eqref{antifields} required invoking a connection). Since the Poisson structure corresponds to a bivector field on $\ms F_{BV}$, we are essentially forced to study the tangent space $T\ms F_{BV}$ and identify its constituents with sections of suitable vector bundles.

In order to achieve this, we need to again use the connection on $\ms S\to \ms M\times \ms H^*$, but now in addition also a connection on the tangent bundle of the base $\ms M\times \ms H^*$ (see the chain of identifications \eqref{identification-chain}). The latter we take to be the natural connection coming from the facts that the tangent space $T_\gm\ms M$ can be naturally identified with $\Gamma(C_+\otimes C_-)$ and we have an identification of ``nearby'' $C_\pm$; this induces a connection on $T(\ms M\times\ms H^*)$, which turns out to be torsion-free, see appendix \ref{app:connectiontangentspaceoffieldspace}.

With this in mind, the Poisson bracket takes the form
\begin{equation}
  \begin{aligned} \label{eq:CME-with-curvature}
        \frac12\{S,S\}&=\int_M\frac{\delta S}{\delta{\gm^*_{a\alpha}}}\frac{\delta S}{\delta{\gm^{a\alpha}}}+\frac{\delta S}{\delta{\sigma^*}}\frac{\delta S}{\delta\sigma}+\frac{\delta S}{\delta{\bar\rho^*}}\frac{\delta S}{\delta\rho} +\frac{\delta S}{\delta{\bar\psi^*_\alpha}}\frac{\delta S}{\delta{\psi^\alpha}}+\frac{\delta S}{\delta{\xi^*}}\frac{\delta S}{\delta\xi}+\frac{\delta S}{\delta{\bar e^*}}\frac{\delta S}{\delta e}+\frac{\delta S}{\delta{f^*}}\frac{\delta S}{\delta f}\\
        &\qquad\quad+\frac1{16}\frac{\delta S}{\delta \gm^*_{a\alpha}}\frac{\delta S}{\delta \gm^*_b{}^\alpha}(\bar \rho^*\gamma_{ab}\rho+\bar\psi^*_\alpha\gamma_{ab}\psi^\alpha+\bar e^*\gamma_{ab}e)+\frac14\frac{\delta S}{\delta \gm^*_{a\alpha}}\frac{\delta S}{\delta \gm^{*a}_\beta}(\bar\psi_\alpha^*\psi_\beta),
      \end{aligned}
    \end{equation}
or more simply as
\begin{equation} \label{eq:CME-with-curvature-concise}
    \frac12\pb{S,S} = \int_M \frac{\delta S}{\delta \phi^*_i} \frac{\delta S}{\delta \phi^i} + \frac12 \!\!\!\!\!\!\!\sum_{\phi \in \{\rho,\psi,e\}} \!\!\!\!\!\! \af{\bar{\phi}} F\left(\frac{\delta S}{\delta \af \gm}, \frac{\delta S}{\delta \af \gm}\right) \phi,
  \end{equation}
where the first term 
is simply the shorthand for the first line on the RHS of \eqref{eq:CME-with-curvature}.
In calculating the Poisson brackets on $\ms F_{BV}$ one can thus \emph{almost} treat \eqref{antifields} as conjugate to \eqref{fields}, except that one has to supplement the naive Poisson bracket with terms coming from the curvature of $\ms S$\footnote{It is for this reason that we adopted here the notation $\phi$, to distinguish it from $\Phi$ used in appendix \ref{sec:review-BV} (whose conjugate is simply $\Phi^*$).} (and in general also by terms corresponding to the torsion of the connection on the base, which however vanishes in the present case). See appendix \ref{app:cotangent} for a full derivation of this fact in the case of ordinary non-graded geometry, as well as appendix \ref{app:graded-conventions} for the details of our conventions.

\subsection{BV action of supergravity}
Since the algebra of supersymmetries \eqref{eq:closurefermions} only closes on-shell, it follows (see appendix \ref{sec:review-BV}) that the BV formulation of $\ms N=1$, $D=10$ supergravity needs to have rank at least 2. We claim that the rank is in fact precisely $2$ and the BV action has the following form \cite{Kupka:2025hln}:
\begin{align*}
    S&=\smash{\int_M}\mc R\sigma^2+\bar\psi_{\alpha}\slashed D\psi^{\alpha}+\bar\rho\slashed D\rho+2\bar\rho D_{\alpha}\psi^{\alpha}\\
      &\qquad\quad-\tfrac1{768}\sigma^{-2}(\bar\psi_{\alpha}\gamma_{abc}\psi^{\alpha})(\bar\rho\gamma^{abc}\rho)
      -\tfrac1{384}\sigma^{-2}(\bar\psi_{\alpha}\gamma_{abc}\psi^{\alpha})(\bar\psi_{\beta}\gamma^{abc}\psi^{\beta})\\
        &\qquad\quad+\sigma^*[\ms L_\xi \sigma-\tfrac18 \sigma^{-1}(\bar \rho e)]
        +\gm^*_{a\beta}[(\ms L_\xi\gm)^{a\beta}+\tfrac12\sigma^{-2}(\bar e\gamma^a\psi^{\beta})]\\
        &\qquad\quad+\bar\rho^*[\ms L_\xi \rho+\di e+\tfrac1{192}\sigma^{-2}(\bar\psi_\beta\gamma_{abc}\psi^\beta)\gamma^{abc}e]\\
        &\qquad\quad+\bar\psi^*_{\beta}[(\ms L_\xi \psi)^{\beta}+D^{\beta}e+\tfrac18\sigma^{-2}(\bar\psi^{\beta}\rho)e-\tfrac18\sigma^{-2}(\bar\psi^{\beta}\gamma_ae)\gamma^a\rho]\lbl{eq:bv-action}\\
        &\qquad\quad+\bar e^*[\ms L_\xi e+\tfrac1{16}\sigma^{-2}(\bar e\gamma_a e)\gamma^a\rho]
        +\la \xi^*,\ms D\!f+\tfrac12\ms L_\xi\xi\ra-\tfrac18\xi^{*}_a\sigma^{-2}(\bar e\gamma^a e)\\
        &\qquad\quad+\tfrac12f^*(\ms L_\xi f+\tfrac18\sigma^{-2}(\bar e\gamma_a e)\xi^a-\tfrac16\la \xi,\ms L_\xi\xi\ra)\\
        &\qquad\quad-\tfrac1{64}\sigma^{-2}(\bar e\gamma_a e)(\bar\psi^*_\beta\gamma^a\psi^{*\beta})-\tfrac1{32}\sigma^{-2}(\bar e\psi^*_{\beta})(\bar e\psi^{*\beta})-\tfrac1{64}\sigma^{-2}(\bar e\gamma_a e)(\bar\rho^*\gamma^a\rho^*).
\end{align*}
  We recall that $\sigma$, $\gm$, $\rho$, $\psi$ are the physical fields of supergravity, $e$ is the supersymmetry ghost, $\xi$ is the generalised diffeomorphism ghost, and $f$ is the ghost for ghost.
  
  Adhering to the general philosophy discussed in appendix \ref{sec:review-BV} we organised the expression in terms of its antifield structure, starting with the classical action $S_0$. This is followed by terms linear in $\G^*, \Gino^*, \dil^*, \dilino^*$, encoding the action of generalised diffeomorphisms \eqref{symdiff} and supersymmetry \eqref{eq:susyvariations}. Note the apparent partial sign discrepancy between the relevant terms in \eqref{eq:bv-action} and those in \eqref{eq:susyvariations} --- this arises from the fact that in the construction we are replacing symmetry parameters with ghosts of opposite parity; performing this transition carefully (cf.\ the first footnote in appendix \ref{sec:review-BV}) one lands with the relative signs in \eqref{eq:bv-action}.
  
  Next, we have terms linear in the antifields $e^*$, $\xi^*$, which encode the symmetry algebra: generalised diffeomorphisms act as generalised Lie derivatives and the commutator of two supersymmetries gives a supersymmetry and a generalised diffeomorphism with field-dependent parameters (the latter two correspond to the two terms containing $\bar e\gamma^a e$). The term $\int_M\la \xi^*,\ms D\!f\ra$ introduces a ghost for ghost $f$ to compensate for the reducibility of generalised diffeomorphisms due to the fact that $\ms L_{\ms D\!f}=0$.
  
  The terms linear in $f^*$ are perhaps slightly less intuitive, as they relate to the ghost-for-ghost. Nevertheless, their form can be guessed from the corresponding expression in dilatonic supergravity, i.e.\ from the case $\gm=1$ whose BV analysis was done in \cite{Kupka:2024tic}.
  
  The term quadratic in $\rho^*$ can similarly be deduced from \cite{Kupka:2024tic}. As explained above, this corresponds to the failure of the off-shell closure of the supersymmetry algebra, namely to the part of the first equation in \eqref{eq:closurefermions} proportional to the equations of motion. Finally, the quadratic $\psi^*$ terms reflect in an analogous fashion the second equation in \eqref{eq:closurefermions}. Note that for chirality reason there can be no $\rho^{*2}$ analogue of the $(\bar e\psi^*_{\beta})(\bar e\psi^{*\beta})$ term.
  
We stress that although highly suggestive, these arguments by themselves do not constitute a full proof of the fact that the action \eqref{eq:bv-action} does indeed satisfy the classical master equation, and in particular that no higher antifield terms are needed in the BV action. 
The full proof of the classical master equation
\begin{equation}
  \{S,S\}=0
\end{equation}
is presented in the following section and takes into account the full form of the Poisson bracket \eqref{eq:CME-with-curvature}.

\section{Classical master equation} \label{sec:non-trivial-checks}
  In this section we verify the classical master equation. We will organise the expressions according to the number of antifields
  \begin{equation}\label{starredbits}
    \sigma^*, \gm^*, \rho^*, \psi^*, e^*, \xi^*, f^*.
  \end{equation}
  For instance, the action \eqref{eq:bv-action} decomposes into
  \begin{equation}
    S = S_0 + S_{\text{lin}} + S_{\text{quad}}.
  \end{equation}
  Using the formula \eqref{eq:CME-with-curvature-concise} for the Poisson bracket we then get
\begin{alignat}{2}
  \tf12 \pb{S,S} &= \int_M \cmeTerm{S_{\text{lin}}}{S_0}\label{eq:CME-zeroth}\\
                 &\quad+  \int_M\cmeTerm{S_{\text{lin}}}{S_\text{lin}} + \cmeTerm{S_{\text{quad}}}{S_0} +\tf12 \!\!\!\!\!\!\!\!\sum_{\phi \in \{\rho,\psi,e\}} \!\!\!\! \af{\bar{\phi}} F\!\left(\frac{\delta S_{\text{lin}}}{\delta \af \gm}, \frac{\delta S_{\text{lin}}}{\delta \af \gm}\right) \phi \label{eq:CME-linear}\\
                 &\quad+  \int_M\cmeTerm{S_{\text{lin}}}{S_{\text{quad}}} + \cmeTerm{S_{\text{quad}}}{S_{\text{lin}}} \label{eq:CME-square}\\
                 &\quad+  \int_M\cmeTerm{S_{\text{quad}}}{S_{\text{quad}}}, \label{eq:CME-cubic}
\end{alignat}
  where as before $\phi^i$ runs over $\sigma, \gm, \rho, \psi, e, \xi, f$, and we wrote terms with the same number of antifields in the same line. Equations \eqref{eq:CME-zeroth}--\eqref{eq:CME-cubic} are the analogues of equations \eqref{justinvariance}--\eqref{consistency2}, taking in particular into account the more subtle form of the Poisson bracket; the subtlety however only manifests itself in the linear part \eqref{eq:CME-linear} of $\{S,S\}$ as $\af \gm$ appears only in $S_{\text{lin}}$.
  
  The vanishing of \eqref{eq:CME-zeroth} thus follows from the invariance of the classical action under the local symmetries, demonstrated in \cite{Kupka:2024vrd}. We also note that \eqref{eq:CME-cubic} vanishes due to the fact the only antifields contained in $S_{\text{quad}}$ are $\rho^*$ and $\psi^*$, neither of which pairs nontrivially with $\sigma$ or $e$, which are the only other ingredients in $S_{\text{quad}}$. It thus suffices to show the vanishing of the linear terms \eqref{eq:CME-linear} and quadratic terms \eqref{eq:CME-square}, to which we now turn.


\subsection{Some simplifications}
Before starting, it will prove useful to introduce the analogue of $V$ from appendix \ref{sec:closurecomputation}, i.e.\ we define $X\in\Gamma(C_+)$ by
\begin{equation}
  X^a = X^a(\dil, e) \coloneqq \tf14 \dil\inv2 (\bar e \gamma^a e).
\end{equation}
A quick computation reveals that $X$ indeed transforms as a $C_+$-vector,\footnote{Note that, once again, in writing $(\ms L_\xi X)^a$ we use the connection to identify $C_+$ and its $\xi$-deformed counterpart.}
\begin{equation}
  \begin{aligned}
    \{S,X^a\}&=\int_M\vari{S_{\text{lin}}}{\af{\phi_i}} \vari{X^a}{\phi^i} = \int_M\vari{S_\text{lin}}{\af \dil} \vari{X^a}{\dil} + \vari{S_\text{lin}}{\bar e^*} \vari{X^a}{e}\\
    &=-\tfrac12\sigma^{-3}(\ms L_\xi \sigma-\tfrac18\sigma^{-1}(\bar \rho e))(\bar e\gamma^a e)+\tfrac12\sigma^{-2}(\ms L_\xi \bar e-\tfrac1{16}(\bar e\gamma_c e)\bar\rho\gamma^c)\gamma^a e\\
    &=\tfrac14(\ms L_\xi\sigma^{-2})(\bar e\gamma^a e)+\tfrac12\sigma^{-2}(\ms L_\xi \bar e)\gamma^a e +\tfrac1{16}\sigma^{-4}[(\bar \rho e)(\bar e\gamma^a e)-\tfrac12(\bar\rho\gamma^c\gamma^a e)(\bar e\gamma_c e)]\\
    &=(\ms L_\xi X)^a+\tfrac1{32}\sigma^{-4}(\bar\rho\gamma^a \gamma^c e)(\bar e\gamma_c e)\\
    &\above{\smash{\eqref{eq:bosonic-f1}}}\;(\ms L_\xi X)^a.
  \end{aligned}
\end{equation}
Next, since \eqref{eq:bosonic-f1} implies $\slashed X e = 0$, from \eqref{eq:CP-spinor-LD} we obtain
\begin{equation} \label{eq:diffeo-on-e}
  \slashed X \slashed D e = 2 \gld_X e.
\end{equation}

Finally, note that the Lie derivative $\ms L_V$ of spinor half-densities is defined with the help of a connection on the field space, i.e.\ it really corresponds to the vertical part of the vector field $\delta_V$ on the field space, which implements the action of a generalised diffeomorphism. Thus, using \eqref{commutator} (cf.\ the discussion regarding the commutator of supersymmetries in section \ref{sec:closure}) one obtains the following formula for any pair of generalised vector fields $V_1,V_2\in\Gamma(E)$ and a spinor half-density $\lambda$:
\begin{equation}
  \ms L_{[V_1,V_2]}\lambda=(\ms L_{V_1}\ms L_{V_2}-\ms L_{V_2}\ms L_{V_1})\lambda-F(\ms L_{V_1}\gm,\ms L_{V_2}\gm)\lambda,
\end{equation}
which in particular gives
\begin{equation}\label{xixi}
  \ms L_{[\xi,\xi]}e=2\ms L_\xi\ms L_\xi e-F(\ms L_\xi\gm,\ms L_\xi\gm)e.
\end{equation}
We will use these observations to simplify the calculations below.

\subsection{Linear terms}
First, we observe that vanishing of the terms in \eqref{eq:CME-linear} linear in $\sigma^*$, $\gm^*$, $\rho^*$, or $\psi^*$ is precisely equivalent to the symmetry algebra \eqref{simplecommutators}, \eqref{susycommutatorbosonic}, \eqref{eq:closurefermions}, whose proof is the content of appendix \ref{sec:closurecomputation}. Note that the curvature term in \eqref{eq:CME-linear} corresponds to the curvature term in \eqref{commutator} (cf.\ the $e^*$ case below). It thus remains to check the vanishing of the terms in \eqref{eq:CME-linear} linear in $e^*$, $\xi^*$, and $f^*$.

Let us start with the terms in \eqref{eq:CME-linear} linear in $\af e$, which are
\begin{equation}
  \int_M\cmeTerm{S_{\text{lin}}}{}\left(\int_M\bar e^*(\ms L_\xi e+\tfrac1{4}\slashed X\rho)\right) +\frac12 \int_M\af{\bar{e}} F\!\left(\frac{\delta S_{\text{lin}}}{\delta \af \gm}, \frac{\delta S_{\text{lin}}}{\delta \af \gm}\right) e=:-\int_M \bar e^* \kappa_e,
\end{equation}
where
\begin{align*}
  \kappa_e&\coloneqq\int_M\cmeTerm{S_{\text{lin}}}{\ms L_\xi e}+\frac14\left(\int_M\vari{S_{\text{lin}}}{\af{\phi_i}} \vari{X^a}{\phi^i}\right)\gamma_a\rho+\frac14\slashed X\left(\int_M\vari{S_{\text{lin}}}{\af{\phi_i}} \vari{\rho}{\phi^i}\right)-\frac12F\!\left(\frac{\delta S_{\text{lin}}}{\delta \af \gm}, \frac{\delta S_{\text{lin}}}{\delta \af \gm}\right) e\\
  &\above{\eqref{eq:spinor-ld-variation}}\,\ms L_{\ms Df+\frac12\ms L_\xi \xi-\frac12X}e-\ms L_\xi(\ms L_\xi e+\tfrac14\slashed X\rho)+F\left(\frac{\delta S_{\text{lin}}}{\delta\gm^*},\ms L_\xi\gm\right)e+\frac14(\ms L_\xi X)^a\gamma_a\rho\lbl{noneedforlabelhere}\\
  &\qquad+\frac14\slashed X\left(\ms L_\xi\rho+\di e+\frac1{192}\sigma^{-2}(\bar\psi_\alpha\gamma_{abc}\psi^\alpha)\gamma^{abc}e\right)-\frac12 F\!\left(\frac{\delta S_{\text{lin}}}{\delta \af \gm}, \frac{\delta S_{\text{lin}}}{\delta \af \gm}\right) e
\end{align*}
Using the notation $\delta_e\gm_{a\alpha}\coloneqq\tfrac12\sigma^{-2}(\bar e\gamma^a\psi^{\beta})$, so that $\delta S/\delta \gm^*=\ms L_\xi \gm+\delta_e\gm$, and using \eqref{eq:diffeo-on-e} and \eqref{xixi}, we get
\begin{align*}
  \kappa_e&=-\tfrac12F(\ms L_\xi\gm,\ms L_\xi\gm)e+F(\ms L_\xi\gm+\delta_e\gm,\ms L_\xi\gm)e+\tfrac1{768}\sigma^{-2}\slashed X(\bar\psi_\alpha\gamma_{abc}\psi^\alpha)\gamma^{abc}e\\
  &\qquad-\tfrac12 F(\ms L_\xi\gm+\delta_e\gm, \ms L_\xi\gm+\delta_e\gm) e\lbl{smthg}\\
  &=\tfrac1{768}\sigma^{-2}\slashed X(\bar\psi_\alpha\gamma_{abc}\psi^\alpha)\gamma^{abc}e-\tfrac12 F(\delta_e\gm, \delta_e\gm) e.
\end{align*}
Vanishing of $\kappa_e$ thus follows from
\begin{align*}
  \tfrac1{768}\sigma^{-2}\slashed X(\bar\psi_\alpha\gamma_{abc}\psi^\alpha)\gamma^{abc}e\;\;&\above{\smash{\eqref{eq:fierzG1_1}}}\;-\tfrac1{32}\sigma^{-2}\slashed X\gamma_a \Gino_\alpha(\bar e \gamma^a \Gino^\alpha)=-\tfrac1{128}\sigma^{-4}\gamma_b\gamma_a \Gino_\alpha(\bar e\gamma^b e)(\bar e \gamma^a \Gino^\alpha)\\
  &\above{\smash{\eqref{eq:bosonic-f1}}}\;\tfrac1{128}\sigma^{-4}\gamma_a\gamma_b \Gino_\alpha(\bar e\gamma^b e)(\bar e \gamma^a \Gino^\alpha)\\
  &\above{\smash{\eqref{eq:bosonic-f2}}} -\tf1{64}\dil\inv4 \gamma_a \gamma_b e (\bar e \gamma^b \Gino_\alpha)(\bar e \gamma^a \Gino^\alpha)\lbl{smthg2}\\
  &=-\tf1{16} \delta_e\G\indices{^b_\alpha} \delta_e \G^{a \alpha} \gamma_a \gamma_b e = \tf1{32} [\delta_e \G, \delta_e\G]^{ab} \gamma_{ab}e\\
  &\above{\smash{\eqref{eq:field-space-curvatures}}}\; \tf12 F(\delta_e\G, \delta_e \G) e.
\end{align*}
This concludes the proof of vanishing of the linear $e^*$ terms.

Next, we turn to the terms linear in $\af\xi$, which are simpler and give
\begin{equation}
  \int_M\cmeTerm{S_{\text{lin}}}{}\left(\int_M\la \xi^*,\ms D\!f+\tfrac12\ms L_\xi\xi-\tfrac12X\ra\right)=:\int_M \la\xi^*, \kappa_\xi\ra,
\end{equation}
with
\begin{align*}
  \kappa_\xi&=\tfrac12\ms D\!\left[\ms L_\xi f+\tfrac12\la X,\xi\ra-\tfrac16\la \xi,\ms L_\xi\xi\ra\right]+\tfrac12\ms L_{\ms D\!f+\frac12\ms L_\xi\xi-\frac12X}\xi-\tfrac12\ms L_\xi(\ms D\!f+\tfrac12\ms L_\xi\xi-\tfrac12X)- \tf12 \gld_\xi X\\
  &=\tfrac12(\ms D\ms L_\xi-\ms L_\xi\ms D) f+\tfrac14(\ms D\la X,\xi\ra-\tfrac14\ms L_X\xi- \tf14 \gld_\xi X)-\tfrac1{12}\ms D\la \xi,\ms L_\xi\xi\ra+\tfrac14\ms L_{\ms L_\xi\xi}\xi-\tfrac14\ms L_\xi\ms L_\xi\xi\lbl{anotherone}\\
  &=-\tfrac1{12}(\ms L_\xi\ms L_\xi\xi+\ms L_{\ms L_\xi\xi}\xi)+\tfrac14\ms L_{\ms L_\xi\xi}\xi-\tfrac14\ms L_\xi\ms L_\xi\xi=\tfrac16\ms L_{\ms L_\xi\xi}\xi-\tfrac13\ms L_\xi\ms L_\xi\xi=0,
\end{align*}
where we used $\ms D \inner{V_1, V_2} = \gld_{V_1}V_2 + \gld_{V_2} V_1$ together with $[\ms D, \gld_\xi] = 0$, and in the last step also the fact that on vectors we have $\gld_\xi \gld_\xi = \tf12 \gld_{\gld_\xi \xi}$.

Lastly, with the help of the same identities we can show the vanishing of the terms linear in $\af f$, i.e.
\begin{equation}
  \int_M\cmeTerm{S_{\text{lin}}}{}\left(\int_M\tfrac12f^*(\ms L_\xi f+\tfrac12\la X,\xi\ra-\tfrac16\la \xi,\ms L_\xi\xi\ra)\right)=:-\tfrac12\int_M f^*\kappa_f,
\end{equation}
with
\begin{align*}
    \kappa_f&= \gld_{\ms D\!f+\frac12 \gld_\xi \xi - \frac12X} f - \gld_\xi ( \tf12 \gld_\xi f + \tf14 \inner{X, \xi} - \tf1{12} \inner{\xi, \gld_\xi \xi}) + \tf12 \inner{\gld_\xi X, \xi} + \tf12 \inner{X, \ms D\!f + \tf12 \gld_\xi \xi - \tf12 X}\\
                                                            &\quad- \tf16\inner{\ms D\! f + \tf12 \gld_\xi \xi - \tf12 X, \gld_\xi \xi} + \tf16 \smash{\inner{\xi, \gld_{\ms D\!f+\frac12 \gld_\xi \xi -\frac12 X} \xi}}  - \tf16 \inner{\xi, \gld_\xi(\ms D\! f + \tf12 \gld_\xi \xi - \tf12 X)}\\
                                                            &= (\tf12 - \tf16) \gld_{\gld_\xi \xi} f - (\tf12 + \tf16) \gld_\xi \gld_\xi f \\
                                                            &\quad-\tf14 \gld_\xi \inner{X, \xi} + (\tf12- \tf1{12}) \inner{\gld_\xi X,\xi} + (\tf14 + \tf1{12}) \inner{X, \gld_\xi \xi}+\tfrac1{12}\la \ms L_X \xi,\xi\ra-\tfrac14\la X,X\ra\\
                                                            &\quad+ \tf1{12}(\gld_\xi \inner{\xi, \gld_\xi \xi} - \inner{\gld_\xi \xi, \gld_\xi \xi} + \inner{\xi, \gld_{\gld_\xi \xi} \xi} - \inner{\xi, \gld_\xi \gld_\xi \xi})\\
                                                            &= \tf13(\gld_{\gld_\xi \xi} f - 2 \gld_\xi \gld_\xi f) -\tfrac14\gld_\xi \inner{X, \xi} +\tfrac5{12}\inner{\gld_\xi X, \xi} +\tfrac13 \inner{X, \gld_\xi \xi}+\tfrac1{12}\la\ms D\la X,\xi\ra-\ms L_\xi X,\xi\ra\\
                                                            &\quad-\tfrac1{64}\sigma^{-4}(\bar e\gamma_a e)(\bar e\gamma^a e)+ \tf1{12}(\gld_\xi \inner{\xi, \gld_\xi \xi} - \inner{\gld_\xi \xi, \gld_\xi \xi} + \inner{\xi, \gld_\xi \gld_\xi \xi})\\
                                                            &\above{\smash{\eqref{eq:bosonic-f1}}}\;\; -\tf13(\gld_\xi \inner{X, \xi} - \inner{\gld_\xi X, \xi} - \inner{X, \gld_\xi \xi})= 0.
  \end{align*}
This concludes the proof of the vanishing of all the linear antifield terms in $\{S,S\}$.

  \subsection{Quadratic terms}\label{sec:quad}
  We now turn to the last remaning part of the proof of the classical master equation $\{S,S\}=0$, i.e.\ to the vanishing of the quadratic antifield terms \eqref{eq:CME-square}. Knowing that all other terms vanish, we can simply write
  \begin{equation}
    \begin{aligned}
      \frac12\{S,S\}&=\int_M\cmeTerm{S_{\text{lin}}}{S_{\text{quad}}} + \cmeTerm{S_{\text{quad}}}{S_{\text{lin}}}\\
      &=\int_M\frac{\delta S_{\text{lin}}}{\delta \sigma^*}\frac{\delta S_{\text{quad}}}{\delta \sigma}+\frac{\delta S_{\text{lin}}}{\delta \bar e^*}\frac{\delta S_{\text{quad}}}{\delta e}+\frac{\delta S_{\text{quad}}}{\delta \bar\rho^*}\frac{\delta S_{\text{lin}}}{\delta \rho}+\frac{\delta S_{\text{quad}}}{\delta \bar\psi^*_\alpha}\frac{\delta S_{\text{lin}}}{\delta \psi^\alpha}
    \end{aligned}
  \end{equation}
  It will be slightly more convenient for us to use $v^a:=\bar e\gamma^a e$ instead of $X$ from the preceding section. With its help we get
  \begin{align*}
      \tfrac12\{S,S\}&=\smash{\int_M} [\ms L_\xi \sigma-\tfrac18 \sigma^{-1}(\bar \rho e)]\sigma^{-3}[\tfrac1{32}\bar\psi^*_\beta\slashed v\psi^{*\beta}+\tfrac1{16}(\bar e\psi^*_{\beta})(\bar e\psi^{*\beta})+\tfrac1{32}\bar\rho^*\slashed v\rho^*]\\
      &\qquad\quad+[\ms L_\xi \bar e+\tfrac1{16}\sigma^{-2}\overline{\slashed v\rho}]\sigma^{-2}[-\tfrac1{32}\gamma_a e(\bar\psi^*_\beta\gamma^a\psi^{*\beta})-\tfrac1{16}\psi^*_{\beta}(\bar e\psi^{*\beta})-\tfrac1{32}\gamma_a e(\bar\rho^*\gamma^a\rho^*)]\\
      &\qquad\quad-[\tfrac1{32}\sigma^{-2}\overline{\slashed v\rho^*}][\tfrac18\sigma^*\sigma^{-1}e-\ms L_\xi \rho^*+\tfrac18\sigma^{-2}\psi^{\beta}(\bar\psi^*_{\beta}e)+\tfrac18\sigma^{-2}\gamma^c\psi^*_{\beta}(\bar\psi^{\beta}\gamma_ce)-\tfrac1{16}\sigma^{-2}\slashed ve^*]\\
      &\qquad\quad+\left[-\tfrac1{32}\sigma^{-2}\overline{\slashed v\psi^{*\alpha}}+\tfrac1{16}\sigma^{-2}(\bar e\psi^{*\alpha})\bar e\right][-\tfrac12\sigma^{-2}\gm^*_{a\alpha}\gamma^ae+\tfrac1{96}\sigma^{-2}\gamma_{abc}\psi_\alpha(\bar\rho^*\gamma^{abc}e)\\
      &\qquad\quad\quad -(\ms L_\xi \psi^*)_\alpha+\tfrac18\sigma^{-2}\rho(\bar\psi^*_{\alpha}e)-\tfrac18\sigma^{-2}\gamma_ae(\bar\psi^*_{\alpha}\gamma^a\rho)].\lbl{test1}
  \end{align*}
  It is easy to see that the terms containing the Lie derivative cancel after integrating by parts in $\int_M(\ms L_\xi\sigma)(\dots)$. Similarly, the two terms containing $\gm^*$ cancel each other out since
  \begin{equation}
    \tfrac1{32}\sigma^{-2}(\bar e\gamma_e e)(\bar\psi^{*\alpha}\gamma^e\gamma_ae)+\tfrac1{16}\sigma^{-2}(\bar e\psi^{*\alpha})(\bar e\gamma_ae)=-\tfrac1{32}\sigma^{-2}(\bar e\gamma_e e)(\bar\psi^{*\alpha}\gamma_a\gamma^ee)\;\;\above{\eqref{eq:bosonic-f1}}\;\;0.
  \end{equation}
  In addition, the term containing $e^*$ vanishes as a consequence of $\slashed v\slashed v=\la v,v\ra\;\;\above{\smash{\eqref{eq:bosonic-f1}}}\;\;0$. We assemble the rest as
  \begin{equation}
    \frac12\{S,S\}=:\frac1{128}\int_M\sigma^{-4}\kappa_{\text{quad}},
  \end{equation}
  where
  \begin{align*}
      \kappa_{\text{quad}}&=-\tfrac12(\bar \rho e)(\bar \rho^*\slashed v\rho^*)-\tfrac12(\bar\rho e)(\bar\psi^* \slashed v\psi^*)-(\bar\rho e)(\bar e\psi^*)(\bar e\psi^{*})+\tfrac14(\bar \rho \slashed v \gamma_a e)(\bar\rho^*\gamma^a\rho^*)\\
      &\qquad+\tfrac14(\bar \rho \slashed v \gamma_a e)(\bar\psi^*\gamma^a\psi^*)+\tfrac12(\bar \rho\slashed v\psi^*)(\bar e\psi^*)+\tfrac12(\bar\psi^* e)(\bar\psi\slashed v \rho^*)-\tfrac12(\bar\psi\gamma_a e)(\bar\psi^*\gamma^a\slashed v\rho^*)\\
      &\qquad+\tfrac1{24}(\bar\rho^*\gamma_{abc}e)(\bar\psi\gamma^{abc}\slashed v\psi^*)-\tfrac1{12}(\bar\rho^*\gamma^{abc}e)(\bar\psi\gamma_{abc}e)(\bar e\psi^*)+\tfrac12(\bar\rho\slashed v\psi^*)(\bar\psi^* e)\\
      &\qquad-(\bar\rho e)(\bar\psi^* e)(\bar e\psi^*)+\tfrac12(\bar\psi^*\gamma^a\rho)(\bar e\gamma_a\slashed v\psi^*)-(\bar\psi^*\slashed v\rho)(\bar e\psi^*)\lbl{kappaquad}\\
      &=-\tfrac12(\bar \rho e)(\bar \rho^*\slashed v\rho^*)-\tfrac12(\bar\rho e)(\bar\psi^* \slashed v\psi^*)+\tfrac14(\bar \rho \slashed v \gamma_a e)(\bar\rho^*\gamma^a\rho^*)+\tfrac14(\bar \rho \slashed v \gamma_a e)(\bar\psi^*\gamma^a\psi^*)\\
      &\qquad+\tfrac12(\bar\psi^* e)(\bar\psi\slashed v \rho^*)-\tfrac12(\bar\psi\gamma_a e)(\bar\psi^*\gamma^a\slashed v\rho^*)+\tfrac1{24}(\bar\rho^*\gamma_{abc}e)(\bar\psi\gamma^{abc}\slashed v\psi^*)\\
      &\qquad-\tfrac1{12}(\bar\rho^*\gamma^{abc}e)(\bar\psi\gamma_{abc}e)(\bar e\psi^*)+\tfrac12(\bar\psi^*\gamma^a\rho)(\bar e\gamma_a\slashed v\psi^*)-(\bar\psi^*\slashed v\rho)(\bar e\psi^*).
  \end{align*}
  Using again \eqref{eq:bosonic-f1} to get
  \begin{equation}\label{vezero}
    \slashed ve=\bar e\slashed v=0
  \end{equation}
  as well as $\slashed v\gamma_ae=2v_ae$ and $\bar e\gamma_a\slashed v=2\bar ev_a$, the first four terms and the last two terms in the resulting expression in \eqref{kappaquad} cancel, leaving us with
  \begin{align*}
      \kappa_{\text{quad}}&=\tfrac12(\bar\psi^* e)(\bar\psi\slashed v \rho^*)-\tfrac12(\bar\psi\gamma_a e)(\bar\psi^*\gamma^a\slashed v\rho^*)+\tfrac1{24}(\bar\rho^*\gamma_{abc}e)(\bar\psi\gamma^{abc}\slashed v\psi^*)-\tfrac1{12}(\bar\rho^*\gamma^{abc}e)(\bar\psi\gamma_{abc}e)(\bar e\psi^*)\\
      &\above{\eqref{yetanothereq:bosonic-fierz}}-(\bar\psi^* e)(\bar\psi\slashed v \rho^*)-\tfrac12(\bar\psi\gamma_a e)(\bar\psi^*\gamma^a\slashed v\rho^*)+\tfrac1{24}(\bar\rho^*\gamma_{abc}e)(\bar\psi\gamma^{abc}\slashed v\psi^*)\\
      &\above{\eqref{the_other_bosonic_fierz}} -(\bar\psi^* e)(\bar\psi\slashed v \rho^*)-\tfrac12(\bar\psi\gamma_a e)(\bar\psi^*\gamma^a\slashed v\rho^*)-\tf12 (\bar \dilino^* \gamma_a \Gino)(\bar e \gamma^a \slashed v \af \Gino) + \tf12 (\bar e \gamma_a \Gino)(\bar \dilino^* \gamma^a \slashed v \af \Gino)\lbl{r}\\
      &\above{\eqref{vezero}}-(\bar\psi^* e)(\bar\psi\slashed v \rho^*)-\tfrac12(\bar\psi\gamma_a e)(\bar\psi^*\gamma^a\slashed v\rho^*) -(\bar\rho^*\slashed v\psi)(\bar e\psi^*)+\tfrac1{2}(\bar e\gamma_a\psi)(\bar\rho^*\gamma^a\slashed v\psi^*)\\
    &=-\tfrac12(\bar\psi\gamma_a e)(\bar\psi^*\gamma^a\slashed v\rho^*)+\tfrac1{2}(\bar e\gamma^a\psi)(\bar\rho^*\gamma_a\slashed v\psi^*)\vphantom{\above{()}}\\
    &=-\tfrac12(\bar\psi\gamma_a e)(\bar\psi^*(\gamma^a\slashed v+\slashed v\gamma^a)\rho^*)=-(\bar\psi\slashed v e)(\bar\psi^*\rho^*)=0.\vphantom{\above{()}}
  \end{align*}
  This concludes the full proof of the classical master equation $\{S,S\}=0$ to all orders in antifields.

\section{Conclusions} 
\label{sec:conclusion}
In this paper, we have presented a full demonstration that the classical master action for $\ms N=1$ supergravity in ten dimensions we reported in~\cite{Kupka:2025hln} indeed satisfies the classical master equation. 
This included an explicit check that the algebra of local supersymmetries closes on-shell and the construction of the resulting antifield terms in the BV action.  
Our construction and calculations have relied heavily on the generalised geometry formulation of the theory~\cite{Kupka:2024xur,Kupka:2024vrd} and intuition gained from studying the dilatonic supergravity theory of~\cite{Kupka:2024tic}, which provided a useful toy model. 

We have also elaborated on the geometric structure of the field space of supergravity such that we did not need to introduce the vielbein degrees of freedom and Lorentz transformations of standard treatments. The benefit of this approach here was that it removed one set of symmetry parameters from consideration, which greatly simplifies the BV formulation as these would give rise to additional ghosts and their antifields, resulting in many more terms in the BV action. We have also explained how to understand the action of (generalised) diffeomorphisms on spinorial quantities and more generally how to understand the simultaneous variation of spinorial quantities and the metric tensor. This is subtle as the spinor bundles in which such fields transform depend upon the choice of metric tensor, and we have explained how the construction of a connection on the field space fibration allows one to transport spinorial objects between different spinor bundles in order to define such simultaneous variations. Similar considerations would again apply to any theory of gravity coupled to spinor fields or other objects similarly arising from dynamical $G$ structures on the spacetime manifold.

We also emphasise that our formulation of supergravity does not employ so-called \emph{supercovariant derivatives} \cite{Freedman:1976py}, i.e.\ modifications of the covariant derivative (by adding quadratic fermionic terms) whose supersymmetry transformations do not contain any derivatives of the supersymmetry parameter.
Since the early days of supergravity, the use of these objects was widely thought to provide 
a good organising principle for checking local supersymmetry. 
However, in the generalised geometry formulation we use here, they do not appear and in fact they cannot be constructed. 
In the usual treatments one typically includes terms such as $\bar\uppsi_\mu \gamma_{[\nu} \uppsi_{\rho]}$ or $\bar\uppsi_\rho \gamma_{\mu\dots\nu}{}^{\rho\sigma} \uppsi_\sigma$ in the supercovariant objects. Such terms are not possible to write in generalised geometry as they are incompatible with the generalised-geometric index structure, since the gamma matrices carry a $C_+$ vector index while the non-spinor index on the gravitino is a $C_-$ index. 
Whilst there are a few combinations of terms that one can write down, none seems to provide a supercovariant object in the usual sense. 
Overall, in ten-dimensions the organising principle provided by the enhanced symmetries of generalised geometry is more powerful than that of supercovariantisation (which failed to account for all of the higher fermion terms in~\cite{Bergshoeff:1981um}), and this is in large part what enabled us to formulate a BV action as concise as that appearing in~\eqref{eq:bv-action}. 

Clearly, the framework that we have developed in deriving our main result will have many further applications. Most immediately, one could apply it to extend the generalised geometry construction of type II supergravities~\cite{CSW1} to include the higher order fermion terms and provide a BV formulation of those theories. 
One could also make use of the elegant formulation of consistent truncations in the generalised geometry framework~\cite{Cassani:2019vcl} to make similar statements about lower dimensional supergravity theories, possibly resulting in a substantial reorganisation of the terms involved. 
It would also be interesting to explore the relation between our approach and recent work on the construction of BV actions from the tensor hierarchy~\cite{Cederwall:2023xbj}. 

Our results can also be used to construct twisted supergravity theories in the sense proposed by Costello and Li~\cite{Costello:2016mgj}. These are defined to be supergravity theories in backgrounds where the bosonic ghost field for local supersymmetry is allowed to take a non-zero vacuum expectation value (satisfying appropriate conditions). 
Numerous works have made conjectures as to the nature of these theories~\cite{Costello:2016mgj,Costello:2016nkh,Costello:2019jsy,Raghavendran:2021qbh,Hahner:2023kts,Ashmore:2025fxr} and some free-field limits of them have been derived~\cite{Saberi:2021weg,Eager:2021ufo}. Our result could be used to prove these conjectures from a first principles approach in the (generalised geometry version of the) component field formulation of supergravity directly following the definition of Costello--Li. This would have the advantages of making the geometric structure manifest in the formulation and also being equally applicable to twists around arbitrary supersymmetric backgrounds. In fact, in our framework one easily sees the possibility that the conditions required of the VEVs could be relaxed slightly from those of Costello--Li to include even more general solutions of the equations coming from the BV action, as described in the appendix of~\cite{Kupka:2025hln}. 
One could also construct twists of lower dimensional theories arising from consistent truncations of the ten-dimensional theory we present here. 

Finally, we note that one could also endeavour to study quantum aspects of supergravity via our BV action, as was the originally intended purpose of the BV formalism. In particular (see e.g.~\cite{Barnich:2000zw}), it should be the correct language in which to consider the gauge invariant observables, anomalies, etc. Clearly in the supergravity case, one runs into the issue that the theory is power-counting non-renormalisable. 
However, some formal aspects of its quantisation, such as anomalies, could nonetheless be investigated and may produce interesting insights.


\acknowledgments
CSC~and FV~are supported by an EPSRC New Investigator Award, grant number EP/X014959/1. FV was also supported by the Charles University grant PRIMUS/25/SCI/018. Part of the work was done while FV was in residence at Institut Mittag-Leffler in Djursholm in 2025 during the program Cohomological aspects of Quantum Field Theory, supported by the Swedish Research Council under grant no.\ 2021-06594. The authors would like to thank Institut Mittag-Leffler, as well as the program organisers, for providing a stimulating environment for collaboration and for developing this work. No new data was collected or generated during the course of this research.

\appendix
  \section{General construction of BV actions} \label{sec:review-BV}
In this section we recall the general philosophy and guidelines in constructing BV actions of physical systems, following \cite{Baulieu:1990uv} (see also \cite{Gomis:1994he} for a more in-depth review).

The construction starts from a collection of classical fields $\Phi_{(0)}$ and an action $S_0(\Phi_{(0)})$ that is invariant under a set of symmetry transformations $\delta_\epsilon$. The algebra of symmetries includes field-dependent transformations and in general closes only on-shell,
\begin{equation} \label{eq:field-algebra}
  [\delta_{\epsilon_1}, \delta_{\epsilon_2}]\Phi_{(0)} = \delta_{[\epsilon_1, \epsilon_2]}\Phi_{(0)} + M(\Phi_{(0)}, \epsilon_1, \epsilon_2) \eom {\Phi_{(0)}},
\end{equation}
where $M$ is some local operator measuring the failure of the algebra to close off-shell and on the right-hand side we have some bracket governing the algebra of parameters $\epsilon$ (note that the bracket $[\epsilon_1,\epsilon_2]$ is assumed to be very general as it may not only depend on $\epsilon_1$ and $\epsilon_2$ but also on fields, and it needs to be neither anti-symmetric nor satisfy the Jacobi-identity). We iteratively introduce ghosts $\Phi_{(1)}$ (of \emph{ghost degree 1}) and ghosts for ghosts $\Phi_{(2)}$ (of \emph{ghost degree 2}), etc., and define the \emph{BRST operator} by
\begin{equation}
  \begin{aligned}
    \brst \Phi_{(0)} &\coloneqq \delta_{\Phi_{(1)}}\Phi_{(0)}, \\
    \brst \Phi_{(1)} &\coloneqq \delta_{\Phi_{(1)}} \Phi_{(1)} + \delta_{\Phi_{(2)}} \Phi_{(1)} = \tf12 [\Phi_{(1)}, \Phi_{(1)}] + \delta_{\Phi_{(2)}} \Phi_{(1)},\\
                     & \vdots
  \end{aligned}
\end{equation}
This operator coresponds to replacing the symmetry parameters by ghosts.\footnote{Note that in order to get the correct signs for the replacement $\epsilon \mapsto \Phi_{(1)}$, it is useful to write $\delta_\epsilon \Phi_{(0)} = \epsilon^\alpha \delta_\alpha \Phi_{(0)}$ for some appropriate index $\alpha$ (i.e.\ we put the parameter to the front). We then define $\delta_{\Phi_{(1)}} \coloneqq \Phi_{(1)}^\alpha \delta_\alpha$.} Its action on ghosts realises the symmetry algebra; the ghosts for ghosts in this picture capture the gauge symmetry of the ghosts.

Collecting the classical fields, ghosts, etc.\ into the set of fields $\{\Phi^i\}$,\footnote{Here for simplicity we either work in a finite-dimensional setting or use the DeWitt notation, where the index $i$ includes the spacetime position as well as any discrete labels. It is however \emph{not} correlated with the ghost degree.} we may write the action of the BRST operator as
\begin{equation}
  \brst \Phi^i = \delta_{\Phi} \Phi^i.
\end{equation}
Let us now assume that for the square of the BRST operator we have
\begin{equation} \label{eq:non-nilpotency}
  \brst^2 \Phi^i = \widetilde M(\Phi)\indices{^i^j} \frac{\partial S_0}{\partial \Phi^j},
\end{equation}
for some local functions $\widetilde M$, which in particular depend on $M$ from \eqref{eq:field-algebra}.\footnote{We note that, although suggested by \eqref{eq:field-algebra}, the condition \eqref{eq:non-nilpotency} does not automatically follow from the preceding discussion.} Among other things \eqref{eq:non-nilpotency} contains the information about (the failure of) the Jacobi identity.
Nevertheless, since $Q_B$ is in general not nilpotent, it cannot be used to define a cohomology to encode the physical degrees of freedom.

We may rectify this shortcoming by introducing the \emph{BV operator} $\bv$. For each field $\Phi_{(n)}$ we adjoin the corresponding conjugate antifield $\Phi^*_{(-n-1)}$ and require that when setting the antifields to zero the BV operator acting on fields reduces to the BRST operator
\begin{equation} \label{eq:bv-to-brst}
  \bv|_{\af \Phi\hspace{-.35mm}=0} \Phi^i= \brst \Phi^i.
\end{equation}
The conjugacy means that on this extended space we have a Poisson bracket $\{\slot,\slot\}$ of degree $1$, pairing $\Phi_{(n)}$ with $\Phi^*_{(-n-1)}$.

We assume that the BV operator is induced by an extension $S$ of the classical action $S_0$, the so-called \emph{BV action}, which is required to be of total degree $0$,
\begin{equation}
  \bv = \pb{S, \cdot }=\frac{\partial S}{\partial \Phi^*_i}\frac{\partial}{\partial \Phi^i} + \frac{\partial S}{\partial \Phi^i}\frac{\partial}{\partial \Phi^*_i},\qquad \deg S=0.
\end{equation}
Expanding it in the powers of antifields we get
\begin{equation}
  S(\Phi, \af\Phi) = S_0(\Phi) + \sum_{k=1}^{N} \tf1{k!}\Phi^*_{i_1} \cdots \Phi^*_{i_k} S_{k}^{i_1 \dots i_k}(\Phi).
\end{equation}
We call $N\in\{0,1,2,\dots,\infty\}$ the \emph{rank} of the theory. The nilpotency of the BV operator can be rephrased as the \emph{classical master equation} (CME)
\begin{equation} \label{eq:CME}
  \pb{S, S} = 0.
\end{equation}

It is easy to see that if one starts with a theory $S_0$ with a symmetry which closes off-shell and has a nilpotent BRST operator $Q_B$, one can construct the following BV action of rank 1:
\begin{equation}
  S(\Phi,\af\Phi)=S_0(\Phi)+\Phi^*_i Q_B \Phi^i.
\end{equation}
The classical master equation is in this case automatically satisfied as it corresponds precisely to the invariance of $S_0$ and the nilpotency of $Q_B$.

Consider now a general theory of rank 2:
\begin{equation} \label{eq:rank2-bv-action}
  S(\Phi, \af\Phi) = S_0(\Phi) + \Phi^*_i S_1^i(\Phi) + \tf12 \Phi^*_i \Phi^*_j S_2^{ij}(\Phi).
\end{equation}
Condition \eqref{eq:bv-to-brst} determines $S_1$ in terms of the BRST operator acting on the fields,
\begin{equation}
  \delta_{\Phi}\Phi^i = \brst \Phi^i = (\bv \Phi^i)|_{\af\Phi=0} = \pb{S, \Phi^i}|_{\af \Phi = 0} = S_1^i(\Phi).
\end{equation}
Furthermore, computing $\{S,S\}$ for the rank 2 ansatz \eqref{eq:rank2-bv-action} and isolating the terms of various orders in $\Phi^*$, we find that the CME \eqref{eq:CME} is equivalent to the following system of equations (where we are cavalier regarding the relative signs in order to simplify the presentation)
\begin{align}
  \brst S_0 &= 0,\label{justinvariance}\\
  \brst^2 \Phi^i + (\partial_j S_0) S_2^{ij} &= 0,\label{eq:linear-constraint}\\
  \brst S_2^{ij} + S_2^{k[i} \partial_k(\brst \Phi^{j]}) & =0, \label{consistency1}\\
  S_2^{i[j} \partial_i S_2^{kl]} &= 0,\label{consistency2}
\end{align}
where $\partial_i:=\partial/\partial \Phi^i$ and $[\slot]$ denotes the appropriate (graded) antisymmetrisation.
The first equation \eqref{justinvariance} states simply that the starting action $S_0$ is invariant under the gauge symmetries. Assuming \eqref{eq:non-nilpotency}, the second equation \eqref{eq:linear-constraint} can be satisfied by setting
\begin{equation}\label{ansatz}
  S_2^{ij} \sim \widetilde M (\Phi)^{ij}.
\end{equation}
Thus, following the above procedure, rank 2 systems are completely determined by $S$ and $Q_B$. The remaining two equations \eqref{consistency1} and \eqref{consistency2} then serve as nontrivial consistency checks --- if they are not satisfied, one would need to continue adding yet higher terms in antifields to the BV action and end up with actions of higher rank.

As a short summary, in the case of BV systems of rank 2 the linear term $S_1$ describes the symmetry transformations and algebra, encoded via the BRST operator $Q_B$, while the quadratic term $S_2$ captures the failure of $Q_B$ being nilpotent (and hence the failure of the symmetry algebra to close off-shell or satisfy the Jacobi identity).

  \section{Spinors in 10 dimensions}\label{app:fierz}
    \subsection{Conventions}
      We are working in ten dimensions $D=10$ with a Lorentzian metric $g$ of signature $(-,+,\dots,+)$. We set
      \begin{equation}
        \epsilon_{0\dots9}=-\epsilon^{0\dots 9}=1.
      \end{equation}
      The Clifford relations are defined by
      \begin{equation}
        \{\gamma_a,\gamma_b\}=2g_{ab}.
      \end{equation}
      We choose the intertwiner/charge conjugation matrix $C$ and the gamma matrices $\gamma_a$ to be anti-symmetric
      \begin{equation}
        C\gamma_a C^{-1}=-\gamma_a^T,\qquad C^T=-C.
      \end{equation}
      The Majorana conjugate is defined by
      \begin{equation}
        \bar\psi:=\psi^TC,
      \end{equation}
      and thus the inner product is anti-symmetric on bosonic (even) spinors and symmetric on fermionic (odd) spinors, i.e.\ for the latter we have
      \begin{equation}
        \bar \psi \chi = \bar \chi \psi.
      \end{equation}
      Furthermore, a gamma matrix with $k$ indices $a_1\dots a_k$ satisfies
      \begin{equation}
        C\gamma_{a_1\dots a_k}C^{-1}=(-1)^{\left[\frac{k+1}2\right]} \gamma_{a_1\dots b_k}^T,
      \end{equation}
      implying a flip formula for the bilinear of two odd spinors,
      \begin{equation}\label{flip}
        \bar\psi \gamma_{a_1\dots a_k}\chi=(-1)^{\left[\frac{k+1}2\right]} \bar\chi\gamma_{a_1\dots a_k}\psi.
      \end{equation}
      The chirality element is given by $\gamma_*:=\gamma^0\dots\gamma^9$, so that
      \begin{equation}
        \gamma_{a_1\dots a_k}\gamma_*=(-1)^{\left[\frac k2\right]}\tfrac1{(10-k)!}\sqrt{-g}\epsilon_{a_1\dots a_k b_1\dots b_{10-k}}\gamma^{b_1\dots b_{10-k}}. \label{eq:hodge_dual}
      \end{equation}
      The positive/negative chiral Majorana spinors are then defined by $\gamma_*\psi=\pm\psi$, respectively. We will denote the chirality by $\on{ch}\psi=\pm1$. Note that the pairing of two Majorana spinors of the same chirality automatically vanishes, i.e.\
      \begin{equation}\label{simplevanish}
        \bar\psi\chi=0\qquad\text{if $\on{ch}\psi=\on{ch}\chi$}
      \end{equation}
      and gamma matrices change the chirality,
      \begin{equation}
        \on{ch}(\gamma_{(p)}\psi)=(-1)^p\on{ch}\psi,\label{gammachiralityflip}
      \end{equation}
      where we used the short notation
      \begin{equation}
        \gamma_{(p)}:=\gamma_{a_1\dots a_p},
      \end{equation}
      
      In particular it follows that the only independent bilinear built out of two copies of the same odd Majorana spinor $\lambda$ is
      \begin{equation}
        \bar\lambda\gamma_{(3)}\lambda,
      \end{equation}
      since
      \begin{equation}\label{bilinearvanish}
        \bar\lambda\gamma_{(1)}\lambda=\bar\lambda\gamma_{(5)}\lambda=0
      \end{equation}
      due to the flip formula \eqref{flip}, the bilinears with $\gamma_{(2n)}$ vanish due to \eqref{simplevanish} and \eqref{gammachiralityflip}, and the bilinears with $\gamma_{(p>5)}$ can be related to the ones with $\gamma_{(p<5)}$ via \eqref{eq:hodge_dual}.
      Similarly, the independent bilinears constructed out of two copies of the same \emph{even} Majorana spinor $e$ are $\bar e \gamma_{(1)}e$ and the (anti-)self-dual part of $\bar e \gamma_{(5)}e$.
      
    \subsection{Chiral spinor identities}
      Applying \eqref{eq:hodge_dual} twice one gets that for Majorana spinors of arbitrary chiralities we have
      \begin{equation}\label{eq:flip}
        \tfrac1{p!}(\bar\lambda_1\gamma_{a_1\dots a_nb_1\dots b_p}\lambda_2)(\bar\lambda_3 \gamma^{b_1\dots b_p}\lambda_4)=(\on{ch}\lambda_2)(\on{ch}\lambda_4)(-1)^{1+[\frac{n}2]}\tfrac{1}{q!}(\bar\lambda_1\gamma^{c_1\dots c_q}\lambda_2)(\bar\lambda_3\gamma_{a_1\dots a_nc_1\dots c_q}\lambda_4),
      \end{equation}
      where $q:=10-(n+p)$. In particular
      \begin{equation}\label{eq:mirror}
        \tfrac1{(10-p)!}(\bar\lambda_1\gamma_{(10-p)}\lambda_2)(\bar\lambda_3\gamma^{(10-p)}\lambda_4)=-\tfrac1{p!}(\on{ch}\lambda_2)(\on{ch}\lambda_4)(\bar\lambda_1\gamma_{(p)}\lambda_2)(\bar\lambda_3\gamma^{(p)}\lambda_4).
      \end{equation}

    \subsection{Some gamma matrix manipulations}
      The fundamental relation for gamma matrices is
      \begin{align}
        \gamma_a \gamma^{b_1 \dots b_p} = \gamma\indices{_a^{b_1\dots b_p}} + p \delta_a^{[b_1} \gamma^{b_2\dots b_p]}, \label{eq:gamma_multiplication}
      \end{align}
      which is used in deriving many of the formulas below. Another useful identity is
      \begin{align}
        \{\gamma_a, \gamma^{b_1 \dots b_{2p+1}}\} &= 2(2p+1)\ \delta_{a}^{[b_1} \gamma^{b_2 \dots b_{2p+1}]}. \label{eq:acG1_p}
      \end{align}

    \subsection{Fierz identities}
    In this subsection all spinors are fermionic (odd) and Majorana. In addition, we adopt a notation where all $\psi_i$ are of the same chirality, while the $\lambda_j$ are of opposite chirality to the $\psi$'s. The starting identity, following from the completeness relation, is
      \begin{equation}
        \lambda\bar\psi=\frac1{32}\sum_{p=0}^{10}\frac{(-1)^{p+1}}{p!}\gamma_{(p)}\psi\bar\lambda\gamma^{(p)}.
      \end{equation}
      Using \eqref{eq:mirror} this produces the master Fierz identity
      \begin{align}
        (\bar\lambda_1\psi_1)(\bar\lambda_2\psi_2)&=\tfrac1{16}(\bar\lambda_1\gamma_{(1)}\lambda_2)(\bar\psi_1\gamma^{(1)}\psi_2)+\tfrac1{96}(\bar\lambda_1\gamma_{(3)}\lambda_2)(\bar\psi_1\gamma^{(3)}\psi_2)+\tfrac1{3840}(\bar\lambda_1\gamma_{(5)}\lambda_2)(\bar\psi_1\gamma^{(5)}\psi_2).\label{eq:master-fierz}
      \end{align}
      From this one derives
      \begin{align}
        (\bar\lambda_1\gamma_a\lambda_2)(\bar\lambda_3\gamma^a\lambda_4)&=\tfrac12(\bar\lambda_1\gamma_a\lambda_3)(\bar\lambda_2\gamma^a\lambda_4)+\tfrac1{24}(\bar\lambda_1\gamma_{(3)}\lambda_3)(\bar\lambda_2\gamma^{(3)}\lambda_4),\label{eq:fierzG1_1}\\
        (\bar \lambda_1 \gamma_{(3)} \lambda_2) (\bar \lambda_3 \gamma^{(3)} \lambda_4) &= 18 (\bar \lambda_1 \gamma_a \lambda_3) (\bar \lambda_2 \gamma^a \lambda_4) - \tf12 (\bar \lambda_1 \gamma_{(3)} \lambda_3) (\bar \lambda_2 \gamma^{(3)} \lambda_4) \label{eq:f3_3},
      \end{align}
      In particular, anti-symmetrising \eqref{eq:fierzG1_1} yields
      \begin{equation} \label{eq:anti-sym-fierz}
        (\bar \lambda_{[1} \gamma_{|a} \lambda_{2|})(\bar \lambda_{3]} \gamma^a \lambda_4) = \tf12(\bar \lambda_1 \gamma_a \lambda_3)(\bar \lambda_2 \gamma^a \lambda_4).
      \end{equation}
    Another useful identity, following easily from the representation theory, is \cite{Kupka:2024vrd, Bergshoeff:1981um}
    \begin{align}
      (\bar\lambda\gamma^{abc}\lambda)\bar\lambda\gamma_{abc}&=0\label{eq:3r3}.
    \end{align}

\subsection{Fierz identities for bosonic spinors}
  There are several further useful identities that one can derive from \eqref{eq:fierzG1_1} for four bosonic spinors $e_1, e_2, e_3, e_4$ of the same chirality. First, taking into account the commuting properties of the spinors one has
  \begin{equation}\label{eq:bosonic-fierz}
    (\bar e_1\gamma_a e_2)(\bar e_3\gamma^a e_4)=-\tfrac12(\bar e_1\gamma_a e_3)(\bar e_2\gamma^a e_4)-\tfrac1{24}(\bar e_1\gamma_{(3)} e_3)(\bar e_2\gamma^{(3)} e_4).
  \end{equation}
  Using the antisymmetry of $\gamma_{(3)}$ this implies that for $e_1 = e_3 = e$
  \begin{equation} \label{eq:bosonic-f2}
    (\bar e \gamma_a e_2) (\bar e \gamma^a e_4) = -\tf12(\bar e \gamma_a e) (e_2 \gamma^a e_4).
  \end{equation}
  Setting also $e_2 = e$ then gives
\begin{equation}\label{eq:bosonic-f1}
    (\bar e\gamma_a e)\bar e \gamma^a=0.
  \end{equation}
  Similarly, antisymmetrising on $e_1$ and $e_3$ in \eqref{eq:bosonic-fierz} yields
  \begin{equation}\label{the_other_bosonic_fierz}
    \tfrac1{24}(\bar e_1\gamma_{(3)} e_3)(\bar e_2\gamma^{(3)} e_4)=-\tfrac12(\bar e_1\gamma_a e_2)(\bar e_3\gamma^a e_4)+\tfrac12(\bar e_3\gamma_a e_2)(\bar e_1\gamma^a e_4).
  \end{equation}
  Finally, writing \eqref{eq:f3_3} for bosonic spinors of the same chirality and setting $e_2=e_4=e$ we deduce
  \begin{equation}\label{yetanothereq:bosonic-fierz}
    (\bar e_1\gamma_{(3)}e)(\bar e_3\gamma^{(3)}e)=-18(\bar e_1\gamma_a e_3)(\bar e\gamma^a e).
  \end{equation}
  
\section{Conventions in graded geometry} \label{app:graded-conventions}
  \subsection{Conventions and formulas for the BV bracket}
    Here we record the graded-geometric conventions which we use.\footnote{We are thankful to Pavol \v Severa for an enlightening discussion on this topic and for a particularly convenient set of conventions.} First and foremost note that the BV space is $\mb Z_2\times \mb Z$-graded --- the first factor corresponds to the fact that we are describing a classical field theory with both bosons and fermions, while the latter records the usual $\mb Z$-grading/ghost number on the BV space. All the signs which we need will follow from the following assumptions:
    \begin{itemize}
      \item we consistently use the Koszul sign convention with the commuting properties of the individual objects governed by the \emph{sum} of the $\mb Z_2$ and $\mb Z$ parities; similarly when talking about differential forms on the BV space we again consider the sum of the $\mb Z_2$-parity, $\mb Z$-parity, and the parity of the form degree; in either case we will denote the total parity of an object $A$ by
      \begin{equation}
        |A|\in\mb Z_2=\{0,1\}
      \end{equation}
      \item we only ever use \emph{left} derivatives; in particular, the operators $d$ and $i_X$ (for any vector field $X$) act on the left and have parities (mod 2)
      \begin{equation}
        |d|\equiv1,\qquad |i_X|\equiv-1+|X|
      \end{equation}
      \item the Lie derivative satisfies
      \begin{equation}
        L_X=[d,i_X],
      \end{equation}
      where $[A,B]:=AB-(-1)^{|A|\cdot|B|}BA$ is the graded commutator; in particular $|L_X|=|X|$; on~functions we require
      \begin{equation}
        L_Xf=Xf
      \end{equation}
      \item for a function $H$ we define the Hamiltonian vector field $X_H$ by
      \begin{equation}
        i_{X_H}\omega=-dH,
      \end{equation}
      the symplectic form has the local Darboux form
      \begin{equation}
        \omega=dp_i\wedge dq^i,
      \end{equation}
      and the Poisson bracket of two functions $H$ and $H'$ is defined by
      \begin{equation}
        \{H,H'\}=X_HH'.
      \end{equation}
    \end{itemize}
    The de Rham differential of a function is thus
    \begin{equation}
      df=dx^i \frac{\partial f}{\partial x^i}.
    \end{equation}
    It also follows from the above assumptions that for any function $f$ and vector field $X$ we have
    \begin{equation}\label{strangesign}
      i_Xdf=(-1)^{|X|}L_X f=(-1)^{|X|}Xf,
    \end{equation}
    which can be interpreted as ``$X$ swaps with $d$, producing a Koszul sign; and then $d$ cancels with $i$''.
    
    As another application, let us calculate the Hamiltonian vector field $X_H$, assuming that $\omega$ is \emph{odd} and $H$ is \emph{even} (as is the case in calculating the classical master equation in the BV framework). Dropping the $i$-indices for clarity, it follows that for the Darboux coordinates we have
    \begin{equation}
      |q|\equiv|p|+1\mod2
    \end{equation}
    and the vector field $X_H=A\partial_q+B\partial_p$ is odd. We thus calculate
    \begin{equation}
    \begin{aligned}
      i_{X_H}\omega&=dp (i_{A\partial_q}dq)+(i_{B\partial_p}dp)dq\;\;\smash{\above{\eqref{strangesign}}}\;-dp\,A-B\,dq,\\
      dH&=dq(\partial_qH)+dp(\partial_pH)=(\partial_qH)dq+dp(\partial_pH),
    \end{aligned}
    \end{equation}
    where we used that $dq$ and $\partial_qH$ have opposite parity and hence commute. Comparing and substituting the indices back we get
    \begin{equation}
      X_H=\frac{\partial H}{\partial p_i}\frac{\partial}{\partial q^i}+\frac{\partial H}{\partial q^i}\frac{\partial}{\partial p_i}.
    \end{equation}
    In particular we recover the following useful formula for two even functions:
    \begin{equation}\label{cmeeven}
      \{H,H'\}=\frac{\partial H}{\partial p_i}\frac{\partial H'}{\partial q^i}+\frac{\partial H}{\partial q^i}\frac{\partial H'}{\partial p_i}.
    \end{equation}
    Performing the calculation in more generality (while still assuming $\omega$ to be odd), one sees that for functions $F$, $F'$ of arbitrary degree we have
    \begin{equation}
      \{F,F'\}=(-1)^{|F|\cdot|p_i|}\frac{\partial F}{\partial p_i}\frac{\partial F'}{\partial q^i}+(-1)^{|F|\cdot|q^i|}\frac{\partial F}{\partial q^i}\frac{\partial F'}{\partial p_i}.
    \end{equation}

  \subsection{A remark on the BV bracket of spinors}\label{app:spinorcme}
  Consider any Majorana--Weyl spinor $\lambda$ of arbitrary chirality and parity, and its antifield $\lambda^*$ (of opposite chirality and parity). First, we record here the form of the BV bracket of two \emph{even} functions, which is (cf.\ \eqref{cmeeven})
\begin{equation}\label{spinorscme}
  \left\{H,H'\right\}=\int_M \frac{\delta H}{\delta \bar\lambda^*}\frac{\delta H'}{\delta\lambda}+\frac{\delta H'}{\delta \bar\lambda^*}\frac{\delta H}{\delta\lambda}.
\end{equation}

Let us now consider the case where $H$ and $H'$ are even bilinears of the form $H:=\int_M \bar\lambda^*\chi$ and $H':=\int_M \bar\lambda\varphi$, where $\chi$ and $\varphi$ have the opposite and the same parity as $\lambda$, respectively, and are both independent of $\lambda$, $\lambda^*$.
Adopting a temporary notation $x,y,\dots$ for spinor indices, we have $(\bar\lambda^*)_y=\lambda^{*x}C_{xy}$ and hence
\begin{equation}
  \frac{\delta H}{\delta (\bar\lambda^*)_y}=\int_M\frac{\delta H}{\delta (\lambda^*)^x} \frac{\delta (\lambda^*)^x}{\delta (\bar\lambda^*)_y}=\frac{\delta H}{\delta (\lambda^*)^x} (C^{-1})^{yx}.
\end{equation}
This allows us to rewrite \eqref{spinorscme} as
\begin{equation}
  \int_M \frac{\delta H}{\delta (\bar\lambda^*)_y}\frac{\delta H'}{\delta\lambda^y}=(C^{-1})^{yx}\int_M \frac{\delta H}{\delta (\lambda^*)^x} \frac{\delta H'}{\delta\lambda^y}=(C^{-1})^{yx}\int_M (C_{xz}\chi^z)(C_{yw}\varphi^w)=\int_M \chi^yC_{yw}\varphi^w.
\end{equation}

Thus we obtain the incredibly useful formula
\begin{equation}\label{incredible_spinor_formula}
  \left\{\int_M \bar\lambda^*\chi,\int_M \bar\lambda\varphi\right\}=\int_M\bar\chi\varphi,
\end{equation}
which holds for spinors of any chiralities and parities, as long as both bilinears $\bar\lambda^*\chi$ and $\bar\lambda\varphi$ are \emph{even} (and $\chi$ and $\varphi$ are independent of $\lambda$ and $\lambda^*$).

  \section{Curvature and torsion on the field space}\label{app:curv}
    \subsection{Tautological bundle}
      Consider the space of generalised metrics $\ms M$ on a given Courant algebroid $E\to M$. This carries a tautological vector bundle $\ms C_+\to\ms M$ of infinite rank, where the fiber at $\gm$ is the vector space $\Gamma(C_+)$.
    As discussed in section \ref{subsec:field-space}, $\ms C_+$ carries a natural connection, which identifies the nearby fibers at $\gm$ and $\gm'=\gm+\delta\gm$ by restricting the orthogonal projection $E\to C_+$ to $C_+'\subset E$.
    We claim that the curvature of this connection is given by
    \begin{equation}\label{curvplus}
      F^+(\delta_1\gm,\delta_2\gm)=\tfrac14[\delta_1\gm,\delta_2\gm]\colon\Gamma(C_+)\to\Gamma(C_+).
    \end{equation}
    Before deriving this formula we note that, analogously, we have the second tautological bundle $\ms C_-\to \ms M$ with connection. Its curvature is analogously
    \begin{equation}\label{curvminus}
    F^-(\delta_1\gm,\delta_2\gm)=\tfrac14[\delta_1\gm,\delta_2\gm]\colon \Gamma(C_-)\to\Gamma(C_-).
    \end{equation}
    Note that although the formulas for these two curvatures formally look the same, they are genuinely different, which can be seen easily when writing their action on $u_\pm\in\Gamma(C_\pm)$ explicitly as
    \begin{equation}
    \begin{aligned}
      \left(F^+(\delta_1\gm,\delta_2\gm)u_+\right)^a&=\tfrac14(\delta_1\gm^{a\alpha}\delta_2\gm_{b\alpha}-\delta_2\gm^{a\alpha}\delta_1\gm_{b\alpha})u_+^b\\
      \left(F^-(\delta_1\gm,\delta_2\gm)u_-\right)^\alpha&=\tfrac14(\delta_1\gm^{a\alpha}\delta_2\gm_{a\beta}-\delta_2\gm^{a\alpha}\delta_1\gm_{a\beta})u_-^\beta
    \end{aligned}
    \end{equation}
    
    Now in order to derive the formula \eqref{curvplus}, we first note that the space $\ms M$ can be seen as a product of the spaces of generalised metrics at all points in $M$ and the tautological bundle $\ms C_+$ is the sum of the corresponding tautological bundles. We also note that the connection on $C_\pm$ uses a pointwise (on $M$) orthogonal projection and hence it can be regarded as the sum of the connections on the individual tautological bundles. It thus suffices to show the formula \eqref{curvplus} at a point (or equivalently in the case of a Courant algebroid with a point base $M=\text{pt}$).
    
    Furthermore, since the bundles $\ms S_0$ and $\ms S$ from sections \ref{subsec:field-space} and \ref{subsec:BV-field-space} are constructed pointwise out of the tautological bundles $\ms C_\pm$ (by taking spinor representations, etc.), one immediately obtains formulas for the curvatures on these bundles as well. For instance, the curvature on $\ms S_0$ acts as
  \begin{align}
    \begin{aligned}
      F(\delta_1\gm,\delta_2\gm)\rho&=\tfrac1{16}[\delta_1\gm,\delta_2\gm]_{ab}\gamma^{ab}\rho,\\
      F(\delta_1\gm,\delta_2\gm)\psi^\alpha&=\tfrac1{16}[\delta_1\gm,\delta_2\gm]_{ab}\gamma^{ab}\psi^\alpha+\tfrac14[\delta_1\gm,\delta_2\gm]^\alpha{}_\beta\psi^\beta.
    \end{aligned}
  \end{align}

  \subsection{Calculating the curvature}
    Let thus $E$ be a Courant algebroid with a point base $E\to\text{pt}$. In other words, the total space $E$ is a Lie algebra with an invariant pairing (the bracket will not in fact enter the generalised metric-related calculations in this section). In particular the space $\ms M$ is now finite-dimensional and the bundle $\ms C_+\to \ms M$ is of finite rank.
      
    Fix an arbitrary reference generalised metric $\hat\gm\in\ms M$. We will aim to calculate the curvature of the connection on $\ms C_+$ at this point. First, we note that any other sufficiently close generalised metric corresponds to
  \begin{equation}
    C_+=\{\hat u_++\varphi(\hat u_+)\mid \hat u_+\in \hat C_+\}
  \end{equation}
  for some linear map $\varphi\colon \hat C_+\to\hat C_-$ (from now on we will use $\gm$, $C_+$, and $\varphi$ interchangeably). We thus have a coordinate system in a neighbourhood $\ms U$ around $\hat\gm$ given by the linear space $\on{Hom}(\hat C_+,\hat C_-)$. Similarly, since the orthogonal projection onto $\hat C_+$ provides a linear isomorphism $C_+\cong \hat C_+$ for any $\gm\in\ms U$, all the fibres of $\ms C_+$ above $\ms U$ can be identified with $\hat C_+$. To summarise, a choice of $\hat\gm$ induces both a coordinate system on $\ms M$ and a local trivialisation of the bundle $\ms C_+$. 
  
  Let us now calculate the connection 1-form $A^+$ on $\ms C_+$ in terms of this trivialisation. In order to do that, suppose we want to perform the parallel transport from $\varphi$ to $\varphi'=\varphi+\delta\varphi$, with $\delta\varphi$ infinitesimal. In our trivialisation this corresponds to the composition of orthogonal projections
  \begin{equation}
    \hat p\circ(p|_{C'_+})^{-1}\circ(\hat p|_{C_+})^{-1}\colon \hat C_+\to\hat C_+,
  \end{equation}
  where $p$ and $\hat p$ are orthogonal projections from $E$ onto $C_+$ and $\hat C_+$, respectively (see the figure).
  
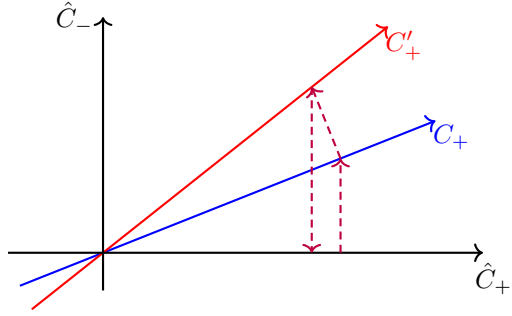
\begin{center}
  \begin{tikzpicture}[scale=2.5]
    \draw[blue, ->, thick] (-.4375,-.175) -- (1.75,.7) node[anchor=north,xshift=6pt,yshift=2pt] {$C_+$};
    \draw[red, ->, thick] (-.375,-.3) -- (1.5,1.2) node[anchor=north,xshift=6pt,yshift=2pt] {$C_+'$};
    \draw[->, thick] (-.5,0) -- (2,0) node[anchor=north,xshift=4pt] {$\hat C_+$};
    \draw[->, thick] (0,-.2) -- (0,1.25) node[anchor=east] {$\hat C_-$};    
    \draw[densely dashed,purple,thick,->] (1.25,0)--(1.25,.5);
    \draw[densely dashed,purple,thick,->] (1.25,.5) -- (145/132,29/33);
    \draw[densely dashed,purple,thick,->] (145/132,29/33) -- (145/132,0);
  \end{tikzpicture}
  \end{center}
  An easy calculation reveals that this is the map
  \begin{equation}
    \hat u_+\mapsto \hat u_+-\varphi^*(1+\varphi\varphi^*)^{-1}\delta\varphi (\hat u_+),
  \end{equation}
  where $\varphi^*\colon \hat C_-\to \hat C_+$ is the transpose of $\varphi$. This implies that at the point $\varphi\in \ms U$ the connection 1-form is
  \begin{equation}
    A^+=\varphi^*(1+\varphi\varphi^*)^{-1}d\varphi,
  \end{equation}
  and hence for the curvature at $\varphi=0$ (i.e.\ at the reference point $\hat\gm$) we have $F^+=d\varphi^*\wedge d\varphi$, i.e.
  \begin{equation}
    F^+(\delta_1\varphi,\delta_2\varphi)=\delta_1\varphi^*\delta_2\varphi-\delta_2\varphi^*\delta_1\varphi\colon \hat C_+\to \hat C_+.
  \end{equation}
  Since for infinitesimal variations we have $\delta\varphi^\beta{}_a=\tfrac12\delta\gm^\beta{}_a$, this yields
  \begin{equation}\label{curvature_pointwise}
    F^+(\delta_1\gm,\delta_2\gm)=\tfrac14[\delta_1\gm,\delta_2\gm]\colon \hat C_+\to\hat C_+. 
  \end{equation}
  
  \subsection{Connection on the space of generalised metrics}\label{app:connectiontangentspaceoffieldspace}
    Now that we have a connection (constructed in a pointwise way) on the bundles $\ms C_\pm$, we also obtain a connection on the tangent bundle to $\ms M$, since (see section \ref{action-symmetries})
    \begin{equation}
      T_\gm\ms M\cong \Gamma(C_+\otimes C_-).
    \end{equation}
    In other words, in order to construct this connection, we can again start with the product of tautological bundles over the space of generalised metrics $\ms M(m)$ at any given point on $m\in M$, and then sum the corresponding connections together.
    
    In particular, the expression for the torsion of the connection on $\ms M$ should be purely algebraic, just as the one for the curvature \eqref{curvplus}. However, there is no invariant way to construct a scalar out of $\otimes^3(C_+\otimes C_-)$, and hence the torsion of this connection must vanish.
    
    Finally, recall from section \ref{subsec:field-space} that the orthogonal projection $E\to C_+$ restricted to an infinitesimally deformed $C_+'$ is in fact an isometry. Thus the connection on the tangent bundle to $\ms M(m)$ preserves the natural metric on this space. Now, in order to define a metric on $\ms M$, we have to fix a half-density $\sigma\in \ms H^*$, so that we can write
    \begin{equation}
      T_\gm\ms M\cong \Gamma(C_+\otimes C_-)\ni u,v \mapsto \int_M\sigma^2\la u,v\ra.
    \end{equation}
    Nevertheless, for any choice of $\sigma$ the inner product on $\ms M$ is simply the sum of inner products over all the points in $M$, with weights provided by $\sigma$. Since the connection on $\ms M$ is constructed in a pointwise manner, it follows that it preserves the metric on $\ms M$ for any choice of the half-density.
    
    We conclude that (for any choice of half-density) the natural connection on the tangent bundle to $\ms M$, constructed out of the connections on the tautological bundles, is Levi-Civita.

    For the purposes of the present text we are however interested in the connection on the tangent space to the entire product $\ms M\times\ms H^*$. This comes as the sum of the above connection on $\ms M$ and the trivial (torsion-free) connection on $\ms H^*$, where the latter arises from the fact that the tangent spaces at all points in $\ms H^*$ are canonically identified with the same space $\Gamma(H)$. Thus the combined connection on $T(\ms M\times\ms H^*)$ is also torsion-free.\footnote{Note that $\ms M\times \ms H^*$ carries a natural metric, in which the norm squared of a tangent vector $u+\tau\in T_\gm\ms M\oplus T_\sigma\ms H^*$ is $\int_M\sigma^2 \la u,v\ra+\kappa\int_M\tau^2$ for some constant $\kappa$, cf.\ \cite{Streets:2024rfo}. The natural torsion-free connection described above however does \emph{not} preserve this metric.}

  \section{Commutator of vector fields on a vector bundle}\label{app:commutator}
  
In this appendix, we demonstrate how the expression for the commutator of vector fields on the total space of a vector bundle picks up terms involving the curvature if one expands them in appropriate bases tangent to the horizontal and vertical subspaces defined by a connection. We do this in a mathematically precise fashion. The takeaway message, in terms which may be preferable to more physically inclined readers, is that if we expand our vector fields in the basis~\eqref{horver}, which is ``covariant" under gauge transformations of the underlying vector bundle, then one has to account for the commutators of the basis elements~\eqref{eq:frame-brackets}. As explained in the main text, one should consider the components of the supersymmetry variations in supergravity to be the expansion of a vector field on field space with respect to such a basis. Hence when computing a commutator of transformations, one needs formulae of the kind presented in this appendix.

  
    \subsection{Statement}
  Suppose that we have a vector bundle $\pi\colon E\to M$ with a connection $\nabla$. This induces a splitting of the tangent space of $E$ to a horizontal and vertical subspace. Thus we have an identification
    \begin{equation}
      TE=T_{\text{hor}}E\oplus T_{\text{ver}}E\cong \pi^*TM\oplus \pi^*E,
    \end{equation}
    where $\pi^*TM$ and $\pi^*E$ denote the pullback vector bundles.
    
    Suppose now that we have a vector field $V$ on the total space, which is given in terms of the above decomposition as a pair
    \begin{equation}
      B\in\Gamma(\pi^*TM),\qquad C\in \Gamma(\pi^*E).
    \end{equation}
    A direct calculation (see below) implies that if we have two such vector fields $V_1$ and $V_2$, encoded by pairs $(B_1,C_1)$ and $(B_2,C_2)$, then the associated pair $(B,C)$ for the commutator $V=[V_1,V_2]$ is
    \begin{equation}\label{commutator}
      B=[B_1,B_2]+C_1B_2-C_2B_1,\qquad C=[C_1,C_2]+\nabla_{B_1}C_2-\nabla_{B_2}C_1-F(B_1,B_2),
    \end{equation}
    where the coordinate-free meaning of the terms is as follows (alternatively, see the next subsection for the coordinate description).
    
    First, take a point $e\in E$ in the fibre over $m\in M$. Let $F\subset E$ be any submanifold which passes through $e$ and which is tangent to the horizontal subspace at $e$. Locally (around $e$) we can thus identify $F\cong M$ via the projection $\pi\colon E\to M$. Under this identification we can see $B|_F$ as a vector field on $M$ and $C|_F$ as a section of $E$. Then
    \begin{itemize}
      \item the terms $[B_1,B_2]$, $\nabla_{B_1}C_2$, and $\nabla_{B_2}C_1$ are understood in terms of this identification
      \item since the bundle $\pi^*E$ becomes trivial over the fibre $\pi^{-1}(m)$,\footnote{i.e.\ at all points in the fibre $\pi^{-1}(m)$, the fibre of the bundle $\pi^*E$ is simply $\pi^{-1}(m)$} it carries a natural action of vertical vector fields on $\pi^{-1}(m)$; this is the meaning of the action of $C$ on $B$ in $C_1B_2$ and $C_2B_1$
      \item $[C_1,C_2]$ is the ordinary commutator of vertical vector fields on $E$
      \item $F(B_1,B_2)\colon E\to E$ is seen as a section of $\pi^*E$ linear in the fibres of $E$.
    \end{itemize}


  \subsection{Proof}
    To show this result, pick local coordinates $x^i$ on $M$, and a local frame $\varepsilon_a$ of $E$, giving linear fibre coordinates $y^a$. The connection one-form $A$ is defined via
        \begin{equation}
          \nabla u=du+Au,\qquad \text{i.e.\ }\quad (\nabla_{\partial_{x^i}} u)^a=\partial_{x^i}u^a+A_i{}^a{}_{b}u^b,
        \end{equation}
        for any $u\in\Gamma(E)$. Thus $\nabla u$ vanishes at $x_0\in M$ iff at that point we have to first order
        \begin{equation}
          u(x_0+\Delta x)=u(x_0)-\varepsilon_aA_i{}^a{}_{b}(x_0)u^b(x_0)\Delta x^i.
        \end{equation}
        The tangent space at $u(x_0)$ to the image of the section $u$ satisfying $\nabla u(x_0)=0$ is therefore spanned by
        \begin{equation}
          \partial_{x^i}-A_i{}^a{}_{b}(x_0)u^b\partial_{y^a}.
        \end{equation}
        We identify this with the horizontal space at $u(x_0)\in E$. Thus the vertical and horizontal subspaces at any $e\in E$ are
        \begin{equation}\label{horver}
          \on{ver}=\text{span}\{e_a\},\quad \on{hor}=\text{span}\{e_i\},\qquad \text{where}\qquad e_a\coloneqq \partial_{y^a},\quad e_i\coloneqq\partial_{x^i}-A_i{}^a{}_{b}(x)y^b\partial_{y^a}.
        \end{equation}
        
        A vector field $V\in\mf X(E)$ is then written in terms of $B$ and $C$ as
    \begin{equation}
      V(x,y)=B^i(x,y)e_i+C^a(x,y)e_a.
    \end{equation}
    A direct calculation then produces
    \begin{equation}\label{eq:frame-brackets}
      [e_a,e_b]=0,\qquad [e_a,e_i]=-A_i{}^b{}_a(x)e_b,\qquad [e_i,e_j]=-F_{ij}{}^a{}_b(x)e_a,
    \end{equation}
    from which we get
    \begin{align*}
      [V_1,V_2]&=(B_1^je_jB_2^i-B_2^je_jB_1^i-C_2^ae_aB_1^i+C_1^ae_aB_2^i)e_i\lbl{commutatorproofresult}\\
      &\quad+(B_1^ie_iC_2^a-B_2^ie_iC_1^a+C_1^be_bC_2^a-C_2^be_bC_1^a+A_i{}^a{}_bB_1^iC_2^b-A_i{}^a{}_bB_2^iC_1^b-B_1^iB_2^jF_{ij}{}^a{}_by^b)e_a,
    \end{align*}
    corresponding to \eqref{commutator}.


\section{Commutator of supersymmetries} \label{sec:closurecomputation}
Here we will calculate the commutator of two supersymmetry transformations
\begin{equation}
  [\delta_1,\delta_2]:=[\delta_{\epsilon_1},\delta_{\epsilon_2}].
\end{equation}
To the best of our knowledge this calculation has never been explicitly shown in the literature for a higher-dimensional supergravity.
It is thus quite remarkable that the whole computation for the $\ms N=1$, $D=10$ case can be presented in only a few pages. This is a consequence of the generalised-geometric formulation and the ``vector bundle'' treatment of the field space (see section \ref{sec:closure}). Recall that the latter essentially means avoiding the introduction of the vielbein as the primary metric degree of freedom. In terms of concrete calculations, the Lorentz-transformations-looking terms will be precisely cancelled by the curvature term in \eqref{commutator}.

To proceed, we will introduce the following name for the bilinear of $\epsilon_1$ with $\epsilon_2$:
\begin{equation}
    V^a \coloneqq \tfrac14 \dil^{-2} (\bar \epsilon_2 \gamma^a \epsilon_1)
\end{equation}
Finally, in order to make the formulas a bit more transparent, we will drop the terms $\delta_{\epsilon_1}\epsilon_2$ and $\delta_{\epsilon_2}\epsilon_1$, which are in general present (cf.\ \eqref{vandzeta}).\footnote{More precisely, recall from section \ref{sec:closure} that $\epsilon_i$ are field dependent, i.e.\ they are sections of a particular vector bundle over the field space $\ms S_0$. The commutator calculation happens at a given field configuration, i.e.\ a point in $\ms S_0$, and so $\epsilon_i$ need to be defined in some neighbourhood of this point. ``Dropping $\delta_{\epsilon_1}\epsilon_2$ terms'' thus simply means that, after choosing the values of $\epsilon_i$ at this point, we extend them to the neighbourhood in a way that at that point the covariant derivative of $\epsilon_1$ along the vector $\delta_{\epsilon_2}\in T\ms S_0$ vanishes, and vice versa.}

\subsection{Bosonic fields}
We first consider the bosonic fields where the computations are relatively tractable and give us a good consistency check for the closure on the fermionic fields. Concretely, we will show that on $\gm$ and $\sigma$ the commutator of two supersymmetries produces the sum of a generalised diffeomorphism and a supersymmetry
\begin{equation}
  [\delta_1,\delta_2]=\delta_V+\delta_\zeta,
\end{equation}
where the supersymmetry parameter is
\begin{equation}
  \zeta:=-\tfrac12\slashed V\rho.
\end{equation}

Starting with the generalised metric $\gm$ we directly calculate
\begin{equation}
  [\delta_1, \delta_2] \G_{a \alpha} = \tfrac12\delta_1(\sigma^{-2}\bar\epsilon_2\gamma_a\psi_\alpha) \minot= -\dil\inv3 (\delta_1 \dil) (\bar \epsilon_2 \gamma_a \Gino_\alpha) + \tf12 \dil\inv2 (\bar \epsilon_2 \gamma_a \delta_1 \Gino_\alpha) \minot.
\end{equation}
Note that the first and second term in the result (including the antisymmetrisation in $1\leftrightarrow2$) correspond in \eqref{commutator} to $[B_1,B_2]$ and $C_1B_2-C_2B_1$, respectively. Continuing, we get
\begin{equation}
  \begin{split}\label{eq:metric_commutator}
   [\delta_1, \delta_2] \G_{a \alpha} &= -\tf18 \dil\inv4 (\bar \dilino \epsilon_1) (\bar \epsilon_2 \gamma_a \Gino_\alpha) + \tf12 \dil\inv2 (\bar \epsilon_2\gamma_a D_\alpha \epsilon_1) + \tf1{16} \dil\inv4 (\bar \Gino_\alpha \dilino) (\bar \epsilon_2 \gamma_a \epsilon_1) \vphantom{\above{\eqref{eq:acG1_p}}}\\
                                     &\quad + \tf1{16}\dil\inv4 (\bar \epsilon_2\gamma_a \gamma^b \dilino)(\bar\Gino_\alpha \gamma_b \epsilon_1 ) \minot \vphantom{\above{\eqref{eq:acG1_p}}}\\
                                     &\above{\eqref{eq:acG1_p}}\;\;\tf12 \dil\inv2(\bar \epsilon_2 \gamma_a D_\alpha \epsilon_1) + \tf14 \dil\inv2 (\bar \Gino_\alpha \dilino) V_a 
                                     + \tf1{16} \dil\inv4 (\bar \epsilon_1 \gamma_b \Gino_\alpha) (\bar \epsilon_2 \gamma^b \gamma_a \dilino) \minot\\
                                     &= \vphantom{\above{\eqref{eq:acG1_p}}}\tf12\dil\inv2[(\bar \epsilon_2 \gamma_a D_\alpha \epsilon_1) + (D_\alpha \bar \epsilon_2 \gamma_a \epsilon_1)] + \tf12 \dil\inv2(\bar \Gino_\alpha \dilino) V_a + \tf18 \dil\inv4 (\bar \epsilon_{[1} \gamma_{|b} \Gino_{\alpha|})(\bar \epsilon_{2]} \gamma^b \gamma_a\rho) \\
  &\above{\eqref{eq:anti-sym-fierz}}\;\; 2 D_\alpha V_a + \tf12 \dil\inv2 (\bar \Gino_\alpha \dilino) V_a - \tf14 \dil\inv2 V_b (\bar \Gino_\alpha \gamma^b \gamma_a \dilino)\\
  &\above{\eqref{eq:acG1_p}} \ \ 2 D_{\alpha} V_{a} + \tfrac14 \dil^{-2} \left(\bar\Gino_\alpha \gamma_a \slashed V \dilino \right) \smash{\overset{\eqref{eq:susyvariations}}=} \delta_V \G_{a \alpha} + \delta_{\zeta } \G_{a \alpha},
  \end{split}
\end{equation}
where in the last step we used the formula \cite{Streets:2024rfo}
\begin{equation}\label{killing}
  (\ms L_X\gm)_{a\alpha}=2(D_{\alpha}X_{a}-D_aX_\alpha),
\end{equation}
valid for any $X\in\Gamma(E)$ and a Levi-Civita connection $D$.

The calculation for the dilaton is easier and gives
\begin{equation}
  \begin{split}
    [\delta_1, \delta_2 ]\dil &=\tfrac18\delta_1[\sigma^{-1}(\bar \rho\epsilon_2)] \minot\\
    &= -\tfrac1{64} \sigma^{-3}(\bar\dilino \epsilon_1)(\bar \dilino \epsilon_2) + \tfrac18\sigma^{-1}(\ol{\di \epsilon_1}\epsilon_2) + \tfrac1{8\times 192} \dil^{-3} (\Gino_\alpha \gamma_{(3)} \Gino^\alpha) (\bar \epsilon_1 \gamma^{(3)} \epsilon_2) \minot \vphantom{\overset{\eqref{eq:divergence}}=}\\
                                    & = \tfrac18 \dil^{-1} (\bar \epsilon_2\di \epsilon_1 - \bar \epsilon_1\di \epsilon_2) = \tfrac12\dil D_a \left( \tfrac14 \dil^{-2} \ol{\epsilon}_2 \gamma^a \epsilon_1\right) \overset{\eqref{eq:divergence}}= \gld_V \dil=\delta_V\dil=\delta_V\dil+\delta_\zeta\dil,
  \end{split}
\end{equation}
where in the last equality we used the fact that
\begin{equation}
  \delta_\zeta \dil = -\tf1{16} \dil\inv1 \bar \dilino \slashed V \dilino = 0,
\end{equation}
due to the symmetry of $\gamma_a$.

\subsection{Fermionic fields}\label{app:commutator-fermoins}
Before proceeding, we record here the variation formulae for the kinetic operators, derived in \cite{Kupka:2024vrd}:
\begin{align} \label{eq:connection-on-spinor-variation}
  \begin{aligned}
    (\delta\di)\dilino&=\tfrac12\delta \G^{\alpha }{}_a\gamma^aD_{\alpha }\dilino+\tfrac14 (D_{\alpha } \delta\G^{\alpha }{}_a)\gamma^a\dilino,\\
    (\delta\di)\Gino^{\alpha }&=\tfrac12\delta \G^{\gamma }{}_a\gamma^aD_{\gamma }\Gino^{\alpha }+\tfrac14 (D_{\gamma } \delta\G^{\gamma }{}_a)\gamma^a\Gino^{\alpha }+(D^{[\alpha }\delta\G^{\gamma ]}{}_a)\gamma^a\Gino_{\gamma },\\
    (\delta D_{\alpha })\dilino&=-\tfrac12\delta\G_{\alpha }{}^aD_a\dilino-\tfrac14(D_b\delta\G_{\alpha c})\gamma^{bc}\dilino-(D_{\alpha }\tfrac{\delta\dil}\dil)\dilino,
  \end{aligned}
\end{align}
Also note that if $V\in\Gamma(C_+)$ then for any spinor half-density $\lambda$ we have
\begin{equation}\label{eq:CP-spinor-LD}
    \gld_V \lambda \;\above{\eqref{eq:diffeo-on-spinor-half-density}}\;\; V^a D_a \lambda + \tfrac12 D_a V_b \gamma^{a b}\lambda + \tfrac12 D_a V^a \lambda=\tf12 ( \di \slashed V + \slashed V \di) \lambda = \tf12 \{\di, \slashed V \} \lambda
\end{equation}
Finally, we recall from section \ref{action-symmetries} that one can write the supersymmetry transformations of $\dilino, \Gino$ as
\begin{equation}\label{rewriting-of-variations}
    \begin{split}
        \delta_\epsilon \dilino &=\di \dilino + \tf14 \delta \G_{a \alpha} \gamma^a \Gino^\alpha,\\
        \delta_\epsilon \Gino_\alpha &= D_\alpha \epsilon + \tf18 \dil\inv2 (\bar \Gino_\alpha \dilino) \epsilon - \tf14 \delta \G_{\alpha a} \gamma^a \dilino,
    \end{split}
\end{equation}
which significantly simplifies the commutator computations below.

\subsubsection{The dilatino}
We now wish to derive
\begin{align}
    [\delta_1, \delta_2] \dilino &= \delta_{V}\dilino + \delta_{\zeta}\dilino -\tf12 \slashed V \eom{\dilino}.
\end{align}
To start, using the second half of the formula \eqref{commutator} (most crucially the curvature) we get
\begin{align*}
    [\delta_1, \delta_2] \dilino &= \delta_1 [\di \epsilon_2 + \tf14 \delta_2 \G_{a \alpha} \gamma^a\Gino^\alpha] \minot -F(\delta_1\gm,\delta_2\gm)\rho\\
    &\above{\smash{\eqref{eq:field-space-curvatures}}}\; (\delta_1\di) \epsilon_2 + \tf14 \delta_1 \delta_2 \G_{a \alpha} \gamma^a \Gino^\alpha + \tf14 \delta_2 \gm_{a \alpha} \gamma^a [D^\alpha \epsilon_1 + \tf18 \dil\inv2 (\bar \Gino ^\alpha \dilino) \epsilon_1 - \tf14 \delta_1 \G\indices{_b^\alpha} \gamma^b \dilino] \minot\\
    &\qquad -\tf1{16}[\delta_1 \G, \delta_2 \G]_{ab} \gamma^{ab} \dilino\lbl{dilatinocommutator}\\
    &=(\delta_1\di) \epsilon_2 - \tf14 \delta_1 \G_{a \alpha} \gamma^a D^\alpha \epsilon_2 + \tf1{32} \dil\inv2 (\bar \Gino^\alpha \dilino) \delta_2 \G_{a \alpha} \gamma^a \epsilon_1 \minot+ \tf14 [\delta_1, \delta_2] \G_{a \alpha} \gamma^a \Gino^\alpha.
\end{align*}
We observe that the first two terms combine in the following way
\begin{equation}
    \begin{split}
       (\delta_1\di) \epsilon_2 - \tf14& \delta_1 \G_{a \alpha} \gamma^a D^\alpha \epsilon_2  \minot \;\;\,\smash{\above{\eqref{eq:connection-on-spinor-variation}}}\;\; \tf14 \delta_1 \G_{a \alpha} \gamma^a D^\alpha \epsilon_2 + \tf14 D_\alpha \delta_1 \G\indices{_a^\alpha} \gamma^a \epsilon_2 \minot\\
       &= \tf14D_\alpha[\delta_1 \G\indices{_a^\alpha} \gamma^a \epsilon_2 \minot] = \tf18 D_\alpha[\dil\inv2 (\bar \epsilon_1 \gamma_a \Gino^\alpha) \gamma^a \epsilon_2 \minot] \\
       &\above{\smash{\eqref{eq:fierzG1_1}}} -\tf18 D_\alpha[\dil\inv2 (\bar \epsilon_2 \gamma_a \epsilon_1) \gamma^a \Gino^\alpha] = -\tf12 D_\alpha(\slashed V \Gino^\alpha) = - \tf12 D_\alpha V_a \gamma^a \Gino^\alpha - \tf12 \slashed V D_\alpha \Gino^\alpha \\
       &= -\tf14 \gld_V \G_{a \alpha} \gamma^a \Gino^\alpha - \tf12 \slashed V D_\alpha \Gino^\alpha.
    \end{split}
\end{equation}
Plugging this back into \eqref{dilatinocommutator} and using \eqref{eq:metric_commutator} we get
\begin{equation}
    \begin{split}
        [\delta_1, \delta_2] \dilino = \tf14 \delta_\zeta \G_{a \alpha} \gamma^a \Gino^\alpha - \tf12 \slashed V D_\alpha \Gino^\alpha + \tf1{32} \dil\inv2[(\bar \Gino_\alpha \dilino) \delta_2 \G\indices{_a^\alpha} \gamma^a \epsilon_1 \minot]\label{dilatinointermediate}.
    \end{split}
\end{equation}
We can rewrite the first term as
\begin{equation}
    \begin{split}
        \tf14 \delta_\zeta \G_{a \alpha} \gamma^a \Gino^\alpha = \delta_\zeta \dilino - \di \zeta = \delta_\zeta \dilino + \tf12 \di \slashed V \dilino \;\above{\smash{\eqref{eq:CP-spinor-LD}}}\ \delta_\zeta \dilino + \delta_V \dilino - \tf12 \slashed V \di \dilino.
    \end{split}
\end{equation}
Next, we investigate the last term. It is helpful to multiply by an auxiliary odd spinor $\bar\chi$, with the same chirality as $\psi$, to keep track of Fierz identities and permutation signs:
\begin{equation}
    \begin{split}
        \tf1{32} \dil\inv2(\bar \Gino_\alpha \dilino) \delta_2 \G\indices{_a^\alpha} (\bar \chi \gamma^a \epsilon_1) \minot &= -\tf1{32} \dil\inv4 (\bar \Gino_\alpha\rho)(\bar \epsilon_{[2} \gamma_a \Gino^\alpha) (\bar \epsilon_{1]}\gamma^a \chi) \\
                                                      &\above{\eqref{eq:anti-sym-fierz}}\ \tf1{64} \dil\inv4 (\bar \Gino_\alpha \dilino) (\bar \epsilon_2 \gamma_a \epsilon_1) (\bar \chi \gamma^a \Gino^\alpha)= \tf1{16} \dil\inv2 (\bar \Gino_\alpha \dilino) (\bar \chi \slashed V \Gino^\alpha)\\
                                                      &\above{\eqref{eq:master-fierz}}\ \tf1{1536} \dil\inv2 (\bar \chi \slashed V \gamma_{(3)} \dilino ) \ginobilu{\alpha}{(3)}.
    \end{split}
\end{equation}
Plugging things back into \eqref{dilatinointermediate} we finally obtain
\begin{equation}
  \begin{split}
    [\delta_1, \delta_2] \dilino &= \delta_V \dilino + \delta_\zeta \dilino - \tf12 \slashed V [\di \dilino + D_\alpha \Gino^\alpha-\tf1{768}\dil\inv2 \Ginobild{\alpha}{(3)} \gamma^{(3)} \dilino ]\\
    &= \delta_V \dilino + \delta_\zeta \dilino - \tf12 \slashed V \eom{\dilino}.
  \end{split}
\end{equation}

\subsubsection{The gravitino}
Finally, we will now show that for the gravitino we have
\begin{equation}\label{psicommutatorappendix}
  [\delta_1, \delta_2] \Gino_\alpha = \delta_{V}\Gino_\alpha + \delta_{\zeta}\Gino_\alpha + (\tf14 \sigma^{-2}\epsilon_{[2} \bar \epsilon_{1]} - \tf12 \slashed V ) \eom{\Gino}_\alpha.
\end{equation}
To do that, we again use \eqref{commutator} and \eqref{rewriting-of-variations} to calculate
\begin{align*}
    [\delta_1, \delta_2] \Gino_\alpha &=  \delta_1 ( D_\alpha \epsilon_2 + \tf18 \dil\inv2 (\bar \Gino_\alpha \dilino)\epsilon_2 - \tf14 \delta_2 \G_{\alpha a} \gamma^a \dilino) \minot -F(\delta_1\gm,\delta_2\gm)\psi_\alpha\\
    &\above{\smash{\eqref{eq:field-space-curvatures}}}\; (\delta_1 D_\alpha) \epsilon_2 - \tfrac1{32}\dil\inv4 (\bar \dilino \epsilon_1) (\bar \Gino_\alpha \dilino) \epsilon_2 + \tf18 \dil\inv2 \bar \dilino [D_\alpha \epsilon_1 + \tf18 \dil\inv2 (\bar \Gino_\alpha \dilino) \epsilon_1 - \tf14 \delta_1 \G_{a \alpha} \gamma^a \dilino]\epsilon_2\\
    &\quad +\tf18 \dil\inv2 \bar \Gino_\alpha(\di \epsilon_1 + \tf14 \delta_1 \G_{a\beta}\gamma^a \Gino^\beta)\epsilon_2 - \tf14 \delta_1 \delta_2 \G_{a \alpha} \gamma^a \dilino - \tf14 \delta_2 \G_{a\alpha}\gamma^a(\di \epsilon_1 + \tf14 \delta_1 \G_{b \beta} \gamma^b \Gino^\beta) \\&\quad\minot - \tfrac1{16}[\delta_1\gm,\delta_2\gm]_{ab}\gamma^{ab}\psi_\alpha-\tfrac14[\delta_1\gm,\delta_2\gm]_{\alpha\beta}\psi^\beta\\
    &\above{\smash{\eqref{bilinearvanish}}}\;\, [(\delta_1 D_\alpha) \epsilon_2 + \tf18\dil\inv2 (\bar \dilino D_\alpha \epsilon_1) \epsilon_2 + \tf18 \dil\inv2 (\bar \Gino_\alpha \di \epsilon_1) \epsilon_2 - \tf14 \delta_2 \G_{a\alpha} \gamma^a \di \epsilon_1 \minot]\\
    &\quad + [\tf{1-2}{64} \dil\inv4 (\bar \Gino_\alpha \dilino)(\bar \dilino \epsilon_1) \epsilon_2+ \tf1{32} \dil\inv2 \delta_1 \G_{a\beta} (\bar \Gino_\alpha \gamma^a \Gino^\beta) \epsilon_2+ \tf1{16} \delta_1 \G_{a \alpha} \delta_2 \G_{b \beta} \gamma^a \gamma^b \Gino^\beta \minot ]\\
    &\quad-\tf14[\delta_1, \delta_2] \G_{a \alpha} \gamma^a \dilino- \tfrac1{16}[\delta_1\gm,\delta_2\gm]_{ab}\gamma^{ab}\psi_\alpha-\tfrac14[\delta_1\gm,\delta_2\gm]_{\alpha\beta}\psi^\beta\lbl{gravitino-start-derivation}\\
    &\above{\smash{\eqref{eq:master-fierz}\&\eqref{eq:metric_commutator}}}\hspace{.6cm} [(\delta_1 D_\alpha) \epsilon_2 + \tf18\dil\inv2 (\bar \dilino D_\alpha \epsilon_1) \epsilon_2 + \tf18 \dil\inv2 (\bar \Gino_\alpha \di \epsilon_1) \epsilon_2 - \tf14 \delta_2 \G_{a\alpha} \gamma^a \di \epsilon_1 \minot\\
    &\hspace{1cm}-\tf14 \gld_V \G_{a \alpha} \gamma^a \dilino]\\
    &\quad + [\tf1{32} \dil\inv2 \delta_1 \G_{a\beta} (\bar \Gino_\alpha \gamma^a \Gino^\beta) \epsilon_2+ \tf1{16} \delta_1 \G_{a \alpha} \delta_2 \G_{b \beta} \gamma^a \gamma^b \Gino^\beta \minot] -\tf14 \delta_\zeta \G_{a \alpha} \gamma^a \dilino\\
    &\qquad-\tf1{3072} \dil\inv 4 \epsilon_{[2}\bar \epsilon_{1]} (\bar \dilino \gamma_{(3)} \dilino) \gamma^{(3)} \Gino_\alpha- \tfrac1{16}[\delta_1\gm,\delta_2\gm]_{ab}\gamma^{ab}\psi_\alpha-\tfrac14[\delta_1\gm,\delta_2\gm]_{\alpha\beta}\psi^\beta
\end{align*}
We will write this result for simplicity as
\begin{equation}
  [\delta_1, \delta_2] \Gino_\alpha=\text{Der}+\text{Alg},
\end{equation}
where Der denotes the first two lines in the result (the terms containing a derivative) and Alg stands for the remainder (purely algebraic terms).

Looking at the two terms separately, we get for the first one
\begin{align*}
        \text{Der} \,\,&\above{\smash{\eqref{eq:connection-on-spinor-variation}}} -\tf12 \delta_1 \G\indices{_\alpha^a}D_a \epsilon_2 - \tf14 D_a \delta_1 \G_{\alpha b} \gamma^{ab}\epsilon_2 - \tf18\dil\inv2 D_\alpha(\bar \dilino \epsilon_1) \epsilon_2 + \tf18 \dil\inv2 (\bar \dilino D_\alpha \epsilon_1) \epsilon_2 \\
        &\quad+ \tf18\dil \inv2(\bar \Gino_\alpha \di \epsilon_1) \epsilon_2 - \tf14 \delta_2 \G_{a \alpha} \gamma^a \di \epsilon_1 \minot -\tf14 \gld_V \G_{a \alpha} \gamma^a \dilino\\
        &\above{\smash{\eqref{killing}}}-\tf12 \delta_1 \G\indices{^a_\alpha} D_a \epsilon_2 - \tf14 \delta_2 \G_{a\alpha} \gamma^a \gamma^b D_b \epsilon_1- \tf14 D_a \delta_1 \G_{b\alpha} \gamma^{ab} \epsilon_2 +\tf18 \dil \inv2 (\bar \Gino_\alpha \di \epsilon_1)\epsilon_2 \\
        &\quad- \tf18 \dil\inv2 \epsilon_2 (\bar \epsilon_1 D_\alpha \dilino) \minot -\tf12 D_\alpha V_a \gamma^a \dilino\lbl{line1}\\
        &=-\tf12 \delta_1 \G\indices{^a_\alpha} D_a \epsilon_2 + \tf14 \delta_1 \G_{a\alpha} \gamma^a \gamma^b D_b \epsilon_2- \tf14 D_a \delta_1 \G_{b\alpha} \gamma^{ab} \epsilon_2 -\tfrac14 D_a \delta_1 \G\indices{^a_\alpha}\epsilon_2 + \tfrac18 \dil\inv2(\bar \epsilon_1 \di \Gino_\alpha)\epsilon_2 \\
        &\quad- \tf18 \dil\inv2 \epsilon_2 (\bar \epsilon_1 D_\alpha \dilino) \minot - \tf12 D_\alpha(\slashed V \dilino) + \tf12 \slashed V D_\alpha \dilino\\
        &=- \tf14 \delta_1 \G_{a\alpha} \gamma^b \gamma^a D_b \epsilon_2- \tf14 D_a \delta_1 \G_{b\alpha} \gamma^{a}\gamma^b \epsilon_2 + \tf18\dil\inv2 \epsilon_2 \bar \epsilon_1 (\di \Gino_\alpha -D_\alpha \dilino) \minot +D_\alpha\zeta + \tf12 \slashed V D_\alpha \dilino\\
        &= -\tf14 D_a(\delta_1 \G_{b \alpha} \gamma^a \gamma^b \epsilon_2) \minot + \tf14 \dil\inv2 \epsilon_{[2}\bar \epsilon_{1]}(\di \Gino_\alpha -D_\alpha \dilino)+D_\alpha\zeta + \tf12 \slashed V D_\alpha \dilino.
\end{align*}
The first term can be rewritten using
\begin{equation}
    \begin{split}
         \delta_1 \G_{b \alpha} \gamma^a \gamma^b \epsilon_2 \minot &= - \dil\inv2 \gamma^a \gamma^b \epsilon_{[2}(\bar \Gino_\alpha \gamma_b \epsilon_{1]}) \,\,\above{\smash{\eqref{eq:anti-sym-fierz}}}\; -\tf12 \dil\inv2 \gamma^a \gamma_b \Gino_\alpha (\bar \epsilon_2 \gamma^b \epsilon_1) = -2 \gamma^a \slashed V \Gino_\alpha,
    \end{split}
\end{equation}
and hence we find that the derivative terms take the form
\begin{equation}\label{derresult}
    \begin{split}
        \text{Der} &= \tf12 \di \slashed V \Gino_\alpha + \tf14 \dil\inv2 \epsilon_{[2}\bar \epsilon_{1]}(\di \Gino_\alpha -D_\alpha \dilino)+D_\alpha\zeta + \tf12 \slashed V D_\alpha \dilino\\
        &\above{\smash{\eqref{eq:CP-spinor-LD}}}\; \gld_V \Gino_\alpha +D_\alpha\zeta + (\tf14 \dil\inv2 \epsilon_{[2}\bar \epsilon_{1]}-\tfrac12\slashed V)(\di \Gino_\alpha -D_\alpha \dilino).
    \end{split}
\end{equation}

Now looking at the terms in Alg, we note that in order to fully assemble the $\delta_\zeta\psi_\alpha$ term on the RHS of \eqref{psicommutatorappendix}, we are still missing the middle term in the second formula of \eqref{rewriting-of-variations}. This can however be written as
\begin{equation}
  \begin{split}
    \tf18 \dil\inv2 (\bar \Gino_\alpha \dilino)  \zeta &= - \tf1{16} \dil\inv2 (\bar \Gino_\alpha \dilino) \slashed V \dilino \;\above{\eqref{eq:master-fierz}} -\tf1{1536}  \dil\inv2\slashed V (\bar \dilino \gamma_{(3)} \dilino) \gamma^{(3)} \Gino_\alpha,
  \end{split}
\end{equation}
where we recognise precisely one of the contributions in the equations-of-motion term of \eqref{psicommutatorappendix} (with opposite sign). We can thus use the creative zero
\begin{align} \label{eq:eom-cancels-susy}
   \tf18 \dil\inv2 (\bar \Gino_\alpha \dilino)  \zeta+\tf1{1536} \dil\inv2 \slashed V (\bar \dilino \gamma_{(3)} \dilino) \gamma^{(3)} \Gino_\alpha  = 0,
\end{align}
to write
\begin{equation}\label{algsemiresult}
    \begin{split}
        \text{Alg} &=\tf18 \dil\inv2 (\bar \Gino_\alpha \dilino)  \zeta  -\tf14 \delta_\zeta \G_{a \alpha} \gamma^a \dilino + (\tf14\dil\inv2 \epsilon_{[2}\bar \epsilon_{1]} - \tf12 \slashed V)[-\tf1{768} \dil \inv2 (\bar \dilino \gamma_{(3)} \dilino) \gamma^{(3)} \Gino_\alpha] \\
        &\quad + [\tf1{16} \delta_1 \G_{a \alpha} \delta_2 \G_{b\beta} \gamma^a \gamma^b \Gino^\beta + \tf1{32} \dil\inv2 \delta_1 \G_{a \beta} (\bar \Gino_\alpha \gamma^a \Gino^\beta)\epsilon_2 \minot]\\
        &\quad - \tfrac1{16}[\delta_1\gm,\delta_2\gm]_{ab}\gamma^{ab}\psi_\alpha-\tfrac14[\delta_1\gm,\delta_2\gm]_{\alpha\beta}\psi^\beta.
    \end{split}
\end{equation}
Comparing \eqref{derresult}$+$\eqref{algsemiresult} with \eqref{eq:fieldeom}, we see that in order to complete the proof of \eqref{psicommutatorappendix} it only remains to show
\begin{equation}\label{theremainder}
  \begin{aligned}
    \tf1{16} \delta_1 \G_{a \alpha}& \delta_2 \G_{b\beta} \gamma^a \gamma^b \Gino^\beta + \tf1{32} \dil\inv2 \delta_1 \G_{a \beta} (\bar \Gino_\alpha \gamma^a \Gino^\beta)\epsilon_2 \minot- \tfrac1{16}[\delta_1\gm,\delta_2\gm]_{ab}\gamma^{ab}\psi_\alpha\\
    &-\tfrac14[\delta_1\gm,\delta_2\gm]_{\alpha\beta}\psi^\beta=(\tf14\dil\inv2 \epsilon_{[2}\bar \epsilon_{1]} - \tf12 \slashed V)[-\tf1{192}\dil\inv2 \Ginobild{\beta}{(3)} \gamma^{(3)} \Gino_\alpha].
  \end{aligned}
\end{equation}

First, using an odd spinor $\lambda$ of opposite chirality to $\Gino$ for book keeping, we derive the identity
\begin{align*}
        -(\bar \epsilon_1 \gamma_a \Gino_\alpha) (\bar \lambda \gamma^b \gamma^a \Gino^\beta) &= (\bar \epsilon_1 \gamma_a \Gino_\alpha)(\ol{\gamma^b \lambda} \gamma^a \Gino^\beta) \;\above{\smash{\eqref{eq:anti-sym-fierz}}}\; (\bar \Gino_\alpha \gamma_a \gamma^b \lambda) (\bar \epsilon_1 \gamma^a \Gino^\beta)-(\bar \epsilon_1 \gamma_a \gamma^b \lambda) (\bar \Gino_\alpha \gamma^a \gamma^b\Gino^\beta) \\
        &= \bar \lambda[ 2 \dil^2\gamma^b \gamma^a \Gino_\alpha  \delta_1 \G_a{}^\beta - \gamma^b \gamma_a \epsilon_1 (\bar \Gino_\alpha \gamma^a \Gino^\beta)],\lbl{auxiliaryidentity}
\end{align*}
Second, we note that for the $F^-$ curvature \eqref{curvminus} we can write
\begin{align}\label{eq:lambda_C_minus}
    \tf14 [\delta_{1}\G, \delta_{2}\G]_{\alpha\beta} = \tf1{16} \dil\inv4 (\bar \epsilon_1 \gamma^a \Gino_\alpha) (\bar \epsilon_2 \gamma_a \Gino_\beta) \minot\ \ \above{\eqref{eq:fierzG1_1}} -\tf14 \dil\inv2 V^a (\bar \Gino_\alpha \gamma_a \Gino_\beta).
\end{align}
We can now rewrite the LHS of \eqref{theremainder} as
\begin{align*}
        \text{LHS}\;\,&\above{\smash{\eqref{eq:acG1_p}}}  -\tf1{32} \delta_2 \G_{b \beta} \dil\inv2 (\bar \epsilon_1 \gamma_a \Gino_\alpha) \gamma^b \gamma^a \Gino^\beta + \tf18 \delta_1 \G_{a \alpha} \delta_2 \G\indices{^a_\beta} \Gino^\beta - \tf1{32} \dil\inv2 \delta_2 \G\indices{^\beta_b} (\bar \Gino_\alpha \gamma^b \Gino_\beta) \epsilon_1 \minot\\
        &\quad- \tfrac1{16}[\delta_1\gm,\delta_2\gm]_{ab}\gamma^{ab}\psi_\alpha-\tfrac14[\delta_1\gm,\delta_2\gm]_{\alpha\beta}\psi^\beta\\
        &\above{\smash{\eqref{auxiliaryidentity}}}-\tf1{32} \delta_2 \G_{b \beta} \dil\inv2 [2 \dil^2 \gamma^{ba} \Gino_\alpha  \delta_1 \G_a{}^\beta - \gamma^b \gamma_a \epsilon_1 (\bar \Gino_\alpha \gamma^a \Gino^\beta)] + \tf18 \delta_1 \G_{a \alpha} \delta_2 \G\indices{^a_\beta} \Gino^\beta \\
        &\quad- \tf1{32} \dil\inv2 \delta_2 \G\indices{^\beta_b} (\bar \Gino_\alpha \gamma^b \Gino_\beta) \epsilon_1 \minot- \tfrac1{16}[\delta_1\gm,\delta_2\gm]_{ab}\gamma^{ab}\psi_\alpha-\tfrac14[\delta_1\gm,\delta_2\gm]_{\alpha\beta}\psi^\beta\\
        &=\tf1{32} \dil\inv2\delta_2 \G_{b \beta}  \gamma^b \gamma_a \epsilon_1 (\bar \Gino_\alpha \gamma^a \Gino^\beta)- \tf1{32} \dil\inv2 \delta_2 \G\indices{^\beta_b} (\bar \Gino_\alpha \gamma^b \Gino_\beta) \epsilon_1 \minot-\tfrac18[\delta_1\gm,\delta_2\gm]_{\alpha\beta}\psi^\beta\\
        &=-\tf1{32}\dil\inv2 \delta_2 \G_{b \beta}  \gamma_a \gamma^b \epsilon_1 (\bar \Gino_\alpha \gamma^a \Gino^\beta) + \tf1{32} \dil\inv2 \delta_2 \G_{a \beta} (\bar \Gino_\alpha \gamma^a \Gino^\beta) \epsilon_1 \minot-\tfrac18[\delta_1\gm,\delta_2\gm]_{\alpha\beta}\psi^\beta\\
        &=\tf1{64}\dil\inv4 \gamma^a \gamma^b \epsilon_1 (\bar \Gino_\beta \gamma_b \epsilon_2)(\bar \Gino_\alpha \gamma_a \Gino^\beta) + \tf1{64}\dil\inv4 (\bar \epsilon_2 \gamma_a \Gino^\beta)(\bar \Gino_\alpha \gamma^a \Gino_\beta) \epsilon_1 \minot-\tfrac18[\delta_1\gm,\delta_2\gm]_{\alpha\beta}\psi^\beta\\
        &\above{\smash{\eqref{eq:anti-sym-fierz}}}\tf1{128} \dil\inv4 \gamma^a \gamma^b \Gino_\beta (\bar \epsilon_1 \gamma_b \epsilon_2) (\bar \Gino_\alpha \gamma_a \Gino^\beta)+ \tf1{64}\dil\inv4 (\bar \epsilon_2 \gamma_a \Gino^\beta)(\bar \Gino_\alpha \gamma^a \Gino_\beta) \epsilon_1 \minot-\tfrac18[\delta_1\gm,\delta_2\gm]_{\alpha\beta}\psi^\beta\\
        &\above{\smash{\eqref{eq:master-fierz}}}-\tf1{16}\dil\inv2 \gamma^a \slashed V \Gino_\beta (\bar \Gino_\alpha \gamma_a \Gino^\beta)-\tf1{768}\dil\inv4 \epsilon_{[2} \bar \epsilon_{1]}\Ginobild{\beta}{(3)} \gamma^{(3)} \Gino_\alpha-\tfrac18[\delta_1\gm,\delta_2\gm]_{\alpha\beta}\psi^\beta\lbl{remainder-calculation}\\
        &\above{\smash{\eqref{eq:acG1_p}}}\,\tf1{16}\dil\inv2 \slashed V \gamma^a \Gino_\beta (\bar \Gino_\alpha \gamma_a \Gino^\beta) - \tf18 \dil\inv2 \Gino_\beta (\bar \Gino_\alpha \slashed V \Gino^\beta)-\tf1{768}\dil\inv4 \epsilon_{[2} \bar \epsilon_{1]}\Ginobild{\beta}{(3)} \gamma^{(3)} \Gino_\alpha\\
        &\quad-\tfrac18[\delta_1\gm,\delta_2\gm]_{\alpha\beta}\psi^\beta\\
        &\above{\smash{\eqref{eq:master-fierz}}}\tf1{384} \dil\inv2\slashed V \Ginobild{\beta}{(3)} \gamma^{(3)} \Gino_\alpha - \tf18 \dil\inv2 \Gino_\beta (\bar \Gino_\alpha \slashed V \Gino^\beta)-\tf1{768}\dil\inv4 \epsilon_{[2} \bar \epsilon_{1]}\Ginobild{\beta}{(3)} \gamma^{(3)} \Gino_\alpha\\
        &\quad-\tfrac18[\delta_1\gm,\delta_2\gm]_{\alpha\beta}\psi^\beta\\
        &\above{\smash{\eqref{eq:lambda_C_minus}}}\ (\tf14\dil\inv2 \epsilon_{[2}\bar \epsilon_{1]} - \tf12 \slashed V)[-\tf1{192}\dil\inv2 \Ginobild{\beta}{(3)} \gamma^{(3)} \Gino_\alpha],
\end{align*}
which establishes \eqref{theremainder} and hence finishes the proof of \eqref{psicommutatorappendix}.


  \section{Poisson bracket on the cotangent bundle of a vector bundle}\label{app:cotangent}

In this appendix, we provide a formula for the Poisson structure on the cotangent bundle of the total space of a vector bundle. As in appendix~\ref{app:commutator}, our presentation is quite mathematical. In simple terms, the essence of the result again comes from wishing to write the symplectic form (and thus Poisson structure) as expanded in a ``covariant" basis on the total space of the cotangent bundle, with respect to ``gauge transformations" of the underlying vector bundle, as in the second line of equation~\eqref{eq:total-space-symp-form} below.

    \subsection{Statement}
    Suppose we have a vector bundle $\pi\colon E\to M$, with a connection $\nabla$ on $E$ and a connection $\bar\nabla$ on $TM$.
      First, note that $\nabla$ allows us to identify
      \begin{equation}\label{identificationcotangent}
        T^*E\cong \pi^*(T^*M\oplus E^*).
      \end{equation}
      In other words, a point in $T^*E$ can be specified as $(m,e,p,e^*)$, with
      \begin{equation}
        m\in M,\quad e\in E_m,\quad p\in T^*_mM,\quad e^*\in E^*_m.
      \end{equation}
      Now notice that $\nabla$ and $\bar\nabla$ induce connections on $T^*M$ and $E^*$, and hence also on $T^*M\oplus E^*$ and $\pi^*(T^*M\oplus E)$. Using this we obtain the identification
      \begin{equation}\label{identification-chain}
        \begin{split}
          T_{(m,e,p,e^*)}T^*E&\cong T_{(m,e,p,e^*)}\pi^*(T^*M\oplus E^*)\\
          &\cong T_{(m,e)}E\oplus \pi^*(T^*M\oplus E^*)_{(m,e)}\\
          &\cong (T_mM\oplus E_m)\oplus (T^*M\oplus E^*)_m\\
          &\cong T_mM\oplus E_m\oplus T^*_mM\oplus E^*_m.
        \end{split}
      \end{equation}
      
      Pick a point $(m,e,p,e^*)\in T^*E$, and choose any basis $\varepsilon_i$ of $T_mM$ and $\varepsilon_a$ of $E_m$.
      We now claim that the canonical Poisson structure on $T^*E$ corresponds under the above identification to
      \begin{equation}\label{poisson_on_bundle}
        \pi=\varepsilon^i\wedge \varepsilon_i+\varepsilon^a\wedge\varepsilon_a-\tfrac12\bar T^k{}_{ij}(m)p_k \varepsilon^i\wedge \varepsilon^j-\tfrac12 F_{ij}{}^a{}_b(m)e^be_a^*\varepsilon^i\wedge\varepsilon^j,
      \end{equation}
      where $\bar T$ is the torsion of $\bar\nabla$ and $F$ is the curvature of $\nabla$.
      
      To derive the formula \eqref{poisson_on_bundle} one has to essentially chase the Poisson bracket through the chain of identifications \eqref{identification-chain}, which is done in the following two subsections.
      
      \subsection{Proof part 1: Poisson structure on the pullback}
      We start with the map \eqref{identificationcotangent}, which can be written as
        \begin{equation}\label{cotantent-identification}
          \pi^*(T^*M\oplus E^*)\ni (m,e,p,e^*)\mapsto (m,e,\pi^*p+ie^*)\in T^*E,
        \end{equation}
        with $i\colon E^*_m\to T^*_{(m,e)}E$ defined by $(ie^*)(v)=e^*(\on{ver}v)$. Here $\on{ver}$ is the projection from $T_{(m,e)}E$ onto the vertical subspace.
        
        Recall that for any space $X$, the symplectic form on $T^*X$ can be written as $d\alpha$, with the potential $\alpha$ defined by
        \begin{equation}
          \alpha(\dot\gamma)=\la\gamma,\pi^X_*\dot\gamma\ra,
        \end{equation}
        where $\pi^X\colon T^*X\to X$ is the natural projection and $\dot\gamma$ is the tangent vector to a path $\gamma$ in $T^*X$.\footnote{In the natural coordinates $(x^A,p_A)$ on the cotangent bundle of $X$ one has $\alpha=p_Adx^A$.}
        Using \eqref{cotantent-identification}, for a curve $\gamma$ in $\pi^*(T^*M\oplus E^*)$ (described by time-dependent $(m,e,p,e^*)$) we then have
        \begin{equation}\label{alphaontstare}
          \alpha(\dot\gamma)=\la (m,e)\dot{},\pi^*p+ie^*\ra=p(\dot m)+e^*(\on{ver}(m,e)\dot{}).
        \end{equation}
        
        Picking local coordinates $x^i$ on $M$ as well as linear coordinates $y^a$ on the fibres, we obtain the coordinates $(x^i,y^a,x_i^*,y_a^*)$ on $\pi^*(T^*M\oplus E^*)$, which at the point $(m,e,p,e^*)$ have values $(m^i,e^a,p_i,e^*_a)$. The derivative of a curve $\gamma$ in $\pi^*(T^*M\oplus E)$ is thus
        \begin{equation}
          \dot\gamma=\dot\gamma^i\partial_{x^i}+\dot\gamma^a\partial_{y^a}+\dot\gamma_i\partial_{x_i^*}+\dot\gamma_a\partial_{y_a^*}.
        \end{equation}
        From \eqref{horver} we get the following explicit formulas for the vertical projection:
        \begin{equation}
          \on{ver}\partial_{x^i}=A_i{}^a{}_{b}(x)y^b\partial_{y^a},\qquad \on{ver}\partial_{y^a}=\partial_{y^a}.
        \end{equation}
        Using these in \eqref{alphaontstare} yields
        \begin{equation}
          \alpha(\dot\gamma)=x_i^*\dot\gamma^i+y^*_a(A_i{}^a{}_{b}(x)y^b\dot\gamma^i+\dot\gamma^a),
        \end{equation}
        and thus
        \begin{equation}
        \begin{aligned}
          \alpha&=(x_i^*+y^*_aA_i{}^a{}_{b}(x)y^b)dx^i+y^*_ady^a \\
         & = x_i^* dx^i+y^*_a( dy^a + A^a{}_{b}(x)y^b) .
         \end{aligned}
        \end{equation}
        It follows that the symplectic form is
        \begin{equation} \label{eq:total-space-symp-form}
        \begin{aligned}
          \omega=d\alpha
          	&=dx_i^*\wedge dx^i+dy_a^*\wedge dy^a+d(y^*_aA_i{}^a{}_{b}(x)y^b)\wedge dx^i \\
	&= dx_i^*\wedge dx^i + (dy_a^* - A^b{}_a(x) y_b^*) \wedge (dy^a + A^a{}_{c}(x)y^c) 
		+ y_a^* F^a{}_b y^b.
         \end{aligned}
        \end{equation}
        Inverting this we get the formula for the Poisson structure on $\pi^*(T^*M\oplus E^*)$:
        \begin{equation}\label{poissonfirst}
          \pi=\partial_{x_i^*}\wedge\partial_{x^i}+\partial_{y_a^*}\wedge\partial_{y^a}+y_b^*A_i{}^b{}_{a}\partial_{x_i^*}\wedge \partial_{y_a^*}-A_i{}^a{}_{b}y^b\partial_{x_i^*}\wedge \partial_{y^a}-\tfrac12y_a^*y^bF_{ij}{}^a{}_b\partial_{x_i^*}\wedge\partial_{x_j^*}.
        \end{equation}
        
    \subsection{Proof part 2: the final change of variables}
      Fixing $(m,e,p,e^*)\in \pi^*(T^*M\oplus E^*)$, the linear map
        \begin{equation}\label{last}
          T_mM\oplus E_m\oplus T^*_mM\oplus E^*_m\to T_{(m,e,p,e^*)}\pi^*(T^*M\oplus E^*)
        \end{equation}
        is geometrically described as
        \begin{equation}\label{individual}
          \begin{aligned}
            T_mM\ni v &\mapsto \lambda^{\pi^*(T^*M\oplus E^*)\to E}_{(m,e)}\lambda^{E\to M}_e v\\
            E_m\ni \tilde e& \mapsto \lambda^{\pi^*(T^*M\oplus E^*)\to E}_{(m,e)}\dt(e+t\tilde e)\\
            T^*_mM\ni \beta&\mapsto \dt(m,e,p+t\beta,e^*)\\
            E^*_m\ni \tilde e^*&\mapsto \dt(m,e,p,e^*+t\tilde e^*),
          \end{aligned}
        \end{equation}
        where $\lambda^{F\to M}_fv$ denotes the horizontal lift of the vector $v$ to a point $f\in F$ in a vector bundle $F\to M$ with connection.
        
        Passing now to the coordinates $x^i$, $x_i^*$, $y^a$, $y_a^*$ and using \eqref{horver} we get explicitly
        \begin{equation}
          \begin{aligned}
            \lambda_e^{E\to M}&\colon \partial_{x^i}\mapsto \partial_{x^i}-A_i{}^a{}_{b}(x)y^b\partial_{y^a}\\
            \lambda_{(p,e^*)}^{T^*M\oplus E^*\to M}&\colon \partial_{x^i}\mapsto \partial_{x^i}+A_i{}^a{}_by_a^*\partial_{y^*_b}+\bar A_i{}^j{}_kx_j^*\partial_{x_k^*},
          \end{aligned}
        \end{equation}
        where $A$ and  $\bar A$ correspond to $\nabla$ and $\bar\nabla$, respectively. Since the total space of the bundle $\pi^*(T^*M\oplus E^*)\to E$ is in our coordinate system described simply by adjoining $y^a$ to the coordinates $x^i$, $x_i^*$, $y_a^*$ on the total space of $T^*M\oplus E^*\to M$, it follows that for the lift from $E$ to $\pi^*(T^*M\oplus E^*)$ we have
        \begin{equation}
          \lambda_{(p,e^*)}^{\pi^*(T^*M\oplus E^*)\to E}\colon \qquad \partial_{x^i}\mapsto \partial_{x^i}+A_i{}^a{}_by_a^*\partial_{y^*_b}+\bar A_i{}^j{}_kx_j^*\partial_{x_k^*},\qquad \partial_{y^a}\mapsto \partial_{y^a}.
        \end{equation}
        
        For the individual maps \eqref{individual} we then get
        \begin{equation}\label{linearidentifications}
          \begin{aligned}
            T_mM\ni \partial_{x^i} &\mapsto  \partial_{x^i}+A_i{}^a{}_by_a^*\partial_{y^*_b}+\bar A_i{}^j{}_kx_j^*\partial_{x_k^*}-A_i{}^a{}_{b}(x)y^b\partial_{y^a}\\
            E_m\ni \varepsilon_a& \mapsto \partial_{y^a}\\
            T^*_mM\ni dx^i&\mapsto \partial_{x^*_i}\\
            E^*_m\ni \varepsilon^a&\mapsto \partial_{y_a^*},
          \end{aligned}
        \end{equation}
        where $\varepsilon_a$, $\varepsilon^a$ are the frame and dual frame corresponding to the coordinates $y^a$ on the fibres of $E$. Setting $\varepsilon_i:=\partial_{x^i}$ and $\varepsilon^i:=dx^i$ (so that the LHS of \eqref{linearidentifications} takes a more unified form), the Poisson structure \eqref{poissonfirst} translated along the map \eqref{last} becomes
        \begin{equation}
          \pi=\varepsilon^i\wedge \varepsilon_i+\varepsilon^a\wedge \varepsilon_a-\bar{A}_i{}^j{}_kx_j^*\varepsilon^i\wedge \varepsilon^k-\tfrac12y_a^*y^bF_{ij}{}^a{}_b\varepsilon^i\wedge\varepsilon^j.
        \end{equation}
        Finally, identifying the third term as the torsion of $\bar\nabla$ via
        \begin{equation}
          \bar A_{[i}{}^j{}_{k]}=\tfrac12\bar T^j{}_{ik},
        \end{equation}
        we obtain the desired formula \eqref{poisson_on_bundle}.

\section{Variation of the generalised Lie derivative of spinors}
  We will now show that for a $V\in\Gamma(E)$, a change $\delta\gm$ in the generalised metric induces the following change of the Lie derivative on \emph{spinor half-densities}:
  \begin{equation} \label{eq:spinor-ld-variation}
    \delta(\ms L_V)=\tfrac1{16}[\delta\gm,\ms L_V\gm]_{ab}\gamma^{ab}=F(\delta\gm,\ms L_V\gm)\qquad\text{on }\Gamma(S(C_+)\otimes H).
  \end{equation}
  First, note that the only place where the metric dependence enters is the generalised Lie derivative of spinors --- the Lie derivative of half-densities is metric independent. Thus it is enough to show $\delta(\ms L_V)=\tfrac1{16}[\delta\gm,\ms L_V\gm]_{ab}\gamma^{ab}$ on $\Gamma(S(C_+))$, which is what we now turn to. We will adopt the same approach as in Appendix B.5 of \cite{Kupka:2024vrd}.
  
  We start with a trivialisation of our vector bundles induced by some orthonormal frame $e_A=\{e_a,e_\alpha\}$. In particular we have
  \begin{equation}\label{trivialisation}
    \Gamma(S(C_+))\cong C^\infty(M)\otimes S_0,
  \end{equation}
  where $S_0$ is the \emph{vector space} of spinors. Changing the generalised metric infinitesimally to $\gm+\delta\gm$ the orthonormal frame gets deformed by
  \begin{equation}
    \delta e_a=\tfrac12\delta\gm_a{}^\alpha e_\alpha,\qquad \delta e_\alpha=-\tfrac12\delta\gm_\alpha{}^ae_a.
  \end{equation}
  The relevant connection coefficients change \cite{Kupka:2024vrd} as
  \begin{equation}
    \begin{aligned}
    \delta(\Gamma_{\alpha bc})&=-\tfrac12\delta\gm_\alpha{}^a\Gamma_{abc}-\partial_{[b}\delta\gm_{c]\alpha}-\Gamma_{[bc]}{}^a\gm_{a\alpha}-\Gamma_{[b|\alpha}{}^\beta\gm_{|c]\beta},\\
    \delta(\Gamma_{[abc]})&=\tfrac12\delta\gm_{[a|}{}^\alpha\Gamma_{\alpha|bc]}.
    \end{aligned}
  \end{equation}
  Here we combined both the nontrivial transformation properties of the connection coefficients under the change of frame, as well as the change of the connection itself under the variation of the generalised metric.
  
  With the identification \eqref{trivialisation} in mind we can now calculate directly that on spinors we have
  \begin{align*}
      \delta(\ms L_V)&=\delta(D_V+\tfrac12(D_aV_b)\gamma^{ab})\\
      &=\delta(V^Aa(e_A)+\tfrac14V^A\Gamma_{Abc}\gamma^{bc}+\tfrac12(a(e_a)V_b+\Gamma_{ab}{}^cV_c)\gamma^{ab})\\
      &=\delta(V^Aa(e_A)+\tfrac12(a(e_a)V_b)\gamma^{ab}+\tfrac14V^\alpha\Gamma_{\alpha bc}\gamma^{bc}+\tfrac34V^a\Gamma_{[abc]}\gamma^{bc})\\
      &=\tfrac12(a(\delta e_a)V_b+a(e_a)\delta V_b)\gamma^{ab}+\tfrac14(\delta V^\alpha\Gamma_{\alpha bc}+V^\alpha \delta \Gamma_{\alpha bc})\gamma^{bc}+\tfrac34(\delta V^a\Gamma_{[abc]}+V^a\delta\Gamma_{[abc]})\gamma^{bc}\\
      &=\tfrac14(\delta\gm_a{}^\alpha\partial_\alpha V_b+\partial_a(\delta\gm_b{}^\alpha V_\alpha))\gamma^{ab}-\tfrac18\delta\gm^\alpha{}_aV^a\Gamma_{\alpha bc}\gamma^{bc}\\
      &\qquad+\tfrac14V^\alpha (-\tfrac12\delta\gm_\alpha{}^a\Gamma_{abc}-\partial_{b}\delta\gm_{c\alpha}-\Gamma_{bc}{}^a\delta\gm_{a\alpha}-\Gamma_{b\alpha}{}^\beta\delta\gm_{c\beta})\gamma^{bc}\\
      &\qquad+\tfrac38(\delta\gm^a{}_\alpha V^\alpha\Gamma_{[abc]}+V^a\delta\gm_{[a|}{}^\alpha\Gamma_{\alpha|bc]})\gamma^{bc}\lbl{calculation-change-derivative}.
  \end{align*}
  The terms $\partial\gm$ cancelling, we expand the antisymmetrisations in the last line to get
  \begin{equation}
    \begin{aligned}
      \delta(\ms L_V)&=\tfrac14(\delta\gm_a{}^\alpha\partial_\alpha V_b+\delta\gm_b{}^\alpha \partial_a V_\alpha)\gamma^{ab}\\
      &\qquad +\tfrac14(-\tfrac12\delta\gm^\alpha{}_aV^a\Gamma_{\alpha bc}-\tfrac12V^\alpha\delta\gm_\alpha{}^a\Gamma_{abc}-V^\alpha\Gamma_{bc}{}^a\delta\gm_{a\alpha}-V^\alpha\Gamma_{b\alpha}{}^\beta\delta\gm_{c\beta}\\
      &\qquad\qquad +\tfrac12\delta\gm^a{}_\alpha V^\alpha\Gamma_{abc}+\delta\gm^a{}_\alpha V^\alpha\Gamma_{bca}+\tfrac12V^a\delta\gm_{a}{}^\alpha\Gamma_{\alpha bc}+V^a\delta\gm_{b}{}^\alpha\Gamma_{\alpha ca})\gamma^{bc}\\
      &=\tfrac14(\delta\gm_b{}^\alpha\partial_\alpha V_c-\delta\gm_b{}^\alpha\partial_c V_\alpha-V^\alpha\Gamma_{b\alpha}{}^\beta\delta\gm_{c\beta}+V^a\delta\gm_{b}{}^\alpha\Gamma_{\alpha ca})\gamma^{bc}\\
      &=\tfrac14\delta\gm_b{}^\alpha (D_\alpha V_c-D_c V_\alpha)\gamma^{bc}=\tfrac18\delta\gm_b{}^\alpha(\ms L_V\gm)_{c\alpha}\gamma^{bc}=\tfrac1{16}[\delta\gm,\ms L_V\gm]_{ab}\gamma^{ab},
    \end{aligned}
  \end{equation}
  where we used \eqref{killing}. This concludes the proof.

  \bibliography{citations}
  \bibliographystyle{JHEP}
\end{document}